\documentclass[twocolumn]{aastex631}
\usepackage{amsmath}

\defcitealias{proxaufObservationsLargescaleSolar2021}{P21}
\defcitealias{greerHELIOSEISMICIMAGINGFAST2015}{GHFT15}
\defcitealias{hottaGenerationSolarlikeDifferential2022}{HKS22}
\defcitealias{hanasogeAnomalouslyWeakSolar2012}{HDS12}

\def\p{\partial}
\def\vv{{\mbox{\boldmath $v$}}}
\def\nab{\mbox{\boldmath $\nabla$}}
\def\oom{\mbox{\boldmath $\Omega$}}
\def\rb{\bar{\rho}}
\def\tb{\bar{T}}

\def\sba{\bar{s}}

\newcommand{\MAlm}[1]{\mathcal{A}^{l_{#1}}_{m_{#1}}}
\newcommand{\MBlm}[1]{\mathcal{B}^{l_{#1}}_{m_{#1}}}
\newcommand{\MClm}[1]{\mathcal{C}^{l_{#1}}_{m_{#1}}}

\newcommand{\sumlm}[1]{\sum_{l_{#1}=0}^{\infty}\sum_{m_{#1} = - l_{#1}}^{l_{#1}}}
\newcommand{\dr}{\partial_r}
\newcommand{\drr}{\partial^2_{rr}}
\newcommand{\dth}{\partial_\theta}
\newcommand{\dphi}{\partial_\varphi}
\newcommand{\bnab}{\boldsymbol{\nabla}}
\newcommand{\er}{\mathbf{e}_r}
\newcommand{\ethe}{\mathbf{e}_\theta}
\newcommand{\ephi}{\mathbf{e}_\varphi}
\newcommand{\rota}{\bnab\times}
\newcommand{\Div}{\bnab\cdot}

\newcommand{\grad}{\bnab}
\newcommand{\gradperp}{\bnab_\perp}

\newcommand{\dint}[2]{\hspace{0.2cm}\mbox{d}^{#2}{#1}}

\newcommand{\Rlm}[1]{\mathbf{R}^{m_{#1}}_{l_{#1}}}
\newcommand{\Slm}[1]{\mathbf{S}^{m_{#1}}_{l_{#1}}}
\newcommand{\Tlm}[1]{\mathbf{T}^{m_{#1}}_{l_{#1}}}
\newcommand{\Ylm}[1]{Y_{l_{#1}}^{m_{#1}}}

\begin{document}

\title{Global Turbulent Solar Convection: a Numerical Path Investigating Key Force Balances\\in the context of the Convective Conundrum}

\correspondingauthor{Quentin NORAZ}
\email{quentin.noraz@astro.uio.no}

\author[0000-0002-7422-1127]{Quentin NORAZ}
\affiliation{Rosseland Centre for Solar Physics, University of Oslo, P.O. Box 1029 Blindern, Oslo, NO-0315, Norway}
\affiliation{Institute of Theoretical Astrophysics, University of Oslo, P.O.Box 1029 Blindern, Oslo, NO-0315, Norway}

\author[0000-0002-1729-8267]{Allan Sacha BRUN}
\affiliation{D\'epartement d'Astrophysique/AIM\\
CEA/IRFU, CNRS/INSU, Univ. Paris-Saclay \& Univ. de Paris\\
91191 Gif-sur-Yvette, France}

\author[0000-0002-9630-6463]{Antoine STRUGAREK}
\affiliation{D\'epartement d'Astrophysique/AIM\\
CEA/IRFU, CNRS/INSU, Univ. Paris-Saclay \& Univ. de Paris\\
91191 Gif-sur-Yvette, France}



\begin{abstract}

Understanding solar turbulent convection and its influence on differential rotation has been a challenge over the past two decades. Current models often overestimate giant convection cells amplitude, leading to an effective Rossby number too large and a shift towards an anti-solar rotation regime. This Convective Conundrum, underscores the need for improved comprehension of solar convective dynamics. We propose a numerical experiment in the parameter space that controls $Ro$ while increasing the Reynolds number ($Re$) and maintaining solar parameters. By controlling the Nusselt number ($Nu$), we limit the energy transport by convection while reducing viscous dissipation. This approach enabled us to construct a Sun-like rotating model (SBR97n035) with strong turbulence ($Re \sim 800$) that exhibits prograde equatorial rotation and aligns with observational data from helioseismology. We compare this model with an anti-solar rotating counterpart, and provide an in-depth spectral analysis to investigate the changes in convective dynamics. We also find the appearance of vorticity rings near the poles, which existence on the Sun could be probed in the future. The Sun-like model shows reduced buoyancy over the spectrum, as well as an extended quasi-geostrophic equilibrium towards smaller scales. This promotes a Coriolis-Inertia (CI) balance rather than a Coriolis-Inertia-Archimedes (CIA) balance, in order to favor the establishment of a prograde equator. The presence of convective columns in the bulk of the convection zone, with limited surface manifestations, also hints at such structures potentially occurring in the Sun.

\end{abstract}



\section{Introduction} \label{sec:intro}

\subsection{Models and observations of global solar convection and its associated differential rotation}

\begin{figure*}
  \centering
  \includegraphics[width=\linewidth]{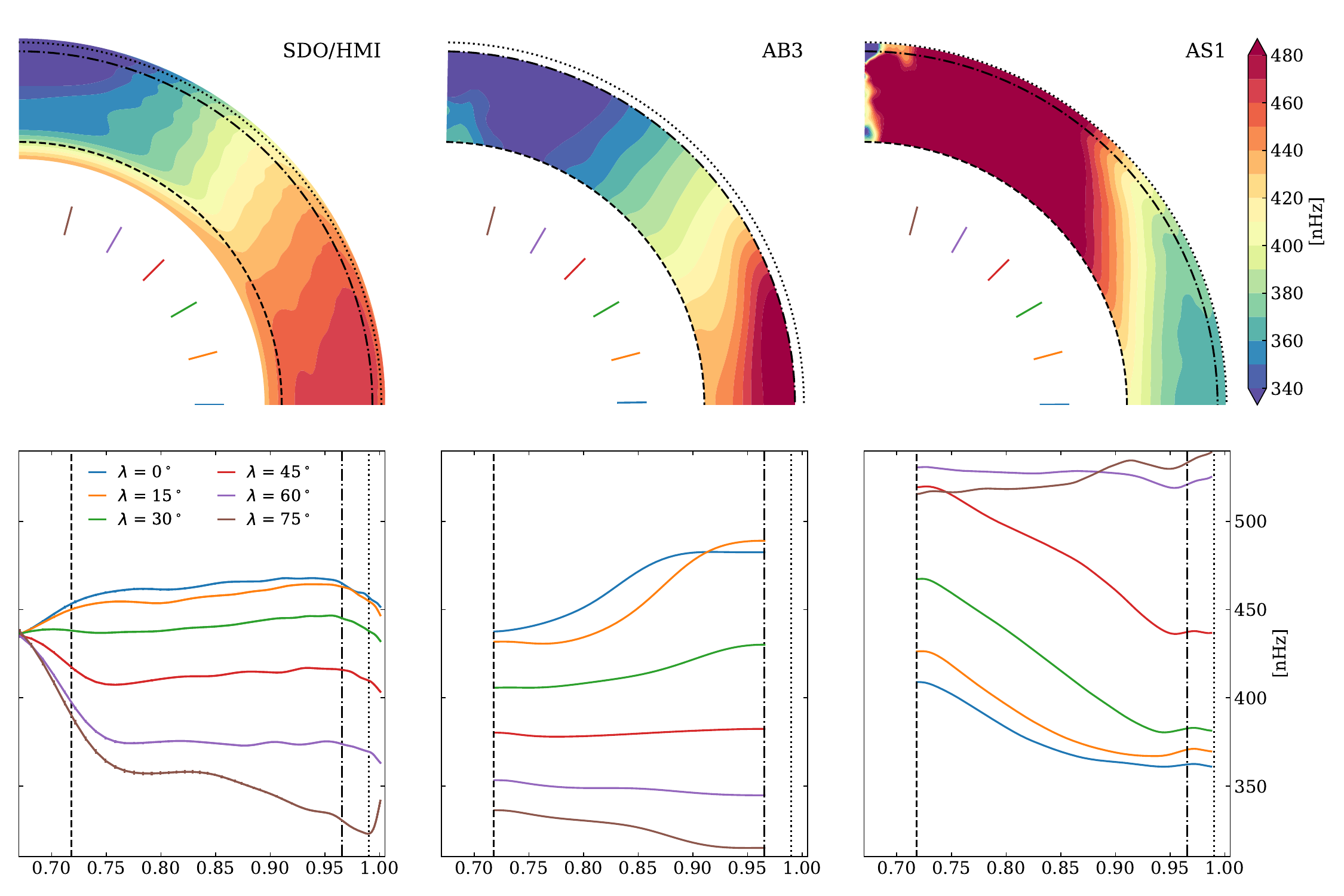}
  \caption{Northern meridional plane (\textit{top}) and radial profiles (\textit{bottom}) of the angular velocity in nHz, inverted from observations of the Helioseismic and Magnetic Imager onboard the Solar Dynamics Observatory satellite, from 0.67 $R_\odot$ to the solar surface (left, \citealt{larsonGlobalModeAnalysisFullDisk2018}). Same quantity taken from the AB3 and AS1 solar convective envelope models, which ran with the ASH code (respectively \textit{middle}, \citealt{mieschSolarDifferentialRotation2006}, and \textit{right}, \citealt{emeriau-viardTurbulencePlasmaDans}). Different color markers indicate in the top panels the selected latitudes for the bottom ones. The dashed line shows the bottom of AB3 and AS1 models, while dot-dashed and dotted lines stand for the top of AB3 (0.965 $R_\odot$) and AS1 (0.99 $R_\odot$) respectively.}\label{fig:Fig1}
\end{figure*}

The solar convection zone (CZ) is differentially rotating, meaning that all latitudes at the surface do not rotate at the same rate. In particular for the Sun, the equatorial region rotates faster than the polar regions, and more specifically, the former (respectively latter) is prograde (respect. retrograde), meaning that it rotates faster (respect. slower) than the mean solar rotation rate $\Omega_\odot$. This was first discovered centuries ago by tracking surface magnetic activity \citep{1630rour.book.....S,1860MNRAS..20..254C}, and then characterized in deeper layers with helioseismology \citep{thompsonDifferentialRotationDynamics1996,howeFirstGlobalRotation2011} as illustrated in the left panel of Figure~\ref{fig:Fig1}. This profile transits towards solid-body rotation when reaching the radiative zone (RZ) around $0.7$~R$_\odot$, where the Schwarzschild criterion \citep{1906WisGo.195...41S} does not hold (see \textit{e.g.} \citealt{brummellTurbulentCompressibleConvection1998,brunMODELINGDYNAMICALCOUPLING2011}). The transition region separating the CZ from the RZ is referred to as the \textit{tachocline} due to the strong velocity gradients present \citep{1992A&A...265..106S,2023SSRv..219...87S}.

Studies of turbulent rotating bodies were historically performed with the complementary effort of linear analysis of convection in a rotating sphere \citep{1961hhs..book.....C,1968RSPTA.263...93R,busseThermalInstabilitiesRapidly1970} and numerical simulations using non-linear global models \citep{1975JAtS...32.1331G,1977GApFD...8...93G,gilmanModelCalculationsConcerning1979}. The incorporation of realistic solar stratification \citep{christensen-dalsgaardCurrentStateSolar1996} and the anelastic approximation in numerical models \citep{gilmanCOMPRESSIBLECONVECTIONROTATING1981,1982ApJ...256..316G} have historically highlighted the role of turbulent convection in redistributing the angular momentum and enabled a more faithful reproduction of the solar rotation profile in numerical simulations, as illustrated in the middle panel of Figure~\ref{fig:Fig1} with the historical AB3 case \citep{mieschThreeDimensionalSpherical2000,brunTurbulentConvectionInfluence2002,mieschSolarDifferentialRotation2006}. However, the spatial resolutions of such global models were moderate and dynamical characteristic time of convection were relatively long in comparison to what is achieved in the solar turbulence regime.

In the relatively large parameter space explored by numerical modelling so far \citep{hindmanMorphologicalClassificationConvective2020,brunPoweringStellarMagnetism2022}, the morphology of large-scale flows has been shown to be strongly dependent on the Rossby number $Ro$, defined as the ratio between the amplitude of non-linear advection and that of the rotational constraint by the Coriolis force (see also \citealt{brunMagnetismDynamoAction2017} and \citealt{hottaDynamicsLargeScaleSolar2023} for reviews). In particular, global models experience a latitudinal reversal of the differential rotation (DR) profile when $Ro$ exceeds a value of the order of unity, leading to the establishment of the so-called \textit{anti-solar} rotation profile, where polar regions rotate faster than the equator \citep{1977GApFD...8...93G,gastineSolarlikeAntisolarDifferential2014,brunDifferentialRotationOvershooting2017,hottaGenerationSolarlikeDifferential2022,kapylaTransitionAntisolarSolarlike2023}. The desire to use the constant improvement in computing power to get closer to the solar regime is now leading the community to increase the degree of turbulence in global convective simulations, quantified by the Reynolds number $Re$. This is generally achieved by increasing spatial resolution, along with decreasing diffusive coefficients, in order to resolve the turbulent dynamics on a larger range of convective scales (see for instance \citealt{mieschStructureEvolutionGiant2008}). However, it also results in an increase of the amplitude of convective velocities, which then increases the Rossby number $Ro$ so much that the vast majority of current turbulent global solar models eventually present an anti-solar DR profile. We illustrate such an example with the AS1 model \citep{emeriau-viardTurbulencePlasmaDans} in the right panels of Figure~\ref{fig:Fig1}, which shows a differential rotation with a slow equator and a fast pole, in contradiction with helioseismic constraints. 

\subsection{Convective conundrum context}
This inability to reproduce the solar DR in the vast majority of global solar turbulent simulations is part of a paradox known as the \textit{convective conundrum} \citep{omaraVelocityAmplitudesGlobal2016, hottaDynamicsLargeScaleSolar2023}. More specifically, the loss of the prograde equator when aiming at getting closer to the solar regime, now questions which scales and regime are important for the establishment of the solar-like rotation profile.

\begin{figure}
  \centering
  \includegraphics[width=\linewidth]{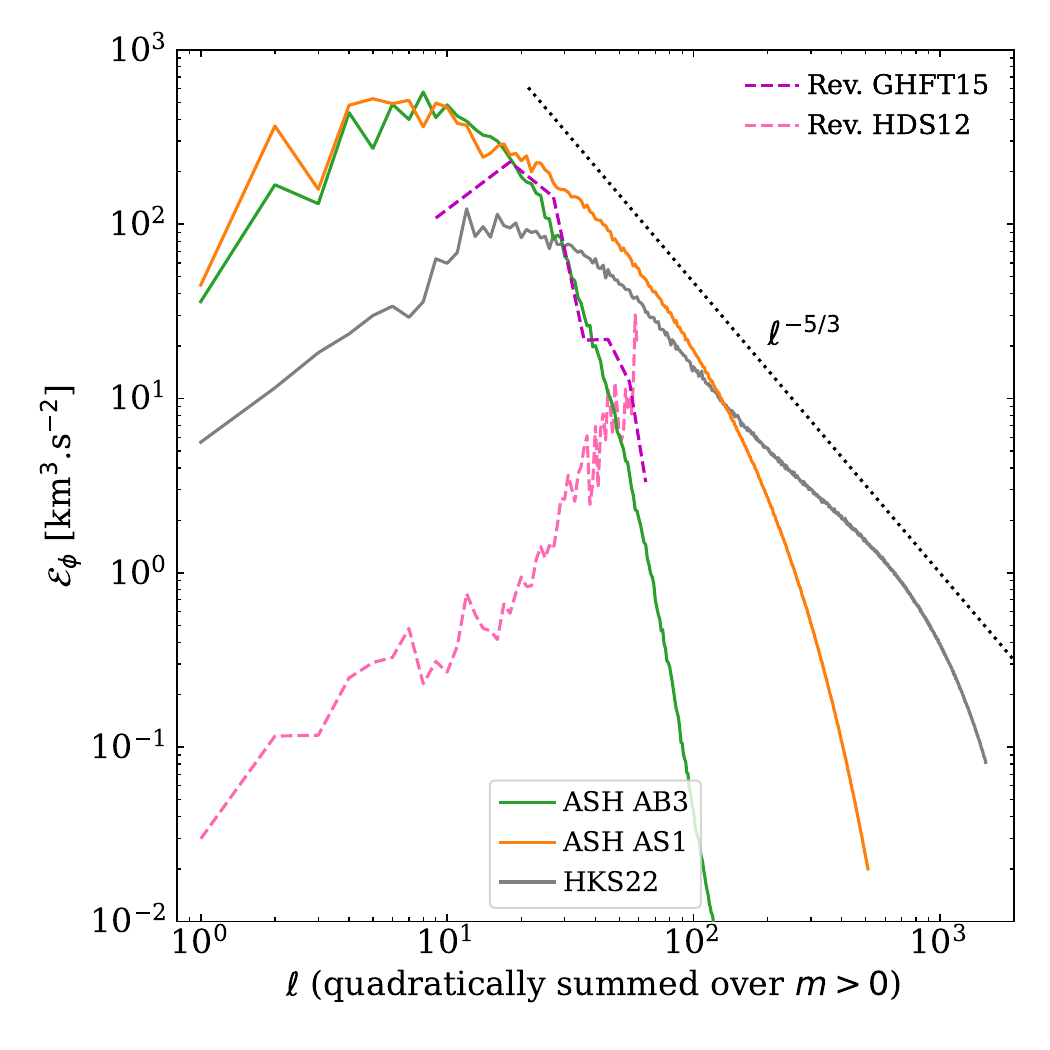}
  \caption{\textit{Left}: Comparison of the non-axisymmetric part of toroidal kinetic spectra per multiplet ${\cal E}_\phi$ at $0.96~R_\odot$, defined following the definition of \cite{gizonHelioseismologyChallengesModels2012}. Solid, and dashed lines represent data from numerical simulations and observations, respectively. Green and orange curves are taken from global anelastic models AB3 and AS1 respectively (see Figure~\ref{fig:Fig1}). The gray line represents the highly resolved global model from \cite{hottaGenerationSolarlikeDifferential2022}, considering magnetic field and compressible effects. Pink and magenta dashed lines show deep-focusing time-distance helioseismic measurements \citep{hanasogeAnomalouslyWeakSolar2012} and multi-ridge fitting ring-diagram analysis \citep{greerHELIOSEISMICIMAGINGFAST2015}, both revised by \cite{proxaufObservationsLargescaleSolar2021}. These measurements are considered today as an upper and lower bound from observations. The weighted degree $\ell_{\rm pond}=\sum {\cal E}_\phi(\ell)*\ell/\sum {\cal E}_\phi(\ell)$ is 14, 35 and 143 for global simulations AB3, AS1 and HKS22 spectra, respectively. Similarly, the integral scale $L_{\rm int}=\sum {\cal E}_\phi(\ell)*r/\sqrt(\ell(\ell+1))/\sum {\cal E}_\phi(\ell)$ is 9.8, 7.5 and 2.9\% of the solar radius, respectively, which corresponds to $\ell_{\rm int}= 9$, 12 and 33. A dotted line representing a theoretical Kolmogorov trend ($\ell^{-5/3}$) is indicated for comparison.}\label{fig:Fig2}
\end{figure}

The energy distribution among the turbulent convective spectrum can be inferred by decomposing velocity maps spectrally. This is straightforward in numerical models where velocities are known everywhere. However, current global modeling does not include the solar surface, as the gap in resolution will be too wide between the solar radius and the convective granulation developed at the surface. Nevertheless, velocities can also be probed few Mm under the surface of the Sun through helioseismic inversions and decomposed spectrally. We thus gather in Figure~\ref{fig:Fig2} convective spectra for $0.96$~R$_\odot$, using dashed curves for observations (see \citealt{proxaufObservationsLargescaleSolar2021}, hereafter P21, for details about them) and solid lines for numerical models (previous AB3 and AS1 models are in green and orange respectively, see \citealt{mieschSolarDifferentialRotation2006} and \citealt{emeriau-viardTurbulencePlasmaDans}).

Despite the convergence of various observations around $\ell\sim 50$, observational constraints diverge and span a large range of amplitudes at larger-scales. Indeed, the revision of \cite{hanasogeAnomalouslyWeakSolar2012} time-distance data (HDS12, dashed pink) reports low amplitudes, which can be down to two orders of magnitude with respect to independent time-distance analysis from \cite{greerHELIOSEISMICIMAGINGFAST2015} (GHFT15, dashed magenta). Both constraints are currently questioned in the observational community: on one hand the 3D-inversion of \citetalias{greerHELIOSEISMICIMAGINGFAST2015} could over-estimate the signal, and on the other hand the noise model of (\citetalias{hanasogeAnomalouslyWeakSolar2012}) could under-estimate it (see also Birch et al. in prep). In that sense, the amplitudes of the largest solar convection scales at $0.96$~R$_\odot$ and below are still debated.

Numerical modeling of the largest scales in global models are then a powerful tool to investigate current observational disagreements. When adding to Figure~\ref{fig:Fig2} spectra from numerical simulations (solid green and orange lines, coming from the solar-like AB3 and anti-solar AS1 models respectively), we see they are close to amplitudes observed by \citetalias{greerHELIOSEISMICIMAGINGFAST2015}, but are most likely also over-estimating large-scale amplitudes, no matter if it develops a solar-like (green line) or anti-solar DR profile (orange line). We also add data from the recent work on global modelling by \cite{hottaGenerationSolarlikeDifferential2022} (HKS22, black solid line), where a solar-like DR profile is achieved in an unprecedented high turbulence regime. Earlier work indeed already underlined that the magnetism can play an important role for the construction of the differential profile \citep{fanSIMULATIONCONVECTIVEDYNAMO2014}. However, its role stays currently unclear, as different conclusions are drawn from the different studies (see also \citealt{warneckeSmallscaleLargescaleDynamos2024}). Hence, we will aim here at finding if sustaining a Sun-like DR is possible at large turbulence degree without invoking magnetic effects.

This overall mismatch between various solar helioseismic inversions, and also with global convection simulations, is inherent to the convective conundrum and questions their role for the construction of the solar-like DR. On one hand, observational inputs show that equatorward angular-momentum transport is likely favored by giant-cells dynamics \citep{hathawayGiantConvectionCells2013}. These large scales were solely resolved in earlier solar-like rotating models (\citealt{mieschSolarDifferentialRotation2006}, see Figure~\ref{fig:Fig1}), which partly explains why they could sustain a prograde equator. But on the other hand, numerical modeling illustrates that small-scale dynamics also plays an important role, as sustaining a prograde equator becomes challenging when the range of scales we resolve is extended. This leads to the following question: what favors the construction of a solar-like DR? Regarding this aspect, several theories are currently debated, and mainly two mechanism are advocated to ensure an equatorward angular-momentum transport: (i) Reynolds stress transport by inertial modes \citep{brunInteractionDifferentialRotation2004,rastDecipheringSolarConvection2020} and (ii) Maxwell stress transport via the \textit{Punching-ball} effect (\citetalias{hottaGenerationSolarlikeDifferential2022}). Nevertheless, all agree on the importance of the rotational constraint through the Coriolis force in the dynamics (see also \citealt{kapylaReynoldsStressHeat2011,moriScaledependentAnalysisAngular2023}). It is therefore important to understand and model the correct solar-like force and power balances along the convective turbulence cascade. The net energy transported by convection in 3D global models has not been much questioned so far, so that the Nusselt number was mostly left as a free parameter. We here propose to control it, in order to recover the solar dynamics despite the limited range of scales accessible with the current computational power.

\subsection{Path approach}\label{sec:pathTheo}

Recently, the geophysics community has proposed to follow a path in parameter space in order to obtain more realistic geodynamo simulations that are closer to the Earth's state \citep{aubertSphericalConvectiveDynamos2017}. They have to some extent succeeded in doing so by rescaling strategically the different diffusivities of the problem, in order to keep the model tractable, while conserving the important force balances happening in the Earth's core \citep{aubertInterplayFastWaves2021}. Independently, we have also designed in this work a numerical path in order to respect what we believe to be the main force balances, as an attempt to build highly turbulent models of solar convection and large-scale flows.

In that sense, we propose a way to increase the degree of turbulence of the solar convection simulation while retaining the key forces balance, \textit{e.g.} that deep convection still feels the influence of rotation (\textit{i.e.} that the dynamics are characterized by a Rossby number smaller than one, \citealt{mattConvectionDifferentialRotation2011,gastineSolarlikeAntisolarDifferential2014,brunDifferentialRotationOvershooting2017}). In that sense, we have chosen to follow a numerical path that maintains the Rossby number $Ro$ constant while increasing the Reynolds number $Re$. These two numbers can easily be defined by comparing the following ratio of time scales. The Rossby number is the ratio of the rotation timescale $\tau_{\Omega}=1/(2\Omega)$ and the advective timescale $\tau_{u}=\ell_{c}/u$ with $\ell_{c}$ a characteristic convection scale and $u$ the convective speed, \textit{i.e.} $Ro = \tau_{\Omega}/\tau_{u}$. Likewise, the Reynolds number is the ratio of the inertia timescale to the viscous diffusion one $\tau_{\nu}= \ell_{c}^2/\nu$, \textit{i.e.} $Re=\tau_{\nu}/\tau_{u}$. So, in order to increase $Re$, we have to ensure that $\tau_{\nu}$ increases faster than any increase in $\tau_{u}$. This is of course what most simulation attempts to model the Sun have been doing \citep{mieschStructureEvolutionGiant2008}. However, increasing $\tau_{\nu}$ has an indirect consequence, it also impacts $u$ and hence $\tau_{u}$.

Stratified turbulent convection experiments show that kinetic energy (KE) increases as the Rayleigh ($Ra$) and Reynolds ($Re$) numbers grow (see for instance Tables 4 to 7 in \citealt{featherstoneSPECTRALAMPLITUDESTELLAR2016}), eventually reaching an asymptote known as the diffusion-free regime. We must consider this when moving toward more turbulent regimes while maintaining a prograde equator. Typically, outside the diffusion-free regime, $u\propto\nu^{-\alpha}$ with $\alpha>0$. Notably, $\alpha$ depends on the viscosity profile and density contrast $N_\rho$ \citep{currieMagnitudeViscousDissipation2017,lanceViscousDissipationDynamics2024}.

Since we aim to maintain $Ro = \tau_{\Omega}/\tau_{u} = \text{const}$, one could then adjust $\tau_{\Omega}$ to decrease it proportionally to the aforementioned $\tau_u$ change. However, the solar rotation rate $\Omega_{\odot}$ is well-known, so ideally $\tau_{\Omega}$ should remain constant. Some studies have simulated in that way faster rotating Suns \citep{brownRapidlyRotatingSuns2008,emeriau-viardTurbulencePlasmaDans,strugarekReconcilingSolarStellar2017,warneckeSmallscaleLargescaleDynamos2024}, but instead we propose here to adjust the convective driving, \textit{i.e.} to offset the decrease of $\tau_u$ due to the increase of $\tau_{\nu}$, in order to keep $Ro$ constant.

In hydrodynamical convection setups, the convective velocity $u$ is mainly controlled by thermal forcing. One straightforward way to reduce it is to reduce the solar luminosity $L_{\odot}$ (as in \citealt{hottaHIGHRESOLUTIONCALCULATIONSOLAR2014,bekkiTheorySolarOscillations2022a}, reduced by a factor of 18 and 20, respectively), but since $L_{\odot}$ is an observational constraint, we prefer to maintain it. We can then control the amount of energy carried by convection. Classical mixing-length theory (MLT, \citealt{1958ZA.....46..108B}) assumes that the convective luminosity $L_c \propto \langle v'T' \rangle$ equals or exceeds $L_{\odot}$ to compensate for the inward kinetic energy luminosity $L_{KE} \propto \rho v^3$ developing in turbulent stratified setups \citep{cattaneoTurbulentCompressibleConvection1991,mieschStructureEvolutionGiant2008}, which yields $L_c \propto L_{KE} \propto \rho v^3$.

The Nusselt number $Nu = (L_{\rm diff} + L_{c})/L_{\rm diff}$ measures the ratio of total energy transport (diffusion and convection) to that by diffusion alone. Most solar convection simulations have assumed large $Nu$ and resulted in anti-solar rotation states due to excessive buoyancy driving at large convection scales. To compensate for the increase of $u$ due to the decreased viscosity (higher $Re$), we can then control $Nu$ such that we lower the energy the convection needs to transport, while keeping $L_{\odot}$ ($\simeq L_{\rm diff} + L_{c} + L_{KE}$, to first order in the CZ). In other words, assuming $u \propto L_c^{1/3} \nu^{-\alpha}$, there exists a $Nu$ that allows $u$ to remain constant while increasing $Re$. This is the parameter space path we propose.\\

In this comparative study, we present a numerical experiment achieving high turbulence in a global rotating solar convection model, while maintaining a solar-like differential rotation profile and convective velocity comparable to solar observations. We also compare this model with an anti-solar rotating control case, in order to investigate changes in convective dynamics and their implications. Section~\ref{sec:Sect2} presents the construction of our simulations. We describe their overall dynamics in Section~\ref{sec:Overview} and discuss our numerical experiment in its solar context in Section~\ref{sec:sol_comp}. In Section~\ref{sec:SpectralAna}, we perform an in-depth spectral analysis to understand the balances achieved along the convective turbulent cascade. We finally discuss possible improvements and prospects in Section~\ref{sec:discuss}, before concluding in Section~\ref{sec:ccl}.

\section{Method \& Models}\label{sec:Sect2}

Simulations performed in this study are computed with the ASH code, described with the set-up and equations in Appendix~\ref{sec:num_method}. We solve the hydrodynamics equations assuming the anelastic approximation, under which the thermodynamic variables are linearized around a spherically symmetric reference state $\bar{x}$, where $\bar{x}$ can be the density $\bar{\rho}(r,t)$, pressure $\bar{P}(r,t)$, temperature $\bar{T}(r,t)$ and specific entropy $\bar{s}(r,t)$. Fluctuations $x$ around this reference state $\bar{x}$ are then denoted by $\rho$, $P$, $T$ and $S$, such as $x_{\rm tot}=\bar{x}(r,t)+x(r,\theta,\phi,t)$.\\

\subsection{Energy transport}\label{sec:energ-transp}

In order to determine and quantify the different processes carrying the energy throughout the stellar interior, we consider here the conservation of the total energy density $E_{\rm tot}=E_k+E_{\rm int}$. The conservation of kinetic energy density $E_k = \rb\vv^2/2$ can be obtained with the scalar product between $\vv$ and the anelastic momentum Equation~\ref{eq:ASHeq_v}, while conservation of internal energy $E_{\rm int} = \rb\tb s$ is directly accessible with the energy evolution Equation~\ref{eq:ASHeq_S}. Combining both gives
\begin{eqnarray}
  \label{eq:EtotEvol}
  \frac{\partial (E_k + E_s)}{\partial t} = \nab\cdot\mbox{\boldmath $\cal F$} = - \nab\cdot\left[ \vv\left( \frac{\rb\vv\cdot\vv}{2} \right) - \vv\cdot\mbox{\boldmath $\cal D$}\right.\nonumber\\ \left. + c_P\rb T\vv - \mbox{\boldmath $q$} + \bar{\rho}\vv\bar{Q}\right],
\end{eqnarray}
(see \citealt{derosaDynamicsUpperSolar2001} or \citealt{mieschLargeScaleDynamicsConvection2005} for the detailed derivation) where \mbox{\boldmath $\cal F$} is the total flux transported, \mbox{\boldmath $\cal D$} the viscous stress tensor, \mbox{\boldmath $q$} the energy flux, and $\bar{Q}$ the reference state heating, defined such as
\begin{equation}
  {\cal D}_{ij} = 2\bar{\rho} \nu \left[ e_{ij} - \frac{1}{3}\nab \cdot \vv \delta_{ij} \right],
\label{eq:tenseurD}
\end{equation}
\begin{eqnarray}
  \label{eq:EnergFlux}
  \mathbf{q} &=& \kappa_{\rm rad} \bar{\rho} c_{p} \nab (\bar{T} + T)\nonumber\\ &+& \kappa\bar{\rho}\bar{T} \nab (s)\Big|_{l>0} + \kappa_{0}\bar{\rho}\bar{T} \frac{\partial (\bar{s} + \langle s\rangle)}{\partial r}\Big|_{l=0} \hat{\bf e}_r,
\end{eqnarray}
\begin{equation}
  \label{eq:EnergFluxQ}
  \nab\cdot\bar{Q}=\frac{d\bar{Q}}{dr}\hat{\bf e}_r=\bar{T}\frac{d\bar{s}}{dr}\hat{\bf e}_r,
\end{equation}
where $\langle\;\rangle$ denotes the spherical average. The energy flux $\mathbf{q}$ is expressed, respectively from left to right, as the sum of the radiative flux, entropy conductive flux and the unresolved flux related to the sub-grid treatment of our LES-SGS approach. Diffusive coefficients $\nu$ and $\kappa$ are respectively the kinematic viscosity and thermal diffusion. They represent the diffusive transport of momentum and heat by microscopic and turbulent interactions that are not resolved by the simulation (see Appendix~\ref{sec:app_num_set} for their detailed parametrization). We also note $\kappa_0$ the effective diffusion coefficient applied on the spherically symmetrical component ($l=0$,$m=0$) of the entropy gradient. It parametrizes with a diffusive approach a net radial flux representing the macroscopic transport by unresolved flows near the surface \citep{gilmanCOMPRESSIBLECONVECTIONROTATING1981,cluneComputationalAspectsCode1999,brunTurbulentConvectionInfluence2002}. Finally, $\kappa_{\rm rad}$ is the radiative diffusion coefficient, parametrizing radiative transfer in the stellar interior.

\begin{figure}
  \centering
  \includegraphics[width=\linewidth]{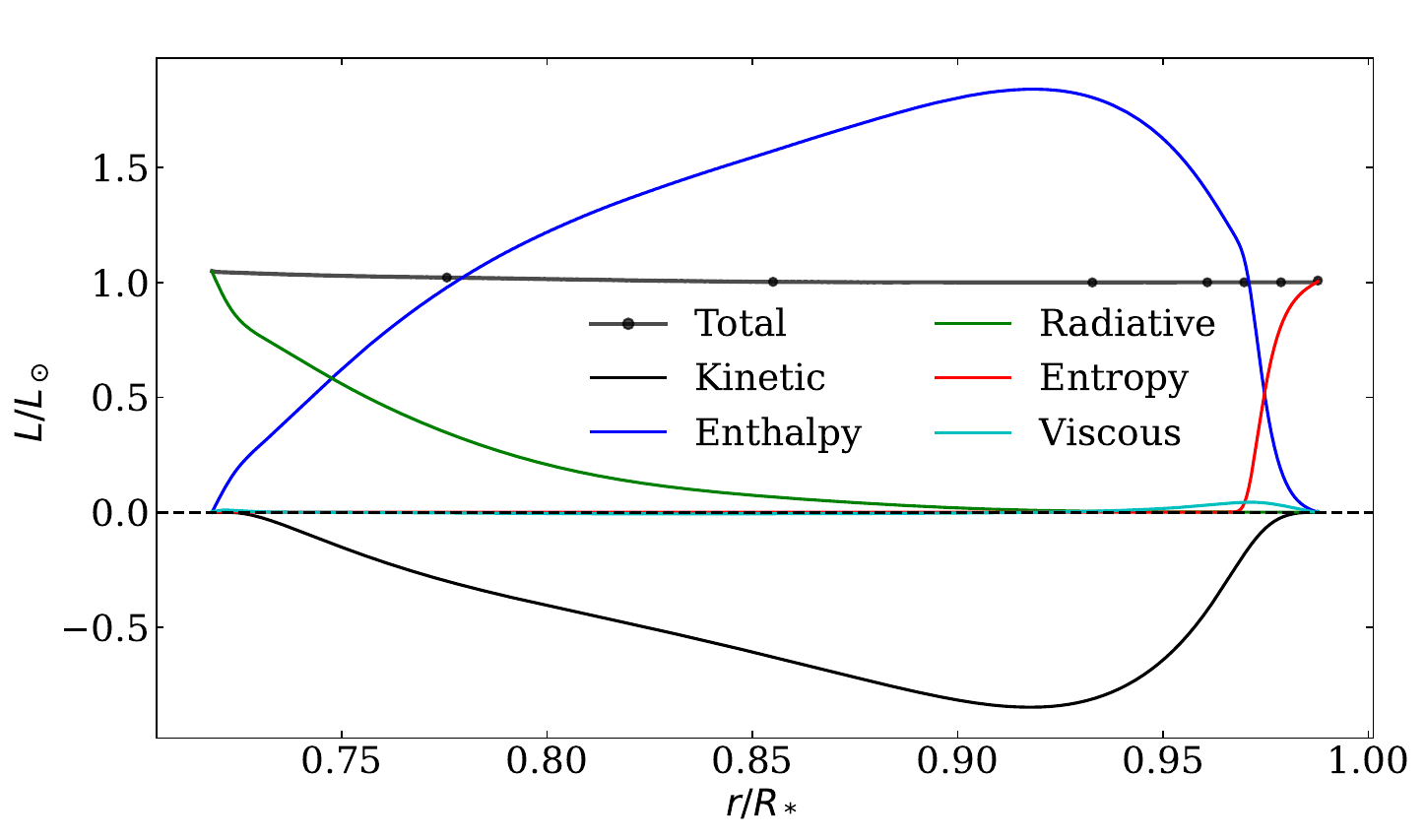}
  \caption{Flux balance of AS1 model, quantifying the different processes contributing to the transport of the energy through the convective envelope. Kinetic, enthalpy, radiative, entropy, viscous flux and the total balance correspond respectively to dark, blue, green, red, cyan solid lines and the dark line with dots (spaced every 100 grid cells). A dashed line indicates the origin of the y-axis.}\label{fig:flux_bal}
\end{figure}

Taking the horizontal average of \mbox{\boldmath $\cal F$} in Equation~\ref{eq:EtotEvol} gives the radial flux balance, ensuring the transport of the stellar luminosity such as
\begin{equation}
  \label{eq:fluxSum}
  \frac{L_*(r)}{4\pi r^2}=F_{\rm ke}+F_{\rm vis}+F_{\rm en}+F_{\rm rad}+F_{\rm irr}+F_{\rm sb},
\end{equation}
where the various fluxes are expressed as
\begin{equation}
  \label{eq:Fke}
  F_{\rm ke}=\frac{1}{2}\bar{\rho} \overline{v^2 v_r},
\end{equation}
\begin{equation}
  \label{eq:Fvis}
  F_{\rm vis}=-\overline{\vv\cdot\mbox{\boldmath $\cal D$}}|_r,
\end{equation}
\begin{equation}
  \label{eq:Fen}
  F_{\rm en}=\bar{\rho} c_P \overline{v_r T},
\end{equation}
\begin{equation}
  \label{eq:Frad}
  F_{\rm rad}=-\kappa_{\rm rad}\bar{\rho} c_P \frac{\partial (\bar{T}+T)}{\partial r}\Big|_{l=0},
\end{equation}
\begin{equation}
  \label{eq:Firr}
  F_{\rm irr}=-\kappa_0\bar{\rho} \bar{T} \frac{\partial (\bar{s}+s)}{\partial r}\Big|_{l=0},
\end{equation}
\begin{equation}
  \label{eq:Fsb}
  F_{\rm sb}=\bar{\rho}\;\overline{v_r}\;\bar{Q},
\end{equation}
with fluxes from the kinetic energy ($F_{\rm ke}$), the enthalpy flux ($F_{\rm en}$), the radiative flux ($F_{\rm rad}$), the viscous flux ($F_{\rm vis}$), the one transported by unresolved motions ($F_{\rm irr}$, \citealt{brunGlobalScaleTurbulent2004}) and the background stratification heating flux ($F_{\rm sb}$). We recall that the system considered is closed, preventing net motion of matter through a spherical surface once the convectively unstable system is relaxed, which leads to $\overline{v_r}=0=F_{\rm sb}$ for the simulations considered in this paper.

We illustrate a classical radial energy balance with the AS1 model \citep{emeriau-viardTurbulencePlasmaDans} in Figure~\ref{fig:flux_bal}. First, transport by radiation carries all the energy from the radiative interior, up to the base of the CZ, and then decreases rapidly with height. In the convective envelope, most of the energy is then transported outward by convection, as evidenced by the rise in enthalpy flux (blue curve). Consequently, a negative kinetic energy flux (dark solid line) is generated by the convective instability, corresponding to inward transport dominated by denser and sinking convective plumes. When reaching the surface, the impenetrable wall boundary condition cancels out convective motion and hence enthalpy flux. At this location, convection should however continue to transport the energy up to the solar surface, but characteristic scales would then become so small that their resolution would become too expensive numerically for the scope of this study. We therefore parametrize them with $F_{\rm irr}$ (red curve), which is expressed as a diffusive transport and is controlled by the coefficient $\kappa_{0}$. Finally, a radial energy equilibrium is reached when the total contribution (black curve with round markers) of the different processes transports the luminosity $L_*$ through the convection zone (CZ) until the surface.

\subsection{Nusselt number: how much energy is carried by convection?}

We can now introduce the dimensionless Nusselt number $Nu$, which compares the respective contribution of enthalpy and radiation/conduction in the transport of the internal energy. In stellar interiors, this can be expressed as
\begin{equation}
  \label{eq:Nus}
  Nu = \frac{F_{\rm rad}+F_{\rm en}}{F_{\rm rad}},
\end{equation}
giving usually $Nu\simeq \nabla_{\rm ad}/\nabla_{\rm rad}$ in adiabatic convection zones \citep{maederPhysicsFormationEvolution2009}. This number decreases as enthalpy transport by convection diminishes, until it tends towards 1 when energy transport is entirely ensured by radiation.

As we saw in Figure~\ref{fig:flux_bal}, the energy transport in the CZ of a classic global solar models is dominated by the enthalpy flux, with a Nusselt number $Nu\sim 20$ in the middle of the CZ. However, we also saw that such models fail in reproducing the solar differential rotation profile when their Reynolds number is increased. This is likely due to a high Rossby number value, resulting in high convective velocities. Therefore, we propose here to voluntary limit them, while conserving a solar rotation rate and luminosity, and while increasing the turbulence degree by decreasing the viscous and thermal diffusions. For that purpose, we propose here to limit the growth of the amplitude of convective flow by controlling the Nusselt number $Nu$.

The ASH code explicitly resolves convection by enabling the convective instability to develop around a reference state. Hence, limiting the amount of energy the convection has to carry, will limit amplitudes of its flows. In the continuity of previous studies \citep{jonesAnelasticConvectiondrivenDynamo2011,kapylaExtendedSubadiabaticLayer2017,kapylaEffectsSubadiabaticLayer2019}, this can be done by increasing the net flux already transported by radiation, which is parametrized with $\kappa_{\rm rad}$ in our model and will be modified in both models we introduce in the next Section. But we could alternatively use $\kappa_0$ to do so. It is the sum of diffusion processes that matters.

\subsection{Numerical set-up}

Both models we present here are similar to AS1 in terms of solar structure (see Appendix~\ref{sec:app_num_set}). As we want to focus on bulk-convection dynamics, we improve boundaries modeling by adding a radiative interior to their base and by extending the domain closer to the solar surface. The number of radial mesh points is increased to $N_r=2000$, spanning from 50 to 99.14\% of the solar radius. This allows to significantly reduce amplitudes of explicit diffusions $\nu$ and $\kappa$, while maintaining a similar Prandtl number $Pr=1/4$, in order to resolve small-scale dynamics. We name these new models with the notation SBR*n\#, * referring to the percentage of the radiative diffusive flux contribution in the balance of Equation~\ref{eq:fluxSum}, and \# referring to the viscosity value at the top of the numerical domain. The higher $F_{\rm rad}$ is, the lower $v$ will be for a given dissipation rate. Recent studies have emphasized that the degree of turbulence can significantly impact system dynamics \citep{hottaGenerationSolarlikeDifferential2022,warneckeNumericalEvidenceSmallscale2023}, so we aim at constructing two models with similar turbulence regimes, which is quantified with the Reynolds number $Re=\tilde{v}\Delta_{\rm conv}/\nu$ where $\tilde{v}$ is the rms velocity and $\Delta_{\rm conv}$ the depth of the modelled CZ. To this end, we set $\nu=3.5\times 10^{11}$ and $1.0\times 10^{12}$ cm$^2$s$^{-1}$ at the top of SBR97n035 and SBR50n1 numerical domains respectively, and set $\kappa_{\rm rad}$ profiles in order to obtain enhanced $F_{\rm rad}$. After implementing these background profiles and a realistic stratification (see Appendix~\ref{sec:app_num_set} for more details), a random 3D fluctuating entropy field is introduced to trigger the convective instability since $\frac{ds}{dr}<0$ in the CZ.

After a linear phase of exponential growth, the non-linear convective instability saturates in terms of energy by redistributing the entropy in the CZ and yields to Rayleigh numbers $Ra=2.051\times 10^7$ and $5.81\times 10^6$ for SBR97n035 and SBR50n1 respectively. These values have to be above a critical Rayleigh value $Ra_{\rm c}$ to make energy transport by convection effective, given the dissipation in our models. In order to quantify the level of supercriticality $Ra*/Ra_{\rm c}$, we compute the modified Rayleigh number $Ra^*$ defined in \cite{takehiroAssessmentCriticalConvection2020}, estimate the critical value by scaling from their results (see Appendix~\ref{sec:app_critic}) and list them in column 8 of Table~\ref{tab:AB2vsAS1vsSunBusyAdim}. All models are supercritical, especially the two new models (SBR97n035 and SBR50n1) where $Ra/Ra_{\rm c}$ exceeds $10^3$. This ensures the development of the convective instability is not close to a marginal state and thus dominates diffusive processes.  

\section{Overview of the models and their dynamics}\label{sec:Overview}

\begin{table*}
  \begin{center}
    \begin{tabular}{lccrccrrrc}
      \hline
      \hline
      Name & $\Delta\Omega$ & $\tau_c$ & $Nu$ & $F_{\rm rad}$ & $Ro_{\rm f}$ & $Re$ & $Ta$ & $Ra$ & $Ra^*/Ra_{\rm c}$\\
      & (nHz) & (days) & & ($F_{\rm tot}$) & & & ($10^6$) & ($10^4$) & \\[0.5ex]
      \hline
      \hline
      AS1      & -130 & 22 & 25.89 & 0.07 & 2.53& 205  & 4.4 & 35.7 & 97 \\[0.5ex]
      SBR97n035 & 106 & 87 & 1.04  & 0.97 & 1.49 & 811  & 1103.9 & 2050.8 & 1611 \\[0.5ex]
      SBR50n1  & -320 & 28 & 2.45  & 0.50 & 4.01 & 860  & 135.2 & 581.1 & 3630 \\[0.5ex]
      \hline
      \hline
    \end{tabular}
    \caption{Global parameters of the 3 solar models. From the left, we list the name of the model and the surface latitudinal DR contrast between the equator and the 60° latitude. Then come the convective turnover time $\tau_c=\Delta_{\rm conv}/\tilde{v}$, the Nusselt number $Nu=(F_{\rm rad}+F_{\rm en})/F_{\rm rad}$, the radiative part of the flux $F_{\rm rad}/F_{\rm tot}$, the fluid Rossby $Ro_{\rm f}=|\nab\times \vv|/(2\Omega)$, Reynolds $Re=\tilde{v}\Delta_{\rm conv}/\nu$, Taylor $Ta = 4\Omega_*^{2} \Delta_{\rm conv}^{4}/\nu^{2}$, and Rayleigh number $Ra=-(\partial\bar{\rho}/ \partial\bar{S}) (\partial S_{\rm tot}/ \partial r) g \Delta_{\rm conv}^4 / (\bar{\rho} \nu \kappa)$, evaluated at the middle of the convective zone, and with the thickness $\Delta_{\rm conv}=r_{\rm top}-r_{\rm BCZ}$. We finally compute the modified Rayleigh number $Ra^\star/Ra_c$ as defined by \citet{takehiroAssessmentCriticalConvection2020}.\label{tab:AB2vsAS1vsSunBusyAdim}}
  \end{center}
\end{table*}

\begin{figure*}
  \centering
  \includegraphics[width=0.49\linewidth]{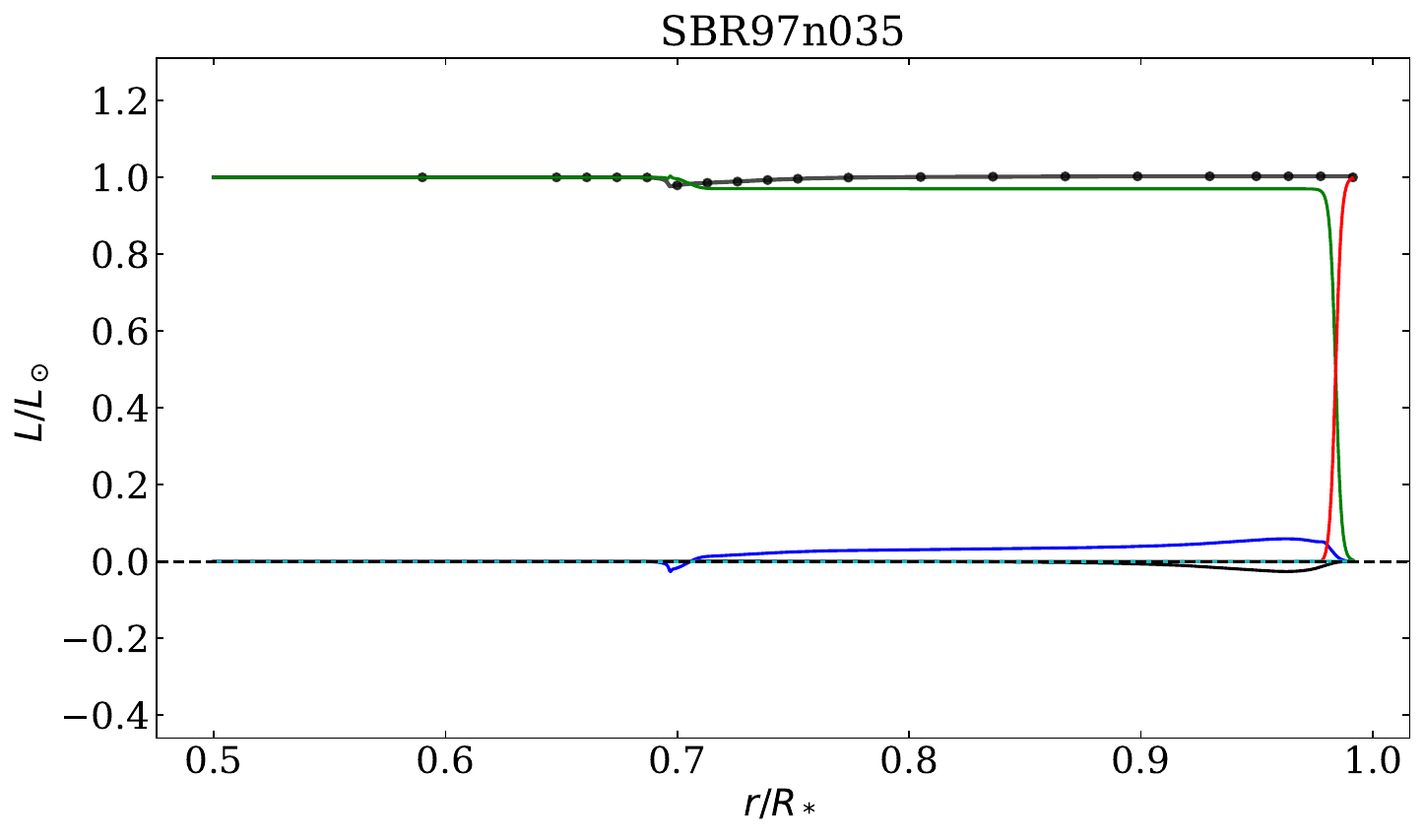}
  \includegraphics[width=0.49\linewidth]{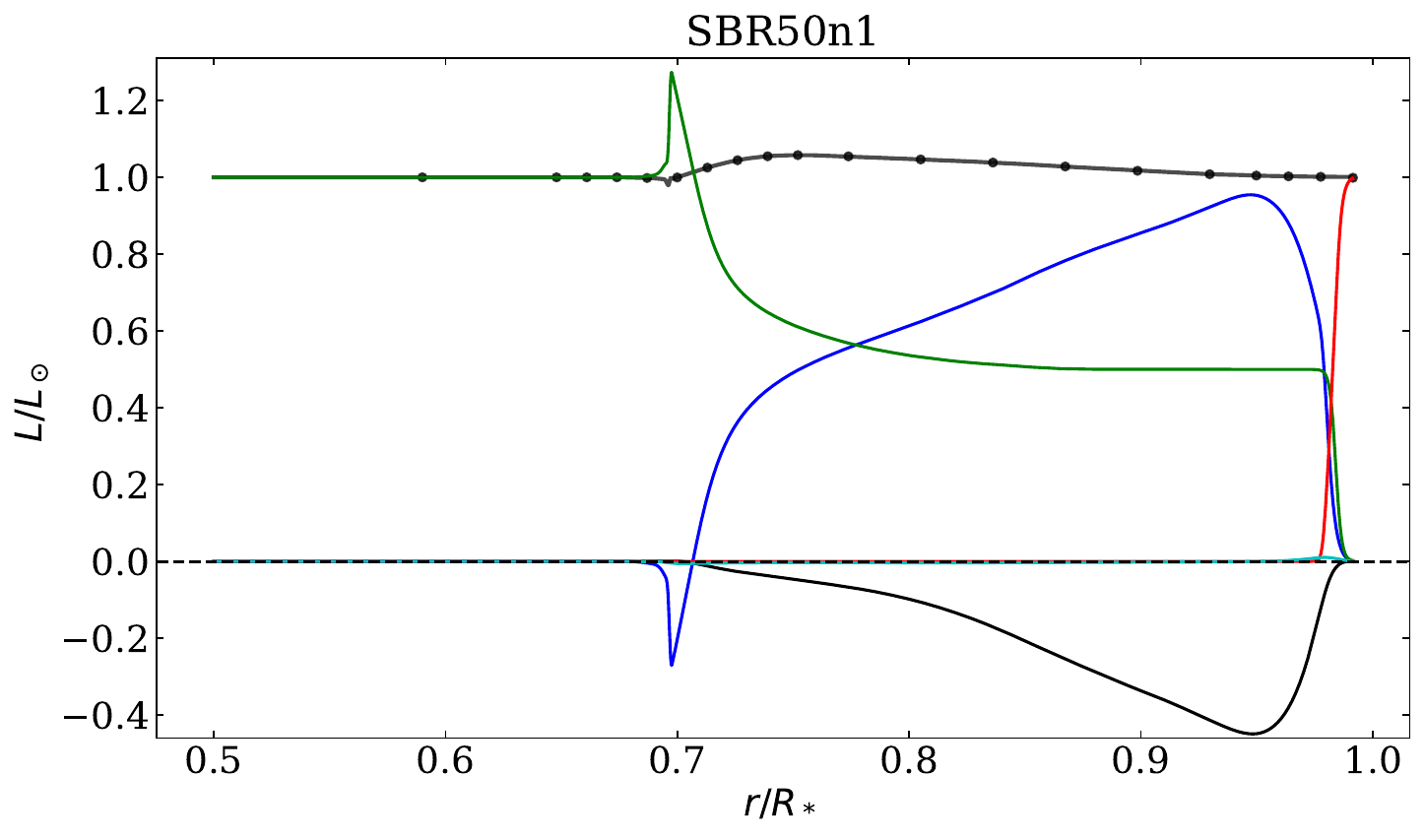}
  \includegraphics[width=0.49\linewidth]{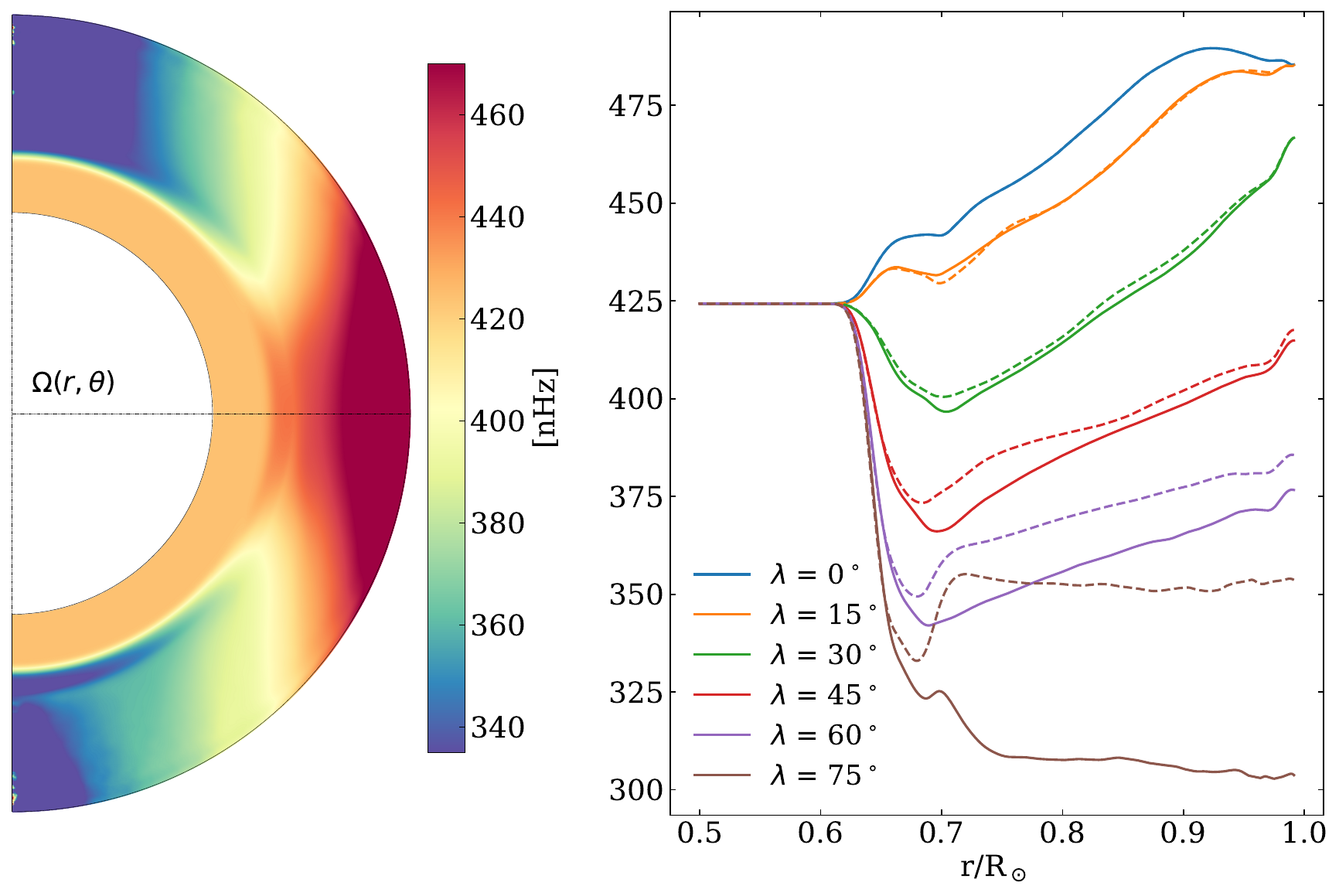}
  \includegraphics[width=0.49\linewidth]{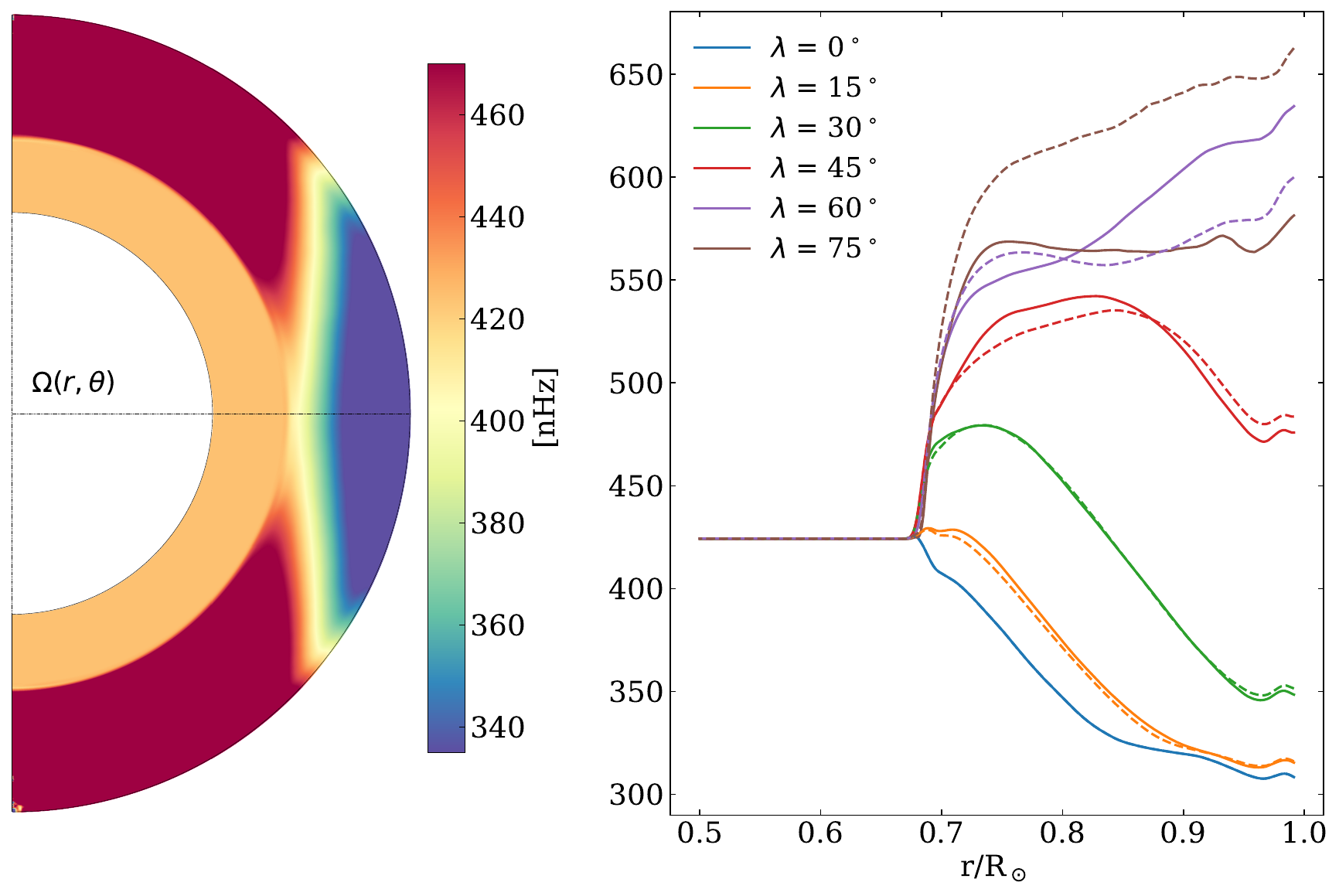}
  \caption{Illustration of different profiles in models SBR97n035 (left panels) and SBR50n1 (right panels). Top panels show the respective flux balances 
  (similar to Figure~\ref{fig:Fig1}, see also Equation~\ref{eq:fluxSum}), where the enthalpy flux of SBR97n035 has been significantly dampened (blue curve of the left panel). Bottom panels show differential rotation profiles of both SBR97n035/SBR50n1 models, showing a Sun-like/anti-solar (left/right), respectively. Radial profiles illustrated with solid/dashed lines are taken from the Northern/Southern Hemisphere, respectively.}
  \label{fig:SBR97n035vsSBR50n1}
\end{figure*}

\begin{figure*}
  \centering
  \includegraphics[width=0.49\linewidth]{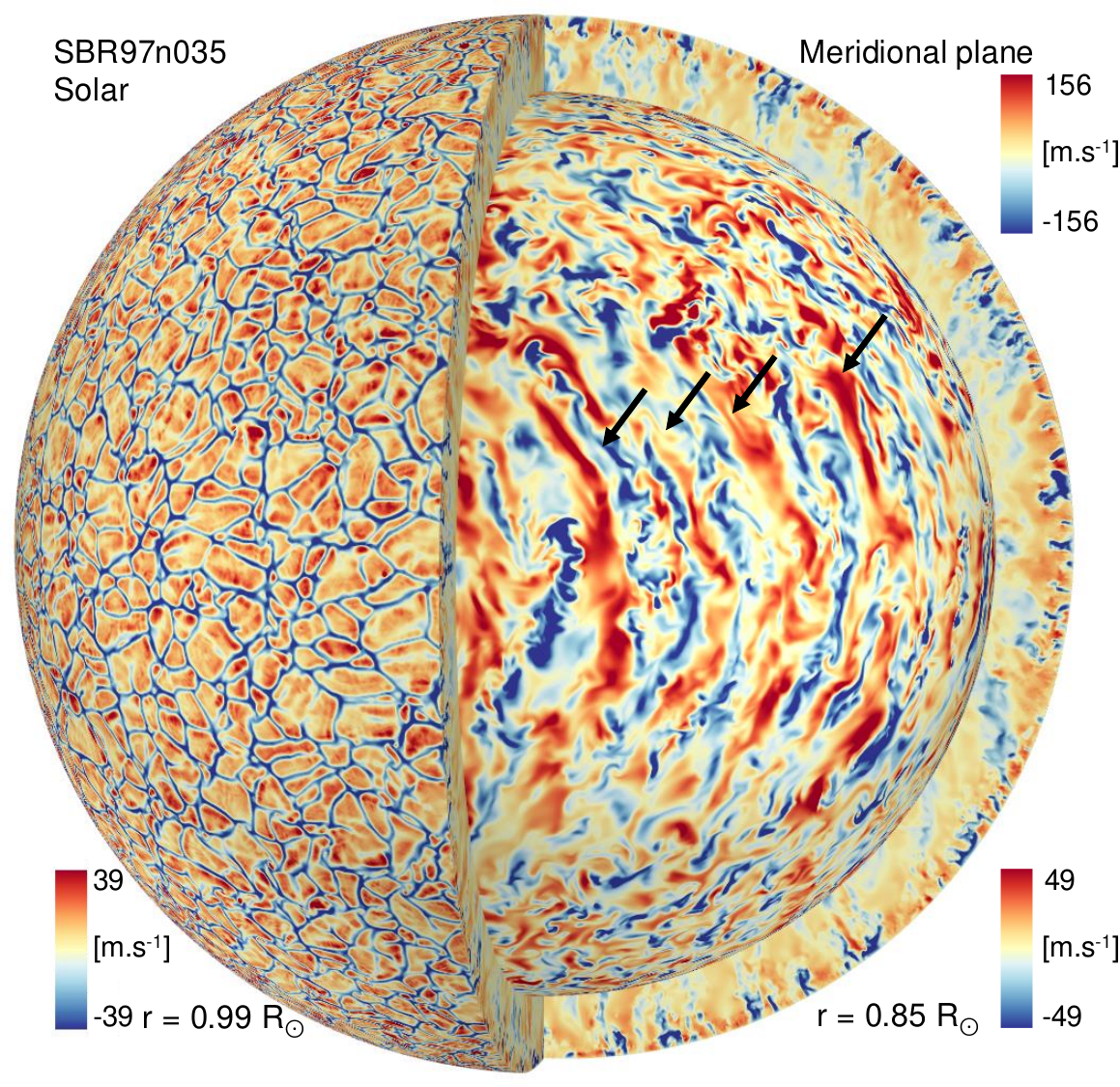}
  \includegraphics[width=0.49\linewidth]{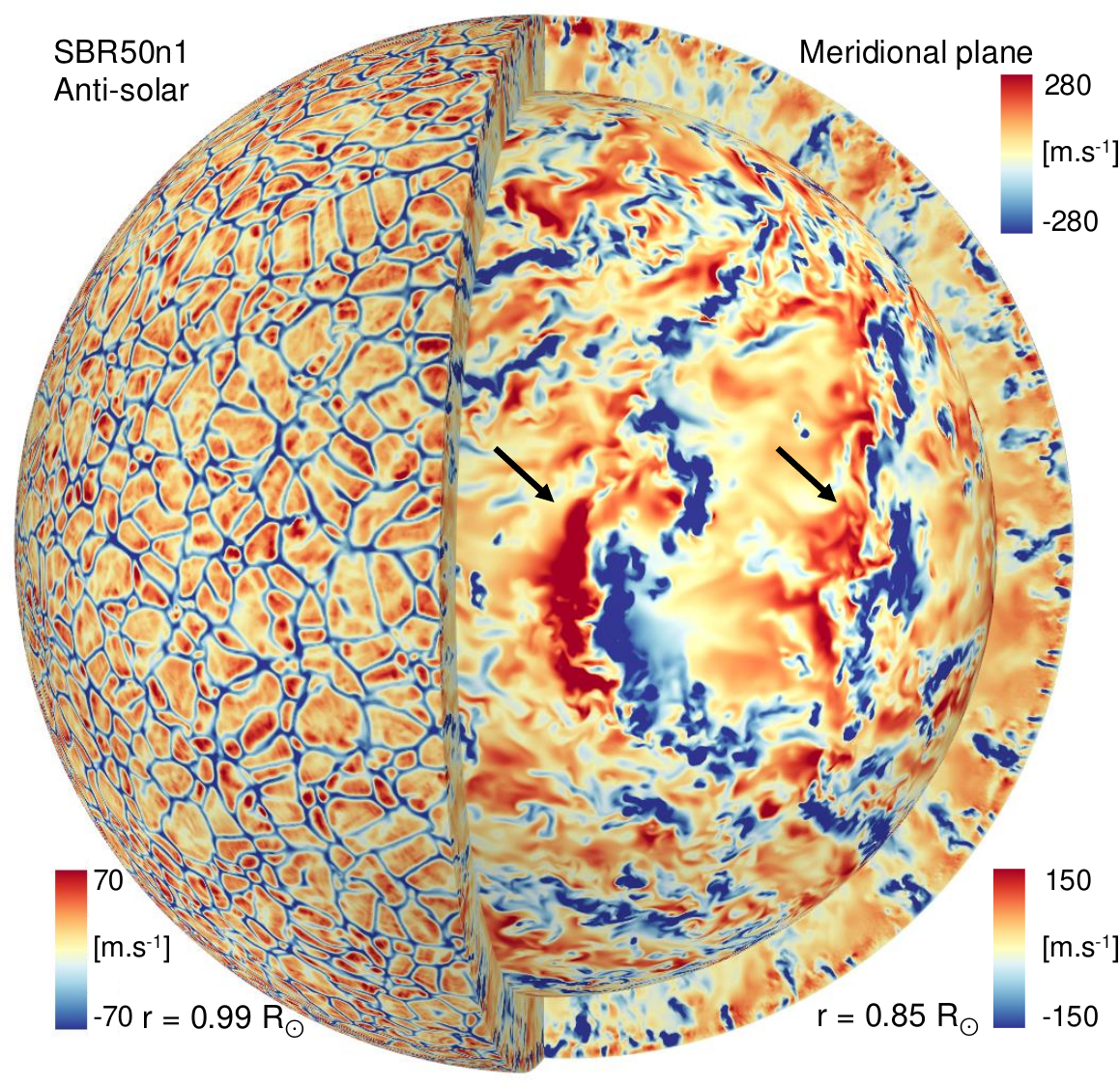}
  \caption{Radial velocity maps of SBR97n035 (left) and SBR50n1 (right), comparing the top of the simulation ($r=0.99~R_\odot$, 10 grid points deep, outer sphere) and the middle of the convective zone ($R_{\rm mid}=0.85$~R$_\odot$, inner sphere). Both depths are illustrated using the bottom-left and -right color bar, respectively. The top-right color bar corresponds to meridional planes. The maximum value of a color bar corresponds to twice the standard deviation of the map it corresponds to. Dark arrows enlighten the presence of columnar patterns (thermal Rossby modes), see \textit{e.g.} \cite{featherstoneEMERGENCESOLARSUPERGRANULATION2016} for schematics. Note the color-scale difference between the two models. Animations are available at\dataset[doi:10.5281/zenodo.14650437]{https://doi.org/10.5281/zenodo.14650437} \citep{norazCcZenodo25}. Movies  covers 17 s, which corresponds to 172 and 90 days in SBR97n035 and SBR50n1, respectively.}\label{fig:SBR97n035vsSBR50n1pv}
\end{figure*}

\subsection{Flux balance}

The radial transport of energy reaches an equilibrium after several convective overturning times and leads to the dynamics quantified in Table~\ref{tab:AB2vsAS1vsSunBusyAdim} and illustrated in Figure~\ref{fig:SBR97n035vsSBR50n1}. This radial balance is quantified with the different fluxes (\ref{eq:fluxSum}) in the top panels. The left-one illustrates their balance in SBR97n035, where $F_{\rm rad}$ is now the main flux, carrying 97\% of the luminosity into the CZ (green curve). On the right panel (model SR50n1), $F_{\rm rad}$ carries 50\% of the luminosity (green curve). Both unresolved flux $F_{\rm ed}$ (red) have been identically constrained as close as possible to the surface. Convective instability has developed to complement the net transport of luminosity, and is quantified with the enthalpy flux (blue). The peak at 0.97 $R_\odot$ in the model SBR97n035 has been significantly damped, as anticipated, decreasing to 7\% of the solar luminosity. Consequently, the kinetic energy flux has also reduced, corresponding to 4\% of the solar luminosity. On the right panel, the enthalpy transport in SBR50n1 is higher, with a similar peak approaching 100\% of the energy transport, and significantly balanced by a negative kinetic flux, coming from strong downward convective plumes. We note the presence of a negative enthalpy flux associated with the overshooting of these plumes, penetrating the top of the radiative interior. Both the dampening of convective velocities and the stronger rotational constraint (see Section~\ref{sec:ConvDynRotCons}) limit the amplitude of this convective overshoot in SBR97n035. In the case of SBR50n1, this requires some adjustment of radiative diffusivity near the base of the convective envelope to balance it and ensure the transport of the whole luminosity.

\subsection{Differential rotation profiles}
The Nusselt number $Nu$ in the canonical model AS1 has a value of 25.89. By the design of our experiments, it has been decreased to 2.45 and 1.04 in models SBR50n1 and SBR97n035 respectively. However, the Reynolds number has significantly increased in both models, reaching 860 and 811 respectively in the middle of the CZ (see Table~\ref{tab:AB2vsAS1vsSunBusyAdim}). Nevertheless, we have kept SBR97n035 in a Rossby number regime that ensures the establishment of a solar-type differential rotation profile with a fast equator. We illustrate it on the bottom-left panel of Figure~\ref{fig:SBR97n035vsSBR50n1} and measure a latitudinal surface contrast of $\Delta\Omega=130$ nHz between the equator and 60° latitude. Despite a similar turbulence degree, SBR50n1 has gone into an anti-solar differential rotation regime (bottom-right panel), and is experiencing a large latitudinal rotation contrast $\Delta\Omega=-320$. This is due to the strong-amplitude convective velocities achieved in this model, which increase its effective Rossby number, and hence diminishes the rotational impact of the Coriolis force on the convective dynamics. This results in a decrease of the equatorward transport by the Reynolds stress, along with enhanced poleward transport by the meridional circulation in the angular momentum balance (see \citealt{brunDifferentialRotationOvershooting2017} for more details).

\subsection{Convective dynamics and their rotational constraint}\label{sec:ConvDynRotCons}
We now look deeper into the convection morphology to understand the impact of a controlled Nusselt number, and illustrates radial velocity maps at different depths in Figure~\ref{fig:SBR97n035vsSBR50n1pv}. We draw attention to the fact that the different layers of a 3D illustration correspond to different color scales.

\begin{figure}
  \centering
  \includegraphics[width=\linewidth]{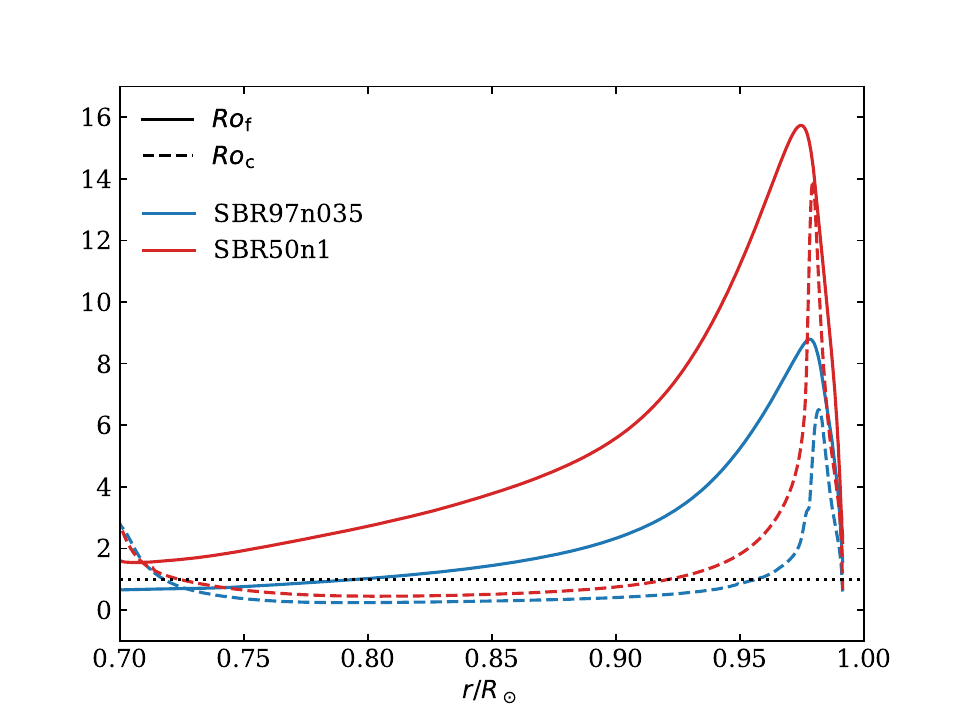}
  \caption{Rossby number radial profiles of SBR97n035 and SBR50n1 models, respectively, in blue and red. The fluid Rossby number $Ro_{\rm f}$ and the convective Rossby number $Ro_{\rm c}$ are respectively shown with solid and dashed lines. The Rossby numbers reach unity when crossing the dotted-horizontal solid line.}\label{fig:Rossby_comp}
\end{figure}

The right parts of these 3D views (inner sphere) show the radial velocity in the middle of the convective zone ($0.85$~R$_\odot$). We can see that the structure of convection in SBR50n1 (right panel) shows a broad range of spatial scales, with an equatorial zone concentrating strong contrasts in amplitude, whose maxima can reach $\pm 400$ m~s$^{-1}$. The left panel shows the same map for the SBR97n035 model, where amplitudes of convective velocities are lower, with maxima reaching $\pm 120$ m~s$^{-1}$, as a consequence of the stronger dampening of $F_{\rm en}$ by $F_{\rm rad}$ in this model. Convective patterns aligned along the north-south axis also appear in the equatorial zone. These are convective rolls swirling around axes parallel to the rotation axis of the star. This phenomenon is well known from global simulations of rotating convective envelopes, known as "banana cells", "Busse columns" or "thermal Rossby waves" (\citealt{1968RSPTA.263...93R,busseThermalInstabilitiesRapidly1970,busseEddingtonSweetCirculations1981}, see also Section 3.4 of \citealt{2007mand.book.....D}). They result from the Coriolis force acting on radial flows in the equatorial plane. Here they appear as a consequence of lower velocity amplitudes, leading locally to a low Rossby number, and therefore to high rotational constraint by the Coriolis force. As expected, these columnar Rossby modes are prograde compared to the local angular velocity in both models. In that sense, one can see in SBR50n1 that the columnar pattern is rotating slower than the solar rotation rate $\Omega_\odot$, but nonetheless rotates faster than the mean angular velocity at the corresponding latitudes (which is notably decreased in the equatorial region of such an anti-solar rotating model, see movies attached to Figure~\ref{fig:SBR97n035vsSBR50n1pv}). Note that the columns also display small scales sub-structures. Clearly, the flow is turbulent in both models with high Reynolds number. 

To quantify the rotational constraint, we illustrate in Figure~\ref{fig:Rossby_comp} radial profiles of the fluid Rossby number, $Ro_{\rm f} = |\nabla \times \vv|/(2\Omega)$, and the convective Rossby number, $Ro_{\rm c} = \sqrt{Ra/(Ta \cdot Pr)}$. These quantify the ratios of inertial forces, due to inertia and buoyancy respectively, relative to the Coriolis force. We first note that $Ro_{\rm f}$ is higher than unity on average at all depth in SBR50n1, and has been significantly reduced in SBR97n035, such that $Ro_{\rm f}\leq 1$ below $0.8$ $R_\odot$. It means that the rotational constraint over the flows is strong at the base of SBR97n035 CZ. We also see the importance of the Coriolis force in the dynamics when looking at $Ro_{\rm c}$. The latter is lower than unity for most of the CZ extend in both models. This explains why some large scale banana cells are also visible in velocity maps of SBR50n1 model (black arrows on right panel of Figure~\ref{fig:SBR97n035vsSBR50n1pv}) as the rotation is still impacting the buoyancy. As buoyancy is the driver of the thermal convective instability, $Ro_{\rm c}$ ratios directly quantify the rotational constraint on the convection morphology. Thinner convective columns are observed as the rotational constraint increase (SBR97n035 in Figure~\ref{fig:SBR97n035vsSBR50n1pv}). We refer the interested reader to Section~\ref{sec:tauEll} for more details. 

When looking at meridional planes in Figure~\ref{fig:SBR97n035vsSBR50n1pv}, we observe that the convective sinking plumes (blue patterns) exhibit smaller scales in SBR97n035 (right panel). Additionally, downdrafts regions in SBR97n035 (left panel) are shallower, whereas they penetrate deeper in SBR50n1 (right panel). Indeed, the reduced velocities and corresponding increased rotational effects in SBR97n035 accentuate the influence of the Coriolis force, which tilts the descending material toward the rotation axis with respect to the local radial direction (see also \citealt{brummellPenetrationOvershootingTurbulent2002}). Turning our attention to SBR50n1 (right panel), the inclination of convective plumes near the poles is indicative of such horizontal deflections imposed by the Coriolis force. The shallower depth of near-surface downdrafts in the equatorial region further supports the increase of the Coriolis force impact over radial motions at low latitudes. Although SBR97n035 (left panel) is more rotationally constrained (lower $Ro$), latitudinal variations in downdraft behavior are not as readily apparent in this Figure, due to the dampening of downdraft velocities.

Overall, we see that a low Rossby number is better than a large one to get the prograde equator, as was expected from previous studies \citep{mattConvectionDifferentialRotation2011,gastineSolarlikeAntisolarDifferential2014,hottaDynamicsLargeScaleSolar2023}.

\subsection{Polar dynamics}\label{sec:PolDyn}

\begin{figure*}
  \centering
  \includegraphics[width=\linewidth]{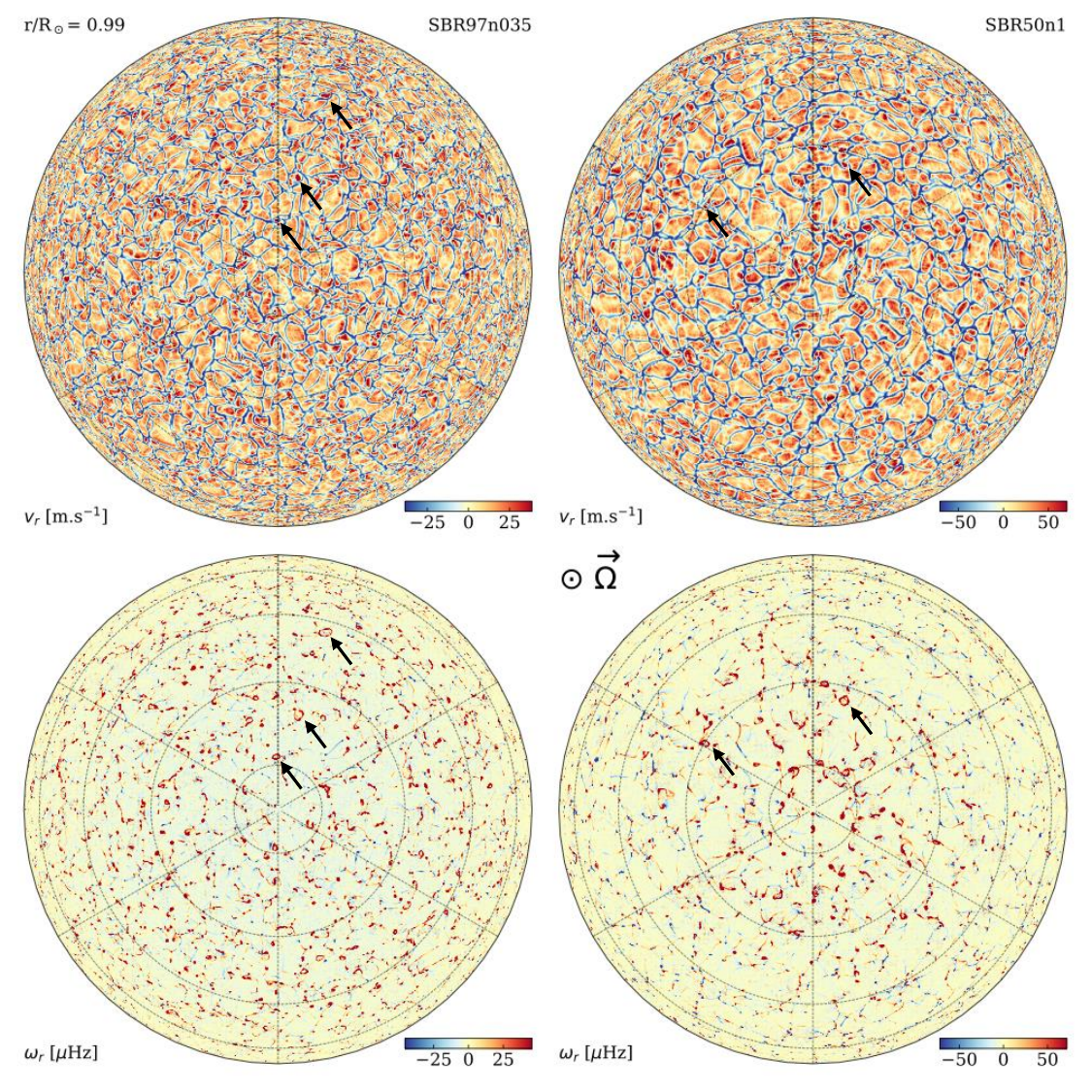}
  \caption{North-Pole view of the near-surface dynamics ($0.99$ R$_\odot$, 10 grid-points below the top) in SBR97n035 (left panels) and SBR50n1 (right panels) models. We show maps of the radial velocity $v_r$ and radial vorticity $\omega_r$ on top and bottom panels, respectively. The maximum value of a color bar corresponds to twice the standard deviation of the map it corresponds to. Concentric grid circle lines correspond to co-latitudes $\theta=10, 30, 50$ and $70^{\circ}$, respectively, from the center to the edges. Dark arrows enlighten the presence of \textit{polar vorticity rings} (see the text). Animations are available at\dataset[doi:10.5281/zenodo.14650437]{https://doi.org/10.5281/zenodo.14650437} \citep{norazCcZenodo25}.}\label{fig:PolarView}
\end{figure*}

We now focus on the polar dynamics of these models, illustrated with Northern Hemisphere in Figure~\ref{fig:PolarView} (we also show Southern Hemispheres in Figure~\ref{fig:PolarView3}). The upper panels show the near-surface radial velocity component $u_r$, as seen from above the North Pole of both SBR97n035 (left) and SBR50n1 (right). The lower panels show the radial component of the vorticity $\omega_r$, where red/blue corresponds to cyclonic/anticyclonic motions, corresponding in the Northern Hemisphere to anti-clockwise/clockwise motion, respectively. Both models show busy small-scale dynamics, coherent with what was discussed previously. It is noteworthy that the morphology we find in the polar dynamics is coherent with the so-called \textit{polar plumes} structures for both models \citep{stellmachApproachingAsymptoticRegime2014,hindmanMorphologicalClassificationConvective2020}. Furthermore, we identify additional interesting patterns in the polar dynamics for which, to our knowledge, no established classification currently exists in the literature.

On the bottom vorticity panels, we note two compact populations of anticyclonic (blue) and cyclonic motions (red), the latter being more intense. The strongest vorticity is indeed found in intergranular lanes, where flows are gathering at the edges of two or more convective cells and then form intense gyres by local conservation of angular momentum, in comparison to up-flows at the center of convection cells. In addition, we find that the strongest radial vorticity values found in intergranular lanes are positive, then corresponding to cyclonic motions in the Northern Hemisphere. This is consistent with the Coriolis-force impact on horizontally converging motions. Such characteristics are typical from \textit{polar-plumes} dynamics. Some plumes are seen to cross much of the convection zone (see meridional cuts in Figure~\ref{fig:SBR97n035vsSBR50n1pv} and also Figure~\ref{fig:PolarView2}).

We also observe the formation of \textit{polar vorticity rings} in both models, as highlighted by the dark arrows. These structures are characterized by small-scale intergranular down-flows that form cyclonic rings around a central up-flow, visible in the radial velocity maps (top panels). Unlike the so-called \textit{polar cells} \citep{hindmanMorphologicalClassificationConvective2020}, these vorticity rings do not maintain their coherence throughout the convective shell and have relatively short lifespans, typically on the order of a few days. Such transient behavior is likely a result of the rapid and turbulent dynamics happening near the surface, along with the significant density stratification ($N_\rho=ln(\rho_{\rm BCZ}/\rho_{\rm top})=5.9$) present in CZ of these models.

From the evolution of these \textit{polar vorticity rings} (see movies attached to Figures~\ref{fig:PolarView} and \ref{fig:PolarView3}), we envision two possible scenarios for their formation: either they are generated by the overturn of rising convective cells at sites of pre-existing intergranular cyclonic activity, leading to the formation of a vorticity ring at the cell periphery, or they originate from pre-existing intergranular vortices that gain enough rotational strength to expand laterally due to a local centrifugal effect. If the latter mechanism is at play, this could signal locally the onset of what is referred to as the \textit{geostrophic turbulence} regime \citep{stellmachApproachingAsymptoticRegime2014}. Distinguishing between these formation scenarios is a challenging task with the currently available data. However, a detailed investigation into the exact dynamics of such features remains open and will be addressed in future work.

In traditional Rayleigh–Bénard convection without rotational constraint (or experiencing sufficiently low Rossby number), vortical signatures are equally weighted between cyclonic and anti-cyclonic structures. When rotation is enhanced, the Coriolis force acts proportionally on the plumes forming in converging flows, making them preferentially cyclonic in the Northern Hemisphere \citep{hindmanMorphologicalClassificationConvective2020}. The asymmetry in the cyclonic/anti-cyclonic distribution we report here is then likely the result of a combination between the Coriolis influence and the enhanced vorticity of converging intergranular lanes due to the local conservation of angular momentum. Furthermore, the impact of the latter on horizontal flows is proportional to the cosine of the co-latitude, due to the vectorial definition of the force. This means that the strongest impact will be felt by polar flows, and will be spread at larger co-latitude $\theta$ (lower latitudes) proportionally to the rotational constraint. This is coherent with what is seen in the bottom panels, where the cyclonic-favored (red) asymmetry is spread at lower latitude in SBR97n035 ($\theta\sim 70^{\circ}$) as it experiences a smaller Rossby number (see Table~\ref{tab:AB2vsAS1vsSunBusyAdim}). In comparison, SBR50n1 model experiences it only in polar regions ($\theta\leq 40^{\circ}$), and shows a rather equally-weighted sign distribution around the equator.

Despite differences observed in the bulk dynamics of the two models, particularly in terms of the thermal Rossby modes near the equator, we note that both models present a remarkably similar morphology in their small-scale near-surface dynamics (as shown in Figures~\ref{fig:SBR97n035vsSBR50n1pv} and top panels of Figure~\ref{fig:PolarView}). To further elucidate this overview, we will now quantify their differences and confront them with solar observations to help contextualize our numerical models within the broader framework of solar dynamics.

\section{Adequation of the Nusselt-controlled approach with helioseismic constraints}\label{sec:sol_comp}

We just presented a global model of solar convection in a rotating spherical shell, which maintains a prograde equator while possessing a high turbulence degree (SBR97n035, $Re\geq 800$). Such models usually tend to over-estimate the large-scale amplitude of convection in comparison to observational constraints (\citetalias{hanasogeAnomalouslyWeakSolar2012}). As a direct consequence, the Rossby number of such models are often over-estimated in comparison to the solar value, leading to the establishment of a retrograde equator (as in SBR50n1 case). It is then particularly problematic that such models lose the characteristic solar large-scale flows when trying to get closer to the solar regime (\textit{c.f.} \textit{Convective Conundrum}, see \textit{e.g.} \citealt{hottaDynamicsLargeScaleSolar2023}).

To circumvent this issue and maintain a prograde equator, we have proposed a theoretical path in the parameter space to control the Nusselt number of the simulation $Nu$ (see Section~\ref{sec:pathTheo}). This yields a Sun-like rotating model that operates at solar luminosity and exhibits a notably high degree of turbulence, which we now aim to compare with solar observations. Helioseismology has provided constraints on the internal dynamics of the Sun for decades \citep{christensen-dalsgaardCurrentStateSolar1996,thompsonDifferentialRotationDynamics1996,hanasogeAnomalouslyWeakSolar2012,greerHELIOSEISMICIMAGINGFAST2015}. We will now further confront this model to the recent ones, regarding large-scale convection, the profiles and amplitudes of inertial modes, and finally in terms of degree of super- and sub-adiabaticity, before diving into more in-depth analysis of the force balances in Section~\ref{sec:SpectralAna}.

\subsection{Energy distribution among convective scales}\label{sec:energySpec}

\begin{figure*}
  \centering
  \includegraphics[width=0.49\linewidth]{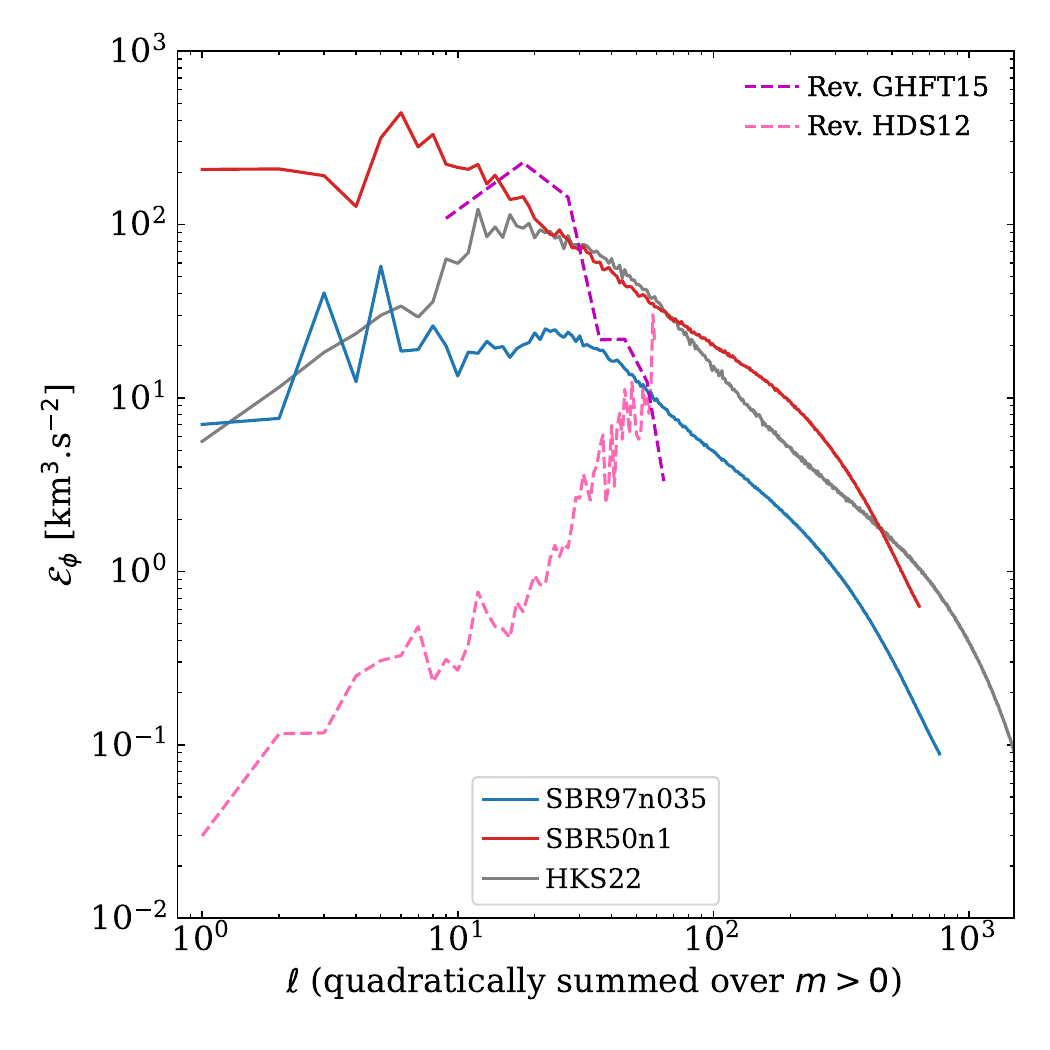}
  \includegraphics[width=0.49\linewidth]{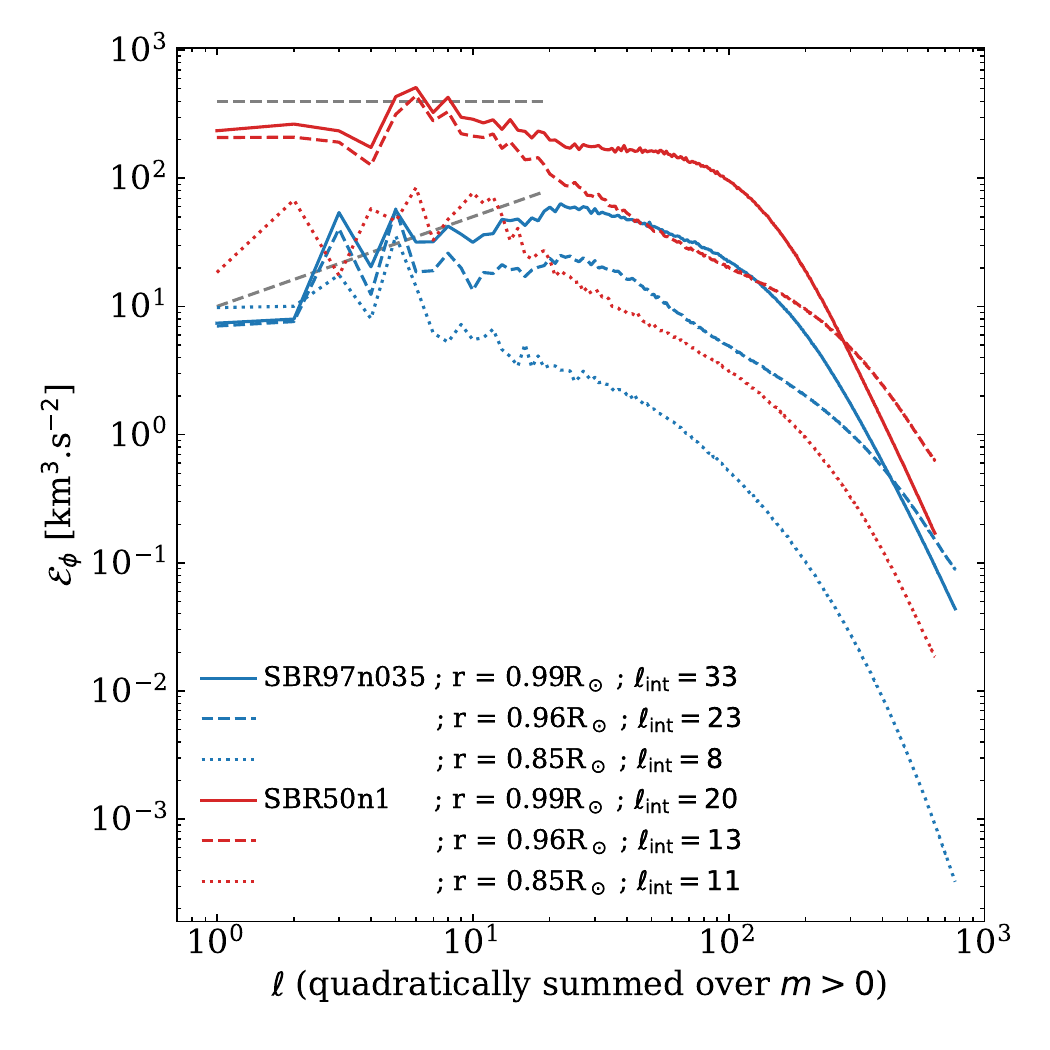}
  \caption{
  \textit{Left}: Comparison of the non-axisymmetric part of toroidal kinetic spectra per multiplet ${\cal E}_\phi$ at $0.96~R_\odot$, defined following the definition of \cite{gizonHelioseismologyChallengesModels2012}. Similar to data presented in Figure~\ref{fig:Fig2}, with blue and red curves taken this time from model SBR97n035 (solar-like DR) and SBR50n1 (anti-solar DR) respectively. The different $\ell_{\rm pond}$ of global models, as defined in Figure~\ref{fig:Fig2}, are 102, 82 and 143 for SBR97n035, SBR50n1 and HKS22, respectively. Similarly, the different integral scales are $L_{\rm int}=4.1$, 7.2 and 2.9\% of the solar radius, respectively, which corresponds to $\ell_{\rm int} = 23$, 13 and 33. \textit{Right}: Similar to left panel, focusing now on SBR97n035 and SBR50n1 models, for longitudinal velocity maps at mid-convection zone ($r=0.85~R_\odot$), close to the enthalpy peak ($0.96~R_\odot$) and near the top of the simulation ($0.99~R_\odot$). We add two indicative trends in gray-dashed lines for comparison. SBR50n1 spectra exhibit a rather flat behavior for $\ell\leq 20$ at all depths, while SBR97n035 spectra show a clear dampening of the large-scales amplitudes when probing near-surface layers.
  }
  \label{fig:SpComp}
\end{figure*}

Under the \textit{Lantz-Braginsky-Roberts}-anelastic approximation (see Appendix \ref{sec:num_method}), the kinetic and internal energy can be defined as \citep{brownENERGYCONSERVATIONGRAVITY2012}

\begin{equation}
  \label{eq:Etot}
  E_{\rm tot}=\rb\Big[ \underbrace{\frac{1}{2}v_r^2}_{{\cal E}_r/r} +\underbrace{\frac{1}{2}v_\theta^2}_{{\cal E}_\theta/r} +\underbrace{\frac{1}{2}v_\phi^2}_{{\cal E}_\phi/r} +\underbrace{\frac{1}{2}\frac{g}{c_P}\left(\frac{{\rm d}\sba}{{\rm d}r}\right)^{-1}s^2}_{{\cal E}_{\rm int}/r} \Big],
\end{equation}
where $(v_r,v_\theta,v_\phi)$ are the spherical coordinates of the velocity field and $g$ the amplitude of the gravitational acceleration. Toroidal convective spectra ${\cal E}_\phi$ at $0.96$~R$_\odot$ are shown in the left panel of Figure~\ref{fig:SpComp} for SBR97n035 (Sun-like DR) and SBR50n1 (anti-solar DR) with blue and red solid lines respectively, along with a summary of the various observational and numerical constraints previously presented in Figure~\ref{fig:Fig2} (dashed and gray-solid lines respectively). 

We first note that the signal of SBR97n035 (blue curve) at low $\ell$ lies between both \citetalias{hanasogeAnomalouslyWeakSolar2012} and \citetalias{greerHELIOSEISMICIMAGINGFAST2015}. This is encouraging to see that such a numerical approach hence goes in the right direction regarding constraints on the large-scale convective amplitude, in addition to exhibiting a prograde equator with this relatively high $Re$ value. At smaller scales, we then see that SBR97n035 converges toward observed values for $\ell\sim50$, which is the only scales range where both \citetalias{hanasogeAnomalouslyWeakSolar2012} and \citetalias{greerHELIOSEISMICIMAGINGFAST2015} agree. Now focusing on SBR50n1 (red curve), amplitudes at large-scales stays too high in comparison to this observational context, which further discriminates this anti-solar rotating model in the solar context.

It is finally interesting to note that the Sun-like rotating model SBR97n035 (blue) exhibits similar convective amplitudes than the Sun-like rotating model from \citetalias{hottaGenerationSolarlikeDifferential2022} at large-scales ($\ell\leq 10$). We note that \citetalias{hottaGenerationSolarlikeDifferential2022} interpret it as a magnetic feedback, whereas in our case it is achieved by controlling the Nusselt number without including magnetic fields. In both cases, the decrease of large-scale amplitude leads to a solar-type differential rotation.

We finally want to remind here that spectra from global convection simulations benefit from the information over the whole sphere without any noise, in contrary to observations, which only see one side of the Sun, miss polar regions, and hence may struggle to catch the lowest $\ell$ signal, even more so when they are based on local techniques.\\

Now looking at the depth dependence of such spectra, we show in the right panel of Figure~\ref{fig:SpComp} this toroidal convective spectrum ${\cal E}_\phi$ at different depths for both SBR97n035 and SBR50n1, which show distinct morphology evolution. On one hand, the shape of SBR50n1 (anti-solar case) spectra of the largest scales (for $\ell\leq 20$) is roughly conserved from the middle to the top of the convection zone. The largest scales dominate the toroidal convective spectrum at all depths and show a maximum at $\ell_{\rm peak}=6$ with a rather flat spectrum for $\ell\leq 20$ (see red curves). On the other hand, SBR97n035 (Sun-like rotating) spectra exhibit a clear change of behavior at these scales throughout the CZ. In the bulk ($0.85$~R$_\odot$), the largest scales also dominate the toroidal convective spectrum, with a maximum at $\ell_{\rm peak}=5$. However, this dominance of large-scales weakens as the surface is approached. Near-surface spectra ($0.96$ and $0.99$~R$_\odot$) even show an increasing trend with $\ell$ at large scales, with $\ell_{\rm peak}=22$ at $0.99$~R$_\odot$.

\begin{figure*}
  \centering
  \includegraphics[width=\linewidth]{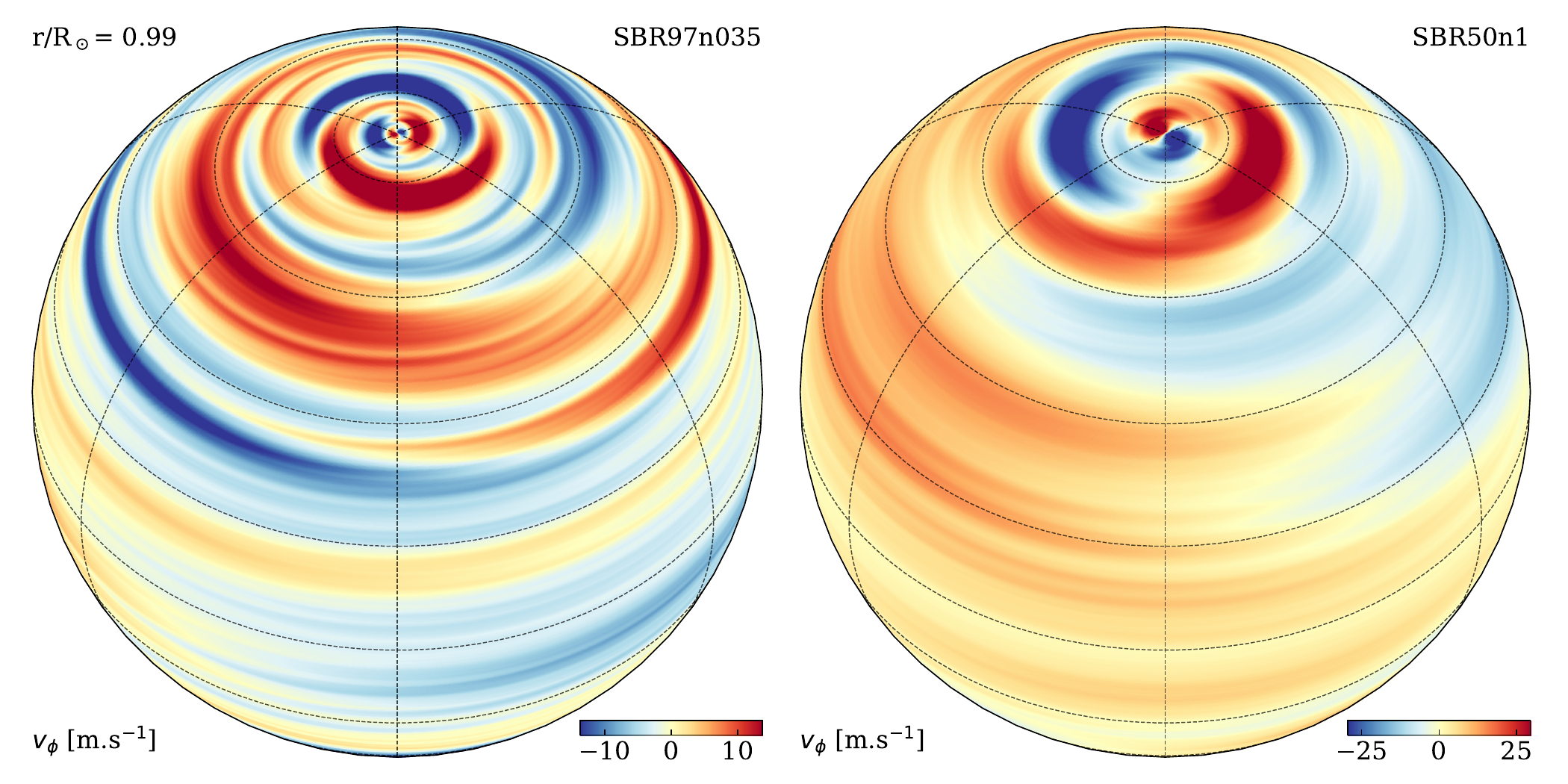}
  \caption{Eigenfunction of the $m=1$ mode taken from the longitudinal velocity $v_\phi$ at $r = 0.99 R_\odot$. \textit{Left and right} panels correspond respectively to SBR97n035 and SBR50n1. The color-bar maximum corresponds to two standard deviations of the given map.
  }
  \label{fig:Vphim1}
\end{figure*}

Such behavior of SBR97n035 is encouraging, as observations of the solar surface report a peak of the spectrum at the supergranular scale ($\ell\sim 120$) reaching $E_\phi = 100$ km$^3$s$^{-2}$ (\citetalias{proxaufObservationsLargescaleSolar2021}). This comparison must be done with great care as our models only reach 99\% of the solar radius, however the shifting of $\ell_{\rm peak}$ to higher $\ell$ values as we get closer to the top is coherent with surface observations.

A possible scenario emerges and correlates well with the schematics in Figure 3 of \cite{featherstoneEMERGENCESOLARSUPERGRANULATION2016}. We emphasize here that the impact of the Coriolis force is greater for large scales (low $\ell$). Indeed, this force becomes dominant in the dynamics of spatial scales $L$ larger than the Rossby deformation radius, \textit{i.e.} for $L>v/2\Omega$. The impact of the Coriolis force is more pronounced on the dynamics of the deep CZ, where the density scale height and thus convective scales are larger, which influences the morphology of the convection if the rotational constraint is strong enough \citep{1961hhs..book.....C,takehiroAssessmentCriticalConvection2020,hindmanMorphologicalClassificationConvective2020}. This is what happens in SBR97n035, where bulk velocities have been dampened, favoring lower amplitude columnar convective structures (banana cells) due to the Coriolis force, and limiting their imprint close to the surface. In the case of SBR50n1, convective velocity amplitudes are larger, which allows the large convective scales formed deeply to imprint more strongly the surface than in SBR97n035, and to be less rotationally constrained.

Finally, we see that the overall amplitude of the spectrum increases with the radius for both cases, in coherence with $v_{\rm rms}$ peaking around $0.98$~R$_\odot$, as can be guessed from Figure~\ref{fig:Rossby_comp}. We also see that the slope at small-scales changes between curves, due to the change in turbulent viscosity value. Indeed, the higher the viscosity is, the steeper the slope will be. Let's remind here that the viscosity in SBR50n1 is higher than for SBR97n035, and that it increases as a function of the radius in both models (see Appendix~\ref{sec:app_num_set} for further details).

In summary, the Sun-like rotating model SBR97n035 exhibits a clear dampening of the large-scales ($\ell\leq 20$) amplitudes of the convection at the surface in comparison to its bulk dynamics. SBR50n1 does not experience such a change, and keep a large-scale dominated spectrum all over the CZ. We aim to explore the underlying reasons for this behavior in Section~\ref{sec:SpectralAna}, a strong rotational constraint can have a strong impact on energy transfers among scales (see for instance \citealt{1992A&A...263..387D}).

\subsection{Inertial modes}

Solar inertial modes result from the interaction of low-frequency waves, the restoring force of which is the Coriolis force \citep{greenspan1968}. They were not observed on the Sun until very recently, due to long-term and high-precision observations required to detect them (\citealt{loeptien2018}, see also \citealt{hottaDynamicsLargeScaleSolar2023}). Recently, \cite{gizonSolarInertialModes2021} reported the detection of several of these modes in data from both SDO/HMI and GONG. We focus our interest here on the $m = 1$ surface mode they report in observed horizontal velocities, because of its relatively high amplitude ($v_\phi\simeq 15$ m~s$^{-1}$). This mode was also detected in previous studies, but not interpreted as such \citep{hathawayGiantConvectionCells2013, bogartEVOLUTIONNEARSURFACEFLOWS2015, howePersistentNearSurfaceFlow2015}. The amplitude of this mode is maximal at high latitudes ($|\lambda|=|90^\circ-\theta| \geq 50^\circ$), where the phase speed becomes similar to the local differential rotation speed, showing a spiral pattern in these polar regions. This mode is a quasi-toroidal mode, which means that its motions mainly lie in the horizontal spherical surface.

We extract this spherical harmonic component of $v_\phi$ at $0.99$~R$_\odot$, average it over time, and show it over a North-polar view for SBR97n035 and SBR50n1 in Figure~\ref{fig:Vphim1}. A South-Pole perspective is also proposed in Figure~\ref{fig:Vphim1South} of the Appendix. In both models, we note that the eigenfunction's is compatible with the mode observed in terms of amplitude. However, only SBR97n035 catch a closer value ($v_\phi\simeq 10$ m~s$^{-1}$), as amplitudes in SBR50n1 are twice higher. We also report for both the characteristic spiral pattern in the polar regions. Notably, in the SBR97n035 model (left panel), the pattern spirals outward (\textit{i.e.} away from the pole) in an anti-clockwise direction, consistent with solar observations. Conversely, in the SBR50n1 model (right panel), the spiral pattern moves outward in a clockwise direction, allowing then a clear distinction of both models using the $m=1$ component signals. To better characterize the Rossby modes, we will continue running these models, as long time series are needed to improve further their analysis.

Using a 2.5D linear eigenvalue solver applied to a differentially rotating CZ model, \cite{gizonSolarInertialModes2021} and \cite{bekkiTheorySolarOscillations2022b} have shown that the direction of the $m=1$ mode spiral pattern is also sensitive to the superadiabaticity amplitude. The compatibility of our model SBR97n035 with solar observations regarding the $m = 1$ mode morphology is then not only the result of a prograde equator, but could therefore also originate from a correct capture of superadiabaticity, which we now turn to.

\subsection{Superadiabaticity}

\begin{figure}
  \centering
  \includegraphics[width=\linewidth]{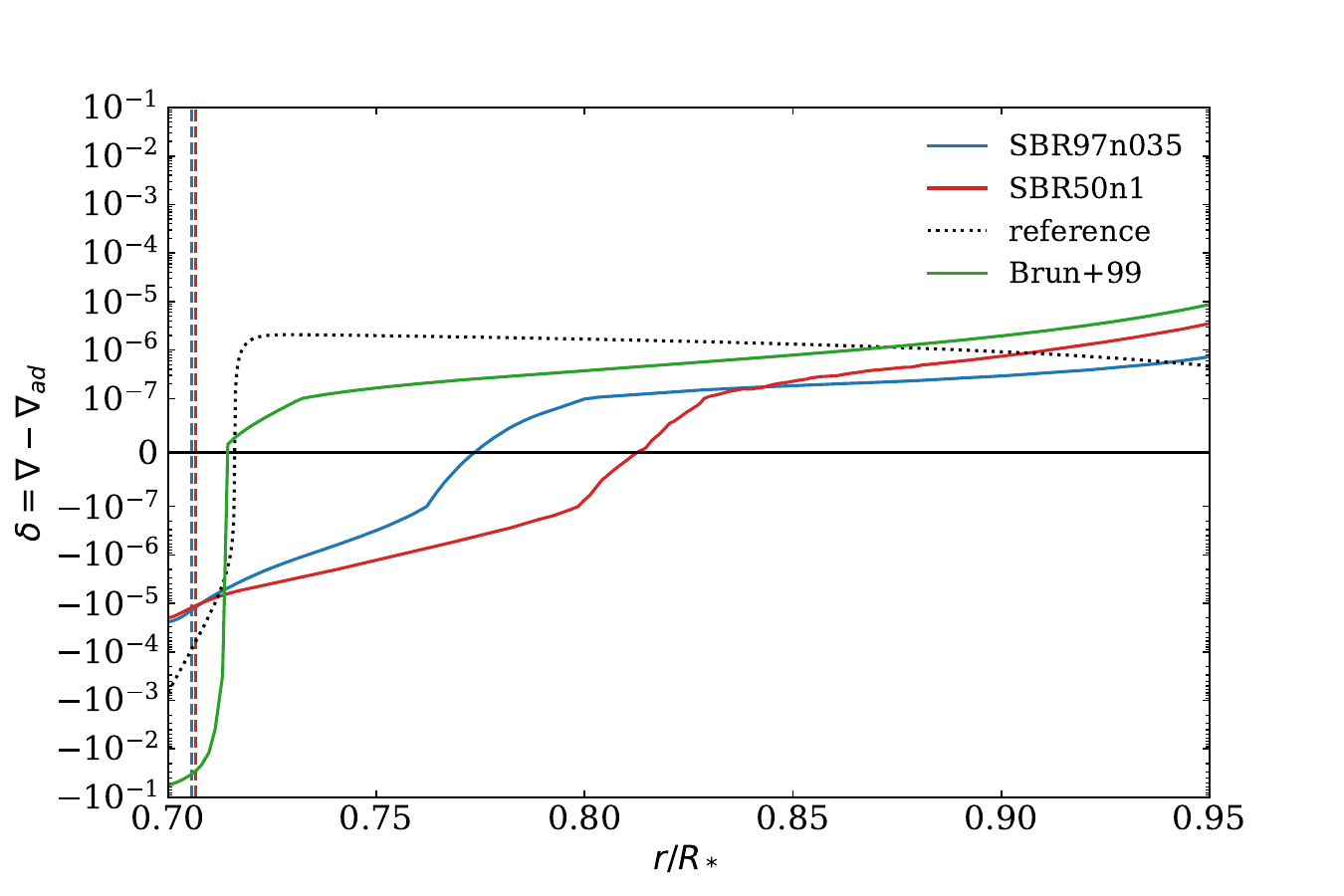}
  \caption{Superadiabaticity index $\delta=\nabla-\nabla_{ad}$ profiles for the different models. The black dotted line represents the initial reference state prescribed as initial condition, along with solid lines representing the respective converged states. We indicate the position of the overshoot, where $F_{\rm en}$ flips its sign with vertical dashed lines (see also top panels of Figures~\ref{fig:SBR97n035vsSBR50n1}). Finally, the green solid line shows the $\delta$ profile taken from the seismically constrained 1D model of \cite{1999ApJ...525.1032B}.
  }
  \label{fig:deltas}
\end{figure}

The influence of the mean entropy stratification on the dynamics is evaluated by the superadiabaticity $\delta$, defined such as
\begin{equation}
    \delta = \nabla - \nabla_{ad} = -\frac{H_P}{c_P}\frac{ds_{\rm tot}}{dr},
    \label{eq:delta}
\end{equation}
where $H_P$ is the pressure scale height, $\nabla=\frac{d\,ln\,T}{d\,ln\,P}$ is the logarithmic derivative of temperature with respect to pressure and $\nabla_{ad}=\frac{\gamma-1}{\gamma}$ is its adiabatic value. The sign of $\delta$ thus indicates if the system is convectively stable ($\delta<0$) or unstable ($\delta>0$). We illustrate in Figure~\ref{fig:deltas} the reference superadiabaticity $\delta$ profile we prescribed as initial condition in both SBR97n035 and SBR50n1 models with the black dotted line. After the convective instability develops and non-linearly saturates, the superadiabaticity  converges to a new profile shown by the blue and red solid lines for the two models. To compare them with the case of the Sun, we also add the helioseismically constrained 1D solar structure model of \cite{1999ApJ...525.1032B} (green line).

First, we note that both SBR97n035 and SBR50n1 models show significant changes in the superadiabaticity profile $\delta$ from its initial state (dotted line). Turbulent convection develops a mean entropy gradient $\langle ds/dr \rangle$, altering the initially prescribed profile $d\bar{s}/dr$. We note that $\delta$ decreases in the lower convection zone and increases in the upper part. This also results in a profile approaching that of the 1D model by \cite{1999ApJ...525.1032B} (green solid line) as the Nusselt number is decreased.  The green line was calibrated to the Sun using helioseismic constraints from SOHO (e.g., \citealt{gabrielPerformanceEarlyResults1997}). Furthermore, recent helioseismic studies suggest subadiabaticity ($\delta < 0$) near the base of the convection zone \citep{gizonSolarInertialModes2021}, although such constraints are still currently improved \citep{bekkiNumericalStudyNontoroidal2024}. Thus, even if the exact superadiabaticity profile in the Sun is still investigated, our models are consistent with current constraints.

Some differences are however present, but nevertheless expected as a result of the turbulent and non-linear 3D dynamics of SBR97n035 and SBR50n1. In particular, we note the appearance of a sub-adiabatic region ($\delta<0$) at the base of the CZ in both models. Such sub-adiabatic layers have previously been reported in various numerical models \citep{1993A&A...277...93R,hottaSolarOvershootRegion2017,kapylaEffectsSubadiabaticLayer2019,hottaGenerationSolarlikeDifferential2022,warneckeSmallscaleLargescaleDynamos2024}. This type of layer is further enhanced in models experiencing a relatively high Prandtl number $Pr=\nu/\kappa$ \citep{bekkiConvectiveVelocitySuppression2017,karakConsequencesHighEffective2018}. The dashed vertical lines show the location where the enthalpy flux $F_{\rm en}$ becomes negative at the top of the overshoot region (OR) for both models. The sub-adiabatic ($\delta<0$) region above this radius is called the Deardorff zone (DZ, in reference to \citealt{1961JAtS...18..540D,1966JAtS...23..503D}). In such zone, the plasma is thermally stable to convection, but non-local transport of heat with positive enthalpy flux is still happening (see Figure~\ref{fig:SBR97n035vsSBR50n1}, also \citealt{brandenburgSTELLARMIXINGLENGTH2016}).

It has been shown that the presence of such sub-adiabatic stable zone at the base of the CZ can help to damp large-scale convective amplitudes, by suppressing thermal instability in depth where the convection driving scale is larger due to the increased pressure scale height \citep{bekkiConvectiveVelocitySuppression2017}. Such mechanism has then been advocated as a possibility to explain the smaller amplitude of large-scale flows in observations, in comparison to current simulations. However, we observe here that the model exhibiting the strongest suppression of large-scale flows, as compared to its whole spectrum, is the one possessing the smallest DZ (SBR97n035, see Figure~\ref{fig:SpComp}). This indicates that the change of the superadiabaticity at the bottom of the CZ does not participate in the suppression of large-scale kinetic energy in our models. This latter is here mainly due to the suppression of buoyancy at these scales, as will now be illustrated in Section~\ref{sec:SpectralAna}.\\

In summary, one can say that our Sun-like rotating model SBR97n035 is in qualitative agreement with current solar observations, regarding the prograde equator, large-scale convection, the $m=1$ Rossby mode and the superadiabaticity profile. In order to further understand how such solar-type dynamical regime can be reached by numerical models, we wish to perform a systematic study of the different power and force balances in the next Section.

\section{Spectral analysis of dynamics equilibria}\label{sec:SpectralAna}

We aim here to quantify the change of convection morphology reported in Section~\ref{sec:Overview}. By decomposing the different terms of equations we solve, we can assess the power and energy balances sustained in both models. Comparing them allows underlining dynamical patterns that are key to construct Sun-like large-scale flows.

\subsection{Power balance}\label{sec:PowerBal}

In order to understand the energy distribution among the different convection scales, we first aim to develop and quantify the energy transfers happening at each scale, and thus building non-axisymmetric spectra showed in Figure~\ref{fig:SpComp}. We refer the reader to Appendix~\ref{app:spec_decomp} for the detailed derivation of the following evolution equations, in which we use and complement the spherical harmonics decomposition developed by \cite{strugarekMAGNETICENERGYCASCADE2013,strugarekModelingTurbulentStellar2016}. 

\subsubsection{Budgets of the kinetic energy spectrum}\label{sec:EkDiag}

\begin{figure*}
  \centering
  \includegraphics[width=0.48\linewidth]{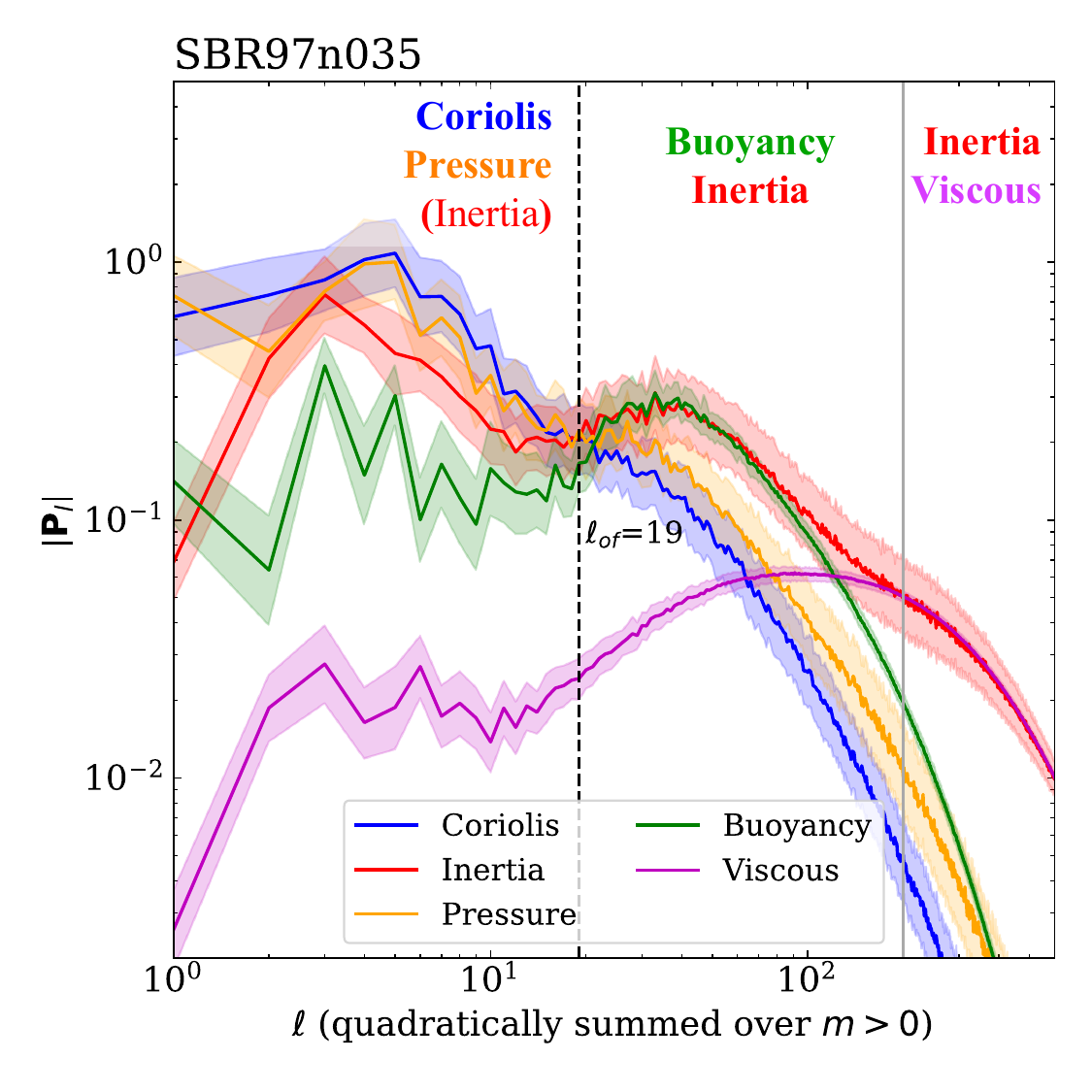}
  \includegraphics[width=0.48\linewidth]{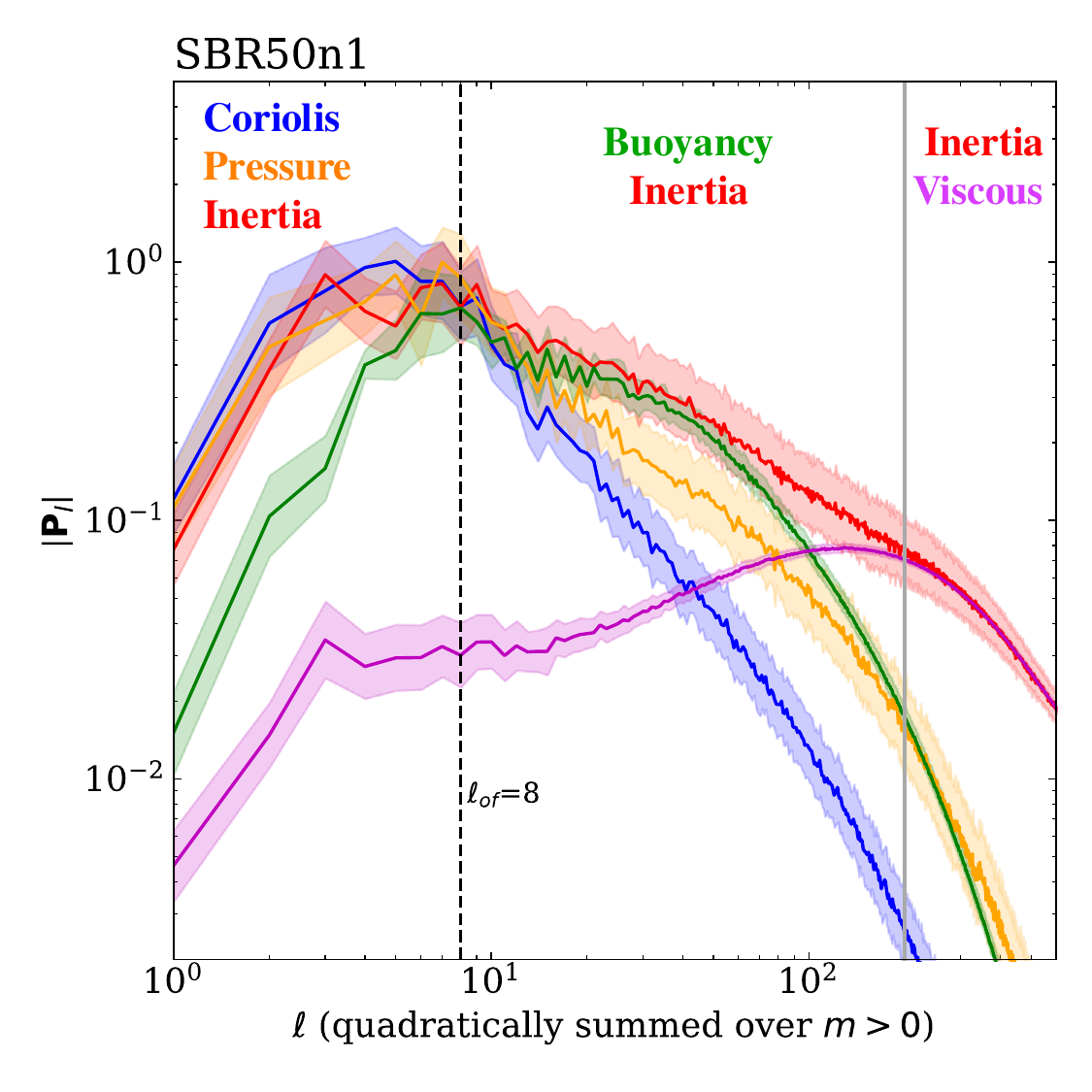}
  \caption{Non-axisymmetric spectrum of the power balance from the kinetic energy evolution of SBR97n035 (\textit{left}) and SBR50n1 (\textit{right}) models, represented as a function of the spherical harmonic degree $\ell$ (from large scales on the left to small ones on the right), and quadratically summed over $m>0$. Each spectrum is normalized to the peak amplitude of the pressure contribution $|{\cal H}_\ell|$, and computed as a root-mean-square (rms), such that $| \textbf{P}_\ell|=\sqrt{\int_t\int_r {\rm {\bf P}}^2_\ell r^2 {\rm d}t{\rm d}r/\int_t\int_r r^2{\rm d}t{\rm d}r}$. Solid lines represent the time/space rms, and corresponding shaded regions represent the standard deviation in time from it (The lower limit has been adapted graphically to match the upper one on a logarithmic scale). The dashed vertical lines targets the lowest scale where the Coriolis (blue) contribution is higher than the advective (red) one. Please note that this scale is smaller (higher degree) in the solar-DR model on the left. We also add a gray vertical line at $\ell=200$ for comparison. We highlight the main equilibria by indicating them.}
  \label{fig:RmsSPdiagKE}
\end{figure*}

We first focus on the evolution of kinetic energy density, whose budgets at a given radius $r$ can be summarized as
\begin{eqnarray}
  \label{eq:KEanaSH}
  \p_t E_\ell^K = \sum \textbf{P}_\ell &=& \overbrace{\sum\limits_{\substack{\ell_1,\ell_2\\|\ell_1-\ell_2|\leq \ell \leq \ell_1+\ell_2}} \big[{\cal R}_\ell(\ell_1,\ell_2)\big]}^{\footnotesize \mbox{\textcolor{red}{Reynolds stress}}}\nonumber\\ &+& \underbrace{{\cal C}_\ell(\ell-1,\ell+1)}_{\footnotesize \mbox{\textcolor{blue}{Coriolis force}}} + \overbrace{{\cal H}_\ell}^{\footnotesize \mbox{\textcolor{orange}{Pressure work}}}\nonumber\\ &+& \underbrace{{\cal B}_\ell}_{\footnotesize \mbox{\textcolor{Green}{Buoyancy}}}  + \overbrace{{\cal V}_\ell}^{\footnotesize \mbox{\textcolor{magenta}{Viscosity}}},
\end{eqnarray}
All right-hand-side power terms $\textbf{P}_\ell$ are detailed in Appendix~\ref{app:spec_decomp}, and the radial density of kinetic energy, \textit{i.e.} the energy of a spherical surface $S$ at radius $r$, is here defined such that $E^K_\ell = \frac{\bar{\rho}}{2}\int_S \vv_\ell\cdot\vv_\ell{\rm d}{\Omega}{}$, where ${\rm d}\Omega$ is the solid angle. At a given radius, the physical length $L(\ell,r)$ characterized by the harmonic degree $\ell$ is given by the relation $L=\frac{r}{\sqrt{\ell(\ell+1})}$. Note that the contribution from Coriolis ${\cal C}_\ell$ cancels when summed over the whole harmonics $\ell$ as it should (the Coriolis force does not do any work),  so that it does not convert energy locally from a reservoir to another, or transport it spatially via a flux. However, the Coriolis force is able to redistribute energy among neighbor spectral scales $\ell-1$ and $\ell+1$ (see Appendix~\ref{app:spec_decomp}, as well as \citealt{augierNewFormulationSpectral2013} and \citealt{strugarekModellingTurbulentStellar2016}), which can be seen physically as a mode conversion.\\

We show in Figure~\ref{fig:RmsSPdiagKE} the non-axisymmetric ($m>0$) power balance operating in both SBR97n035 and SBR50n1 models at each scale, by considering the root-mean-square (rms) value $|\textbf{P}_\ell|$ in time and radius of each right-hand-side terms of Equation~\ref{eq:KEanaSH}. We consider radius from $0.75$ to $0.9$~R$_\odot$ to reveal the actual power balance in the bulk of the convection zone, excluding then boundary-layer effects from the calculation. We observe 3 regimes depending on the scales considered in both models. At small scales (large $\ell$ degrees), we observe the \textit{dissipation regime} where an equilibrium occurs between the inertia by the Reynolds tensor (red), bringing kinetic into these scales, and the viscosity term (purple) dissipating it, as anticipated by turbulence theory \citep{1962JFM....13...82K}. This regime begins at a similar degree $\ell\sim 200$ for both models, which is consistent with both models having similar Reynolds numbers (see Table~\ref{tab:AB2vsAS1vsSunBusyAdim}). At intermediate scales, we observe a \textit{buoyant-inertial regime} where buoyancy (green) is mainly balanced by inertia, which this time withdraws energy from these scales. Finally, at large scales (small degrees $\ell$), we note a \textit{quasi-geostrophic} regime, where the main balance appears between the pressure gradient and the Coriolis force (geostrophy), but also inertia. 

We propose to characterize the influence of the Coriolis force using $\ell_{of}$, that we will now call the Rossby scale, and which represents the highest degree $\ell$ where the Coriolis contribution ${\cal C}_\ell$ is comparable to the contribution of advective transport ${\cal R}_\ell$. In other words, we can define a spectral Rossby number $Ro_{\rm SP}(\ell)={\cal R}^{m>0}_\ell/{\cal C}^{m>0}_\ell\sim\frac{v_\ell}{2\Omega L_\ell}$ with $v_\ell$ and $L_\ell$ respectively the typical velocity and spatial scale corresponding to the degree $\ell$. The Rossby scale $\ell_{of}$ is then defined such that $Ro_{\rm SP}(\ell_{of})= 1$.

We see that $\ell_{of}$ is a good measure to locate the transition from the quasi-geostrophic regime to the buoyant-inertial regime in both cases. We note this occurs at a smaller scale (higher degree $\ell$) in the SBR97n035 case (left panel, $\ell_{of}=19$ in comparison to 8), extending the quasi-geostrophic regime towards smaller scales for that model. As a result, the rotational constraint is felt by fewer scales in the anti-solar case SBR50n1 (right panel). Inversely, the dampening of convective velocities in SBR97n035 makes the dynamics more sensitive to the rotation, especially at large scales, where the regime is even closer to geostrophy in this Sun-like rotating case.

We finally note that the Coriolis (blue) and buoyancy (green) contributions intersect close to $\ell_{of}$, meaning $Ro_{\rm c}\simeq Ro_{\rm f}\simeq 1$ at this scale. At larger scales (smaller $\ell$), the Coriolis term is nearly one order of magnitude higher than the buoyancy, and thus the convective excitation by buoyancy is limited by rotation for these large scales. This impact of rotation is well illustrated for SBR97n035 on the left panel, as we decreased the Nusselt number of this model, and then limited the amplitude of velocities resulting from the convective instability.

\subsubsection{Angular-momentum transporting scales}\label{sec:DRscales}

\begin{figure}
  \centering
  \includegraphics[width=\linewidth]{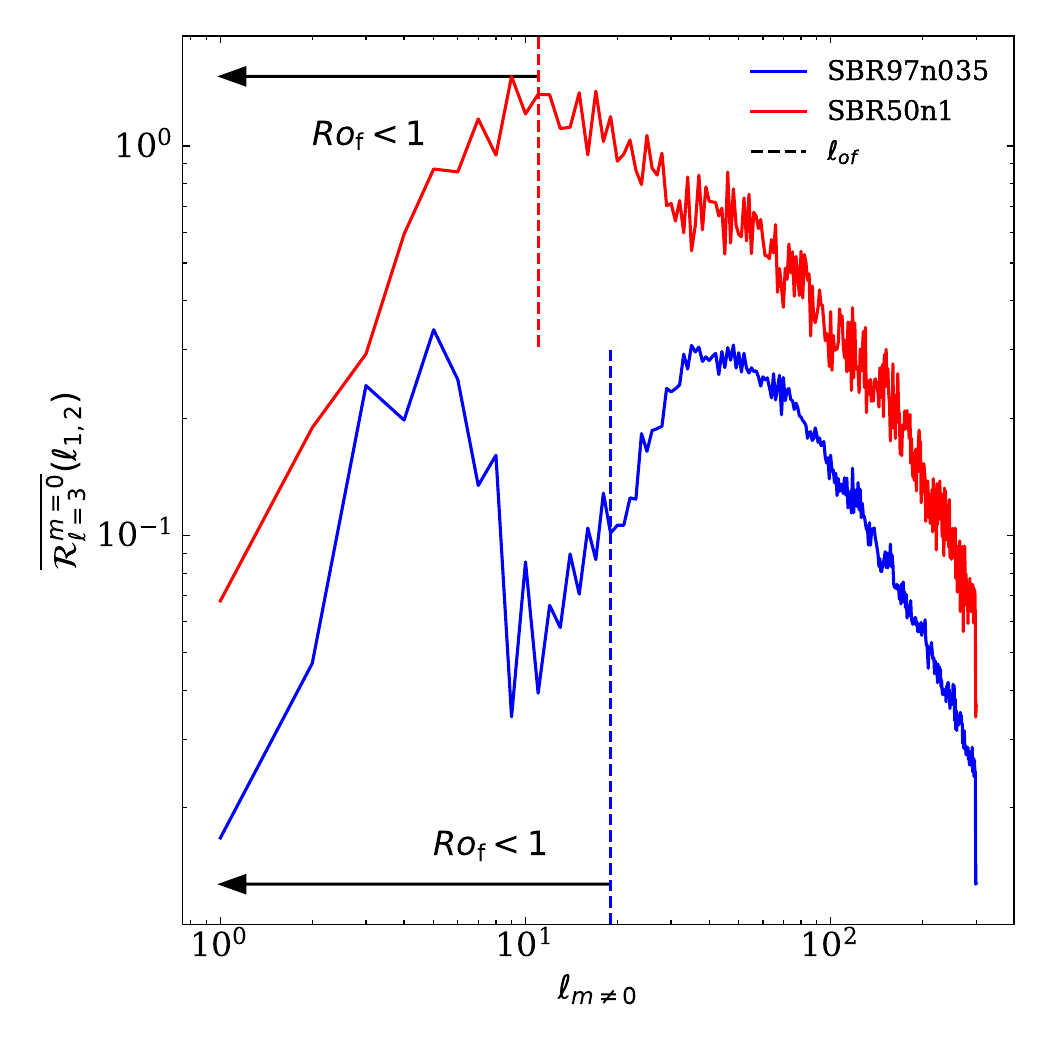}
  \caption{Spectrum $\overline{ {\cal R}_{\ell=3}^{m=0}}(\ell_{m\neq0})$ of the mean transfer of a non-axisymmetric convective scale $\ell_{{m\neq 0}}$ toward the axisymmetric mode of the DR ($\ell_{m=0}=3$). We illustrate SBR97n035 and SBR50n1 in blue and red respectively, with their corresponding $\ell_{of}$ taken from Figure~\ref{fig:RmsSPdiagKE}. In particular, we see that a significant part of transfers have been shifted at larger scales in the Sun-like rotating case (blue), where more scales feels the rotational constraint of the Coriolis force ($Ro_{\rm f}<1$).
  }
  \label{fig:Inject2Dcomp}
\end{figure}

One of the main striking difference, between SBR97n035 and SBR50n1 models, lies in the fact that they present reversed differential rotation profiles, despite reaching a very similar turbulence degree and overall similar regimes of power balance (see previous section). The detailed angular momentum balance can be found in Appendix \ref{sec:AmomBal}. Here, we focus on identifying the convective scales that are responsible for the build-up and maintenance of the differential rotation profile.

We consider the Reynolds tensor ${\cal R}_\ell(\ell_1,\ell_2)$, quantifying the non-linear advection of momentum in Equation~\ref{eq:KEanaSH}. This tensor involves triadic interactions following the triangular selection rule ($|\ell_1-\ell_2|\leq \ell \leq \ell_1+\ell_2$; $m_1 + m_2 = m$) of spherical harmonics. These terms represent how non-linear interactions between two different scales $\ell_1$ and $\ell_2$ can act as a source or a sink of kinetic energy for the scale of interest $\ell$. We choose to consider here how interactions between non-axisymmetric modes transfer energy towards a large-scale axisymmetric mode. In other terms, how interactions between different convective scales play a role in the maintenance of the DR.

For this purpose, we consider $\overline{ {\cal R}_{\ell=3}^{m=0}}(\ell_{1,2})$ the normalized spectrum of mean unsigned non-axisymmetric contributions of each scale $\ell_{1,2}$, towards the axisymmetric mode of the DR ($\ell=3,m=0$) defined in Equation~\ref{eq:R30}. As the tensor ${\cal R}_{\ell=3}(\ell_1,\ell_2)$ of non-axisymmetric interactions $(\ell_1,\ell_2)$ is bi-dimensional, we show in Figure~\ref{fig:Inject2Dcomp} the averaged contribution $\overline{ {\cal R}_{\ell=3}^{m=0}}(\ell_{m\neq0})=\langle\overline{ {\cal R}_{\ell=3}^{m=0}}(\ell_x)\rangle_{x=\{1;2\} }$ (\textit{i.e.} we sum over $\ell_1$ or $\ell_2$), following Equation~\ref{eq:R30}, for SBR97n035 and SBR50n1 in blue and red respectively. We also report the respective Rossby scale defined in Figure~\ref{fig:RmsSPdiagKE} for each case, using vertical dashed lines.

We first see that the morphology of the kinetic energy transfers spectrum has changed. In the anti-solar rotating case (red), the strongest contributions to energy transfer are located at scales of the order of the Rossby scale ($\ell_{of}=8$), where the transition between the quasi-geostrophic and buoyant-inertial regimes happens. For the solar-DR rotating case (blue), the spectrum is now bi-modal, with two distinct peaks of contributions in the spectrum. One of them is located at higher scales (smaller degree $\ell$) in comparison to the anti-solar rotating case SBR50n1. The stronger influence of rotation previously visualized in Section~\ref{sec:Overview} for the Sun-like rotating case SBR97n035 has now shifted a significant part of the transfer at larger scales ($3\leq\ell\leq 7$). In addition, the Rossby scale is now located at a smaller scales ($\ell_{of}=19$ \textit{v.s.} 8) because of the rotation influence, meaning that a larger part of energy transfers towards the large scale axisymmetric differential rotation are now originating from non-axisymmetric convective modes that are significantly influenced by the Coriolis force (\textit{i.e.} experiencing $Ro_{\rm SP}(\ell)<1$). We note that we have also carried out the same analysis for the transfers towards the axisymmetric mode $\ell=5$, and we found the exact same behavior. Hence, we confirm that our $Nu$-controlled case is significantly different regarding the angular momentum transport explaining the solar-DR.

\subsubsection{Budgets of the internal energy spectrum} \label{sec:internSbudget}

\begin{figure*}
  \centering
  \includegraphics[width=0.48\linewidth]{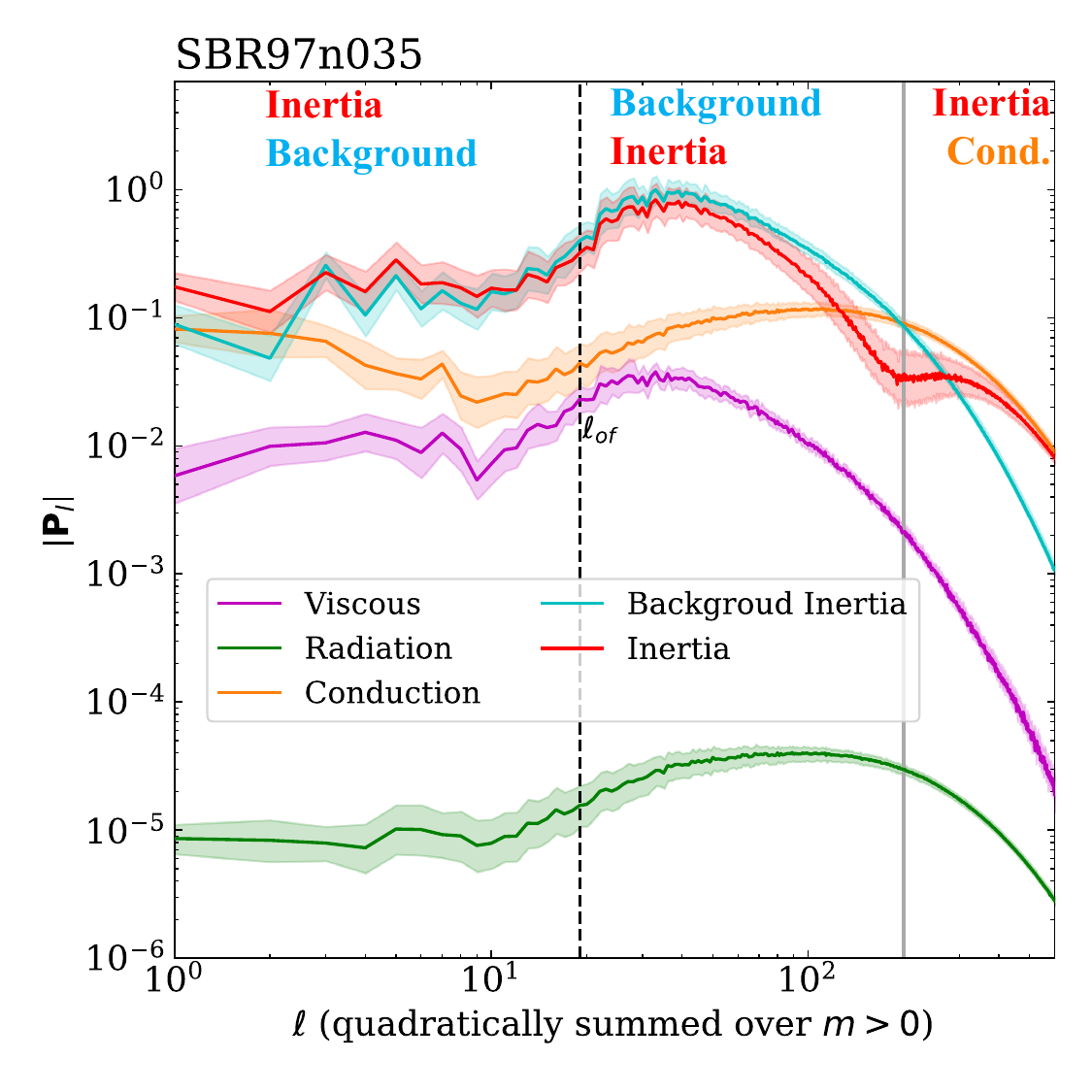}
  \includegraphics[width=0.48\linewidth]{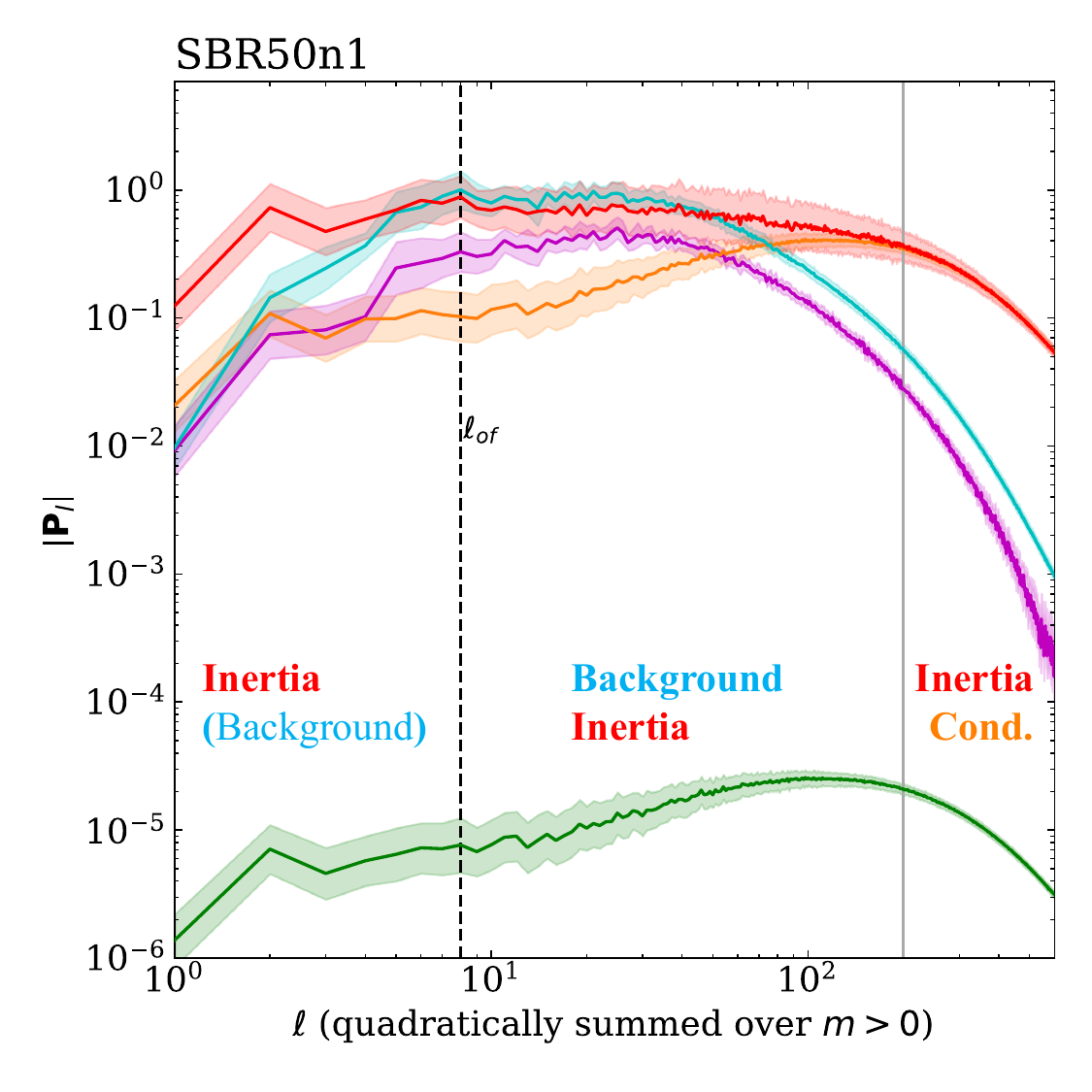}
  \caption{Non-axisymmetric spectrum of the power balance from the internal energy evolution of SBR97n035 (\textit{left}) and SBR50n1 (\textit{right}) models. The Figure is constructed similarly to Figure~\ref{fig:RmsSPdiagKE}, computing this time right-hand-side terms of the internal energy spectrum evolution (\ref{eq:S2anaSH}). The dashed vertical line represent $\ell_{of}$, taken from Figure~\ref{fig:RmsSPdiagKE}. Each spectrum is normalized to the peak amplitude of the background contribution $|{\cal S}^b_\ell|$, and computed as a root-mean-square (rms) over time and radii from $0.75$ to $0.9$~R$_\odot$. We highlight the main equilibria by indicating them.}
  \label{fig:RmsSPdiagS2}
\end{figure*}

Having looked at kinetic energy transfers sustaining the convection, we now turn our attention to the power balance sustaining the internal energy spectrum. To do so, we  assess heat transfers with an analysis of the internal energy balance. Similarly to Equation~\ref{eq:KEanaSH}, we can construct the radial density spectrum of the quadratic entropy ${\cal E}^S_\ell = \frac{1}{2}\int_S S_\ell\cdot S_\ell{\rm d}{\Omega}{}$ decomposed over the spherical harmonics. We can then use it as a proxy to follow the internal energy evolution from Equation~\ref{eq:Etot} such as:
\begin{eqnarray}
  \label{eq:S2anaSH}
  \p_t {\cal E}_\ell^S = \sum \textbf{P}_\ell &=& \overbrace{\sum\limits_{\substack{\ell_1,\ell_2\\|\ell_1-\ell_2|\leq \ell \leq \ell_1+\ell_2}} \big[{\cal A}_\ell(\ell_1,\ell_2)\big]}^{\footnotesize \mbox{\textcolor{red}{Inertia}}} \nonumber\\ 
  &+& \underbrace{{\cal S}_\ell^b}_{\footnotesize \mbox{\textcolor{cyan}{Background Inertia}}} + \overbrace{{\cal T}_\ell}^{\footnotesize \mbox{\textcolor{Green}{Radiation}}} \nonumber\\ &+& \underbrace{{\cal K}_\ell}_{\footnotesize \mbox{\textcolor{orange}{Thermal Dissipation}}} + \overbrace{{\cal V}^S_\ell}^{\footnotesize \mbox{\textcolor{magenta}{Viscosity}}}  \nonumber\\ &+& \underbrace{{\cal Q}_0\delta_{\ell,0}}_{\footnotesize \mbox{Background net flux}},
\end{eqnarray}
where $\delta_{\ell,0}$ is the Kronecker symbol centered on the harmonic degree $\ell=0$, meaning here that ${\cal Q}_0$ includes terms which contribute only to the spherically symmetric ${\ell=0}$ component of the Equation~\ref{eq:S2anaSH}, \textit{i.e.} representing a net radial flux through the shell of interest located at a given radius.

We consider root-mean-square (\textit{rms}) values $|\textbf{P}_\ell|$ of right-hand-side terms in Equation~\ref{eq:S2anaSH} and show the balance of their non-axisymmetric component in Figure~\ref{fig:RmsSPdiagS2}, in order to focus on the convective dynamics. First, we observe 3 regimes in both cases: At small scales (high degree $\ell$), the thermal diffusion (orange) balances the entropy brought at these scales by the advective transport $(\vv\cdot\nab)S$ (red). At intermediate scales, the contribution of the latter switch and becomes negative, due to the transport of entropy by convection. This convective instability is triggered in reaction to the negative $d\sba/dr$ background gradient of entropy in the CZ, represented here by a positive background contribution (cyan). At large scales, a similar balance occurs, however we note a decrease of the amplitude for both the inertial and background contributions. We further notice that the background contribution experiences a particularly strong drop in SBR50n1 for $\ell\leq4$. Overall, we note that the 3 different regimes found here correspond to the \textit{quasi-geostrophic}, \textit{buoyant-inertial} and \textit{dissipation} ones found in Figure~\ref{fig:RmsSPdiagKE}. The large-scale balance thus extends to higher degrees $\ell$ for SBR97n035 (left panel) in this diagnostics due to the stronger rotational influence.

Similarly to Figure~\ref{fig:RmsSPdiagKE}, we also report differences in the shape of spectrums between both models. On one hand, we see that the behavior of the thermal term (orange) does not change. In particular, it becomes the main contributor to the terms balance in both models at $\ell\sim200$, as expected from them sharing a similar Peclet number. The shape of the background contribution (cyan) is also similar for small scales $\ell\geq 30-40$ between both models, but on the other hand, it changes at larger scales and decreases in amplitude for $\ell\leq\ell_{of}$. This emphasizes the influence of the Coriolis force as responsible for such a change. The scale where maximum non-axisymmetric power $|{\bf P}_\ell|$ sustains the internal energy has thus been moved from $\ell\sim 8$ in the anti-solar rotating case, to a more pronounced peak contributions around $\ell\sim 30-40$ in the Sun-like rotating case.

It is interesting to note that the radiative term (green) is negligible over the entire non-axisymmetric spectrum for both models, despite the significant increase of $\kappa_{\rm rad}$ when controlling the Nusselt number $Nu$ of both models. Indeed, this term contributes almost exclusively to the spherical mode $\ell=0$, meaning that it is a net radial transfer contributing globally over the whole shell (see for instance the green contribution of flux balances in Figure~\ref{fig:SBR97n035vsSBR50n1}). Therefore, we underline here the importance to control our exploratory path of the parameter space with a coefficient that does not act on the non-axisymmetric balance. An increase of, for instance, $\kappa$ would have amplified the non-axisymmetric contribution of the thermal term (orange), which is already significantly acting in the power balance here. This would have then impacted arbitrarily all scales by turbulent cascade.

Furthermore, we note that the entropy advective spectrum (red) has changed all over the scales. In the Sun-like rotating case SBR97n035, its contribution to the balance has globally decreased, which is likely due to the dampening of velocities implied by our control of the Nusselt number $Nu$. We note an interesting minimum at the transition towards the dissipation regime ($\ell\sim 200$), making the dynamics in this small scale range resulting from a balance between the thermal dissipation (orange) and the background inertia (cyan), which is not happening in the anti-solar case SBR50n1. We then emphasize here that by impacting the transport of entropy at the largest scales, its nature has also been changed at smaller scales.

In that sense, an increase of the radiative diffusion does not impact directly the non-axisymmetric balance of heat transfers through the amplitude of the radiative transfer term ${\cal T}_\ell$, but still plays a role on the nature of this heat dissipation cascade, by acting indirectly through the velocity field spectra.

\begin{figure*}
  \centering
  \includegraphics[width=0.49\linewidth]{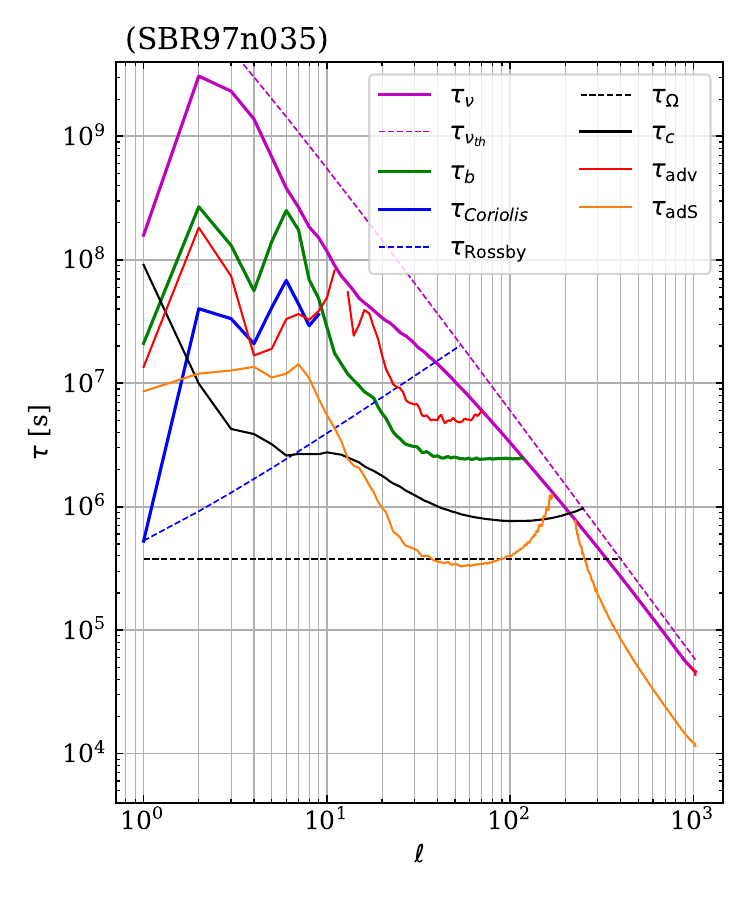}
  \includegraphics[width=0.49\linewidth]{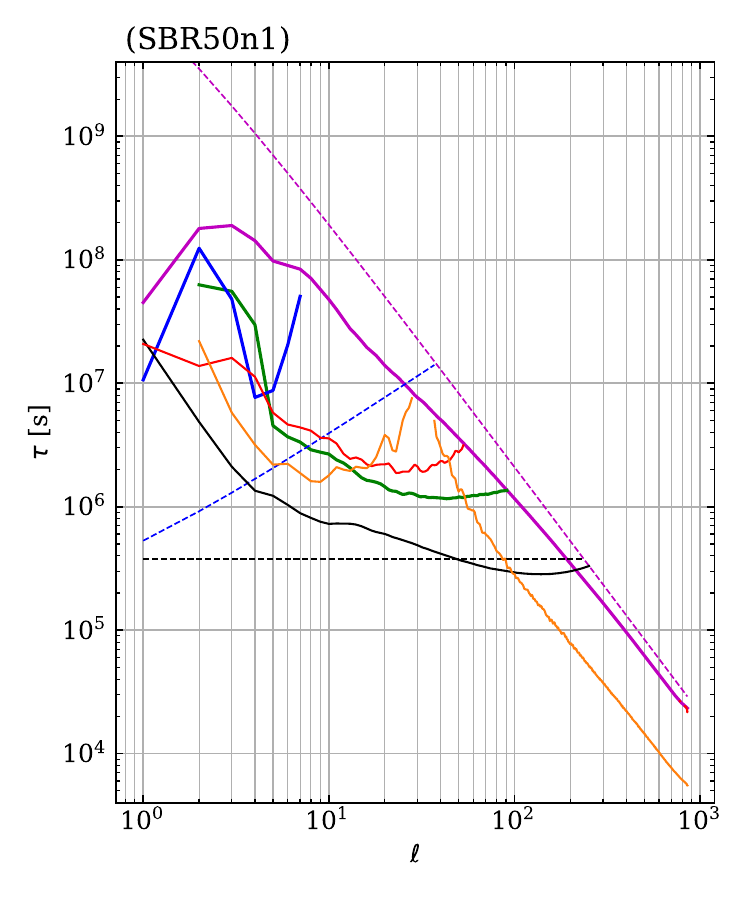}
  \caption{$\tau-\ell$ Diagrams of the different processes in the bulk of the convective zone for the SBR97n035 \textit{left} and SBR50n1 \textit{right}. Solid lines represent different characteristic time-scales extracted from both models. The  viscosity (violet), Coriolis (blue), buoyancy (green), convective turnover (black), advection of momentum (red) and quadratic entropy (orange) time-scales are respectively defined such that $\tau_\nu=E_\ell^K/{\cal V}_\ell$, $\tau_{Coriolis}=E_\ell^K/{\cal C}_\ell$, $\tau_b=E_\ell^K/{\cal B}_\ell$, $\tau_c=L(\ell)/\sqrt{2E_\ell^K/\rho}=L(\ell)/v_\ell$, $\tau_{\rm adv}=E_\ell^K/|{\cal R}_\ell|$ and $\tau_{\rm adS}=E_\ell^S/|{\cal A}_\ell|$. Each spectrum show the non-axisymmetric contribution, averaged in time and space over $r\in [0.75,0.9]$~R$_\odot$. The dashed lines represent the theoretical characteristic times for comparison, such that: the viscous time of each scale, $\tau_{\nu_{th}}=L(\ell)^2/\nu$, the rotation time-scale $\tau_{\Omega}=1/\Omega_\odot$ and $\tau_{Rossby}=r/(\Omega L(\ell))$ the propagation time for a Rossby wave of size $L(\ell)$. At a given radius, the physical length $L(\ell,r)$ characterized by the harmonic degree $\ell$ is given by the relation $L=\frac{r}{\sqrt{\ell(\ell+1})}$.
  }
  \label{fig:CompTauEll}
\end{figure*}

\subsection{Scale by scale time scale : $\tau$- $\ell$ diagrams}\label{sec:tauEll}

After looking at the amplitude of the different energy transfer terms, we now present non-axisymmetric spectra of the characteristic time scales associated with some of them. Inspired by the theoretical work of \cite{natafTurbulenceCore2015}, such $\tau$-$\ell$ \textit{diagrams} allow comparing the different transfer time-scales between the different energy reservoirs (see also the recent \citealt{natafDynamicRegimesPlanetary2023}). This enables to highlight which process(es) predominantly control(s)/influence(s) the dynamics at a given scale, by possessing a time-scale shorter than other processes. To extract such a time-scale from our simulations, we look at the ratio between the kinetic $|E_\ell^K|$ or internal energy $|E_\ell^S|$, and a power term $|\textbf{P}_\ell|$ contributing to it (defined in the right-hand-side term of either Equation~\ref{eq:KEanaSH} or \ref{eq:S2anaSH}). This corresponds to the typical time of energy injection/extraction of this term into/from the corresponding energy reservoir. We look at the corresponding non-axisymmetric spectrum for each ratio, averaged over time and radii ($r\in [0.75,0.9]$~R$_\odot$). For information, $\ell=10$ and 100 correspond to physical lengths of 56 and 5.6 Mm, respectively, in the middle of the convection zone (0.85~R$_\odot$). We show in the left and right panels of Figure~\ref{fig:CompTauEll} $\tau-\ell$ diagrams of the Sun-like (SBR97n035) and anti-solar rotating case (SBR50n1) respectively.

First, we plot the characteristic numerical viscous time-scale $\tau_\nu=E_\ell^K/|{\cal V}_\ell|$ in thick purple lines, found in our simulations. Every other line crossing this viscous border and going at longer time-scales will then be dominated by the viscous contribution. Therefore, we choose to omit the time scales higher than $\tau_\nu$ for the sake of clarity. The dashed purple line represents the theoretical dissipation time $\tau_{\nu_{th}}(\ell)=\int_r r^2L(\ell,r)^2/\nu(r){\rm d}r\;/\;\int_r r^2 {\rm d}r$, spherically averaged over $r\in [0.75,0.9]$~R$_\odot$. We note that the numerical viscous time $\tau_\nu$ is different from the theoretical one $\tau_{\nu_{\rm th}}$ for both cases. Indeed, the latter illustrates a purely isotropic dissipative assumption, which means that the difference $\tau_\nu < \tau_{\nu_{th}}$ likely comes from the anisotropy of velocity gradients.

Then, $\tau_{\rm Rossby}=r/(\Omega_\odot L(\ell,r))$ (dashed-blue diagonal line) represents the theoretical propagation time of a $\ell$-degree Rossby wave moving horizontally on the spherical surface of radius $r$ \citep{natafTurbulenceCore2015}. Here, we show its averaged value over $r\in [0.75,0.9]$~R$_\odot$. Above this line, the regime is supposed to be quasi-geostrophic. We then show the Coriolis contribution $\tau_{\rm Coriolis}=E_\ell^K/|{\cal C}_\ell|$ and the characteristic time of rotation $\tau_\Omega=1/\Omega=P_{\rm rot}/2\pi$, indicated by the blue solid curve and dark horizontal dashed line, respectively. The latter is independent of the scale considered. Similarly, we notice for both models a difference between the Coriolis timescale $\tau_{\rm Coriolis}$ and the theoretical one $\tau_{\rm Rossby}$, as the latter described the idealized bi-dimensional motion of Rossby waves along a spherical plane and the former acknowledges for the three-dimensionality of the convective dynamics we are resolving. Nevertheless, both time-scales have the same trend (decreasing as a function of $\ell$) as the Coriolis force has a stronger influence on large scales $L(\ell)$.

We further show $\tau_{\rm adv}=E_\ell^K/|{\cal R}_\ell|$ and $\tau_{b}=E_\ell^K/|{\cal B}_\ell|$, the characteristic time-scale of inertia and buoyancy (red and green) in the kinetic energy evolution respectively. We see that the decrease of inertia spectrum  at large scales is steeper in SBR97n035, which is coherent with the right panel in Figure~\ref{fig:SpComp}. Indeed, the $\tau_{\rm adv}-\tau_{\rm Rossby}$ crossing happens at larger scale (\textit{i.e.} lower $\ell$) in SBR50n1. More generally, the overall time-scale of the inertia $\tau_{\rm adv}$ in SBR97n035 is longer than in SBR50n1, coinciding with the decrease of convective velocities in this model.

We also illustrate the spectrum of the convective turnover time $\tau_c$ with the solid black curve, averaged over $r\in [0.75,0.9]$~R$_\odot$ such that $\tau_{c}(\ell)=\int_r r^2 L(\ell,r)/\sqrt{2 E_\ell^K(\ell,r)/\bar{\rho}(r)}{\rm d}r\;/\;\int_r r^2 {\rm d}r$. Its intersections with the theoretical $\tau_{\nu_{th}}$ and $\tau_{\rm Rossby}$ dashed lines respectively define the \textit{Kolmogorov} $L(\ell_\nu)$ and \textit{Rhines} $L(\ell_{\rm Rh})$ scales, where $Re=v(\ell_\nu)L(\ell_\nu)/\nu=1$ and $L(\ell_{\rm Rh})\simeq\sqrt{v(\ell_{\rm Rh})/\Omega}$\footnote{More precisely, the Rhines scale is exactly defined as $\sqrt{u\cdot r/(2\Omega {\rm sin}(\theta)).
}$, where $\theta$ is the co-latitude.} respectively (\citealt{1941DoSSR..30..301K,rhinesWavesTurbulenceBetaplane1975}, see also \citealt{natafDynamicRegimesPlanetary2023}). We note that SBR50n1 (right panel) exhibits a shorter convective turnover timescale $\tau_c$ than SBR97n035 (left panel) for a given degree $\ell$, as expected. However, when the intersection of $\tau_c$ and $\tau_{\nu_{\rm th}}$, we see both models possess a similar Kolmogorov scale $\ell_\nu\sim 250$, hence a similar extent of the turbulence cascade despite their difference in numerical resolution ($N_\theta=1024$ for SBR50n1 and 1536 for SBR97n035, see Table~\ref{tab:AB2vsAS1vsSunBusy}). As for the similar Reynolds number $Re$, hence turbulence degree, we reported in Table~\ref{tab:AB2vsAS1vsSunBusyAdim}, this is due to the lower viscosity value $\nu$ of SBR97n035 compensating for its lower velocities amplitudes. 

Another interesting property emerges when examining larger scales where $\tau_c$ crosses $\tau_{\rm Rossby}$. The Rhines scale is larger (smaller $\ell_{\rm Rh}$) in SBR50n1 than in SBR97n035. Although the $\tau_{\rm Rossby}$–$\tau_c$ crossing is only an approximation of the Rhines scale, it provides a reasonable estimate of the spatial extent of thermal Rossby modes (convective columns). This suggests that convective columns are about twice as small in SBR97n035, which is consistent with Figure~\ref{fig:SBR97n035vsSBR50n1pv}. Finally, in SBR50n1, $\tau_c$ crosses the rotation timescale $\tau_\Omega$, but in SBR97n035, the convective turnover time remains longer than $\tau_\Omega$ for all degrees $\ell$, further validating the significant rotational constraint of case SBR97n035.

We also report a particular feature when looking at the advection time of quadratic entropy $\tau_{\rm adS}=E_\ell^S/|{\cal A}_\ell|$ (orange line), which becomes shorter than $\tau_c$ (dark line) in SBR97n035 for intermediate scales ($10\leq\ell\leq 100$). This indicates that despite lower velocity amplitudes, the non-local transport of heat via convection is stronger with respect to the local internal energy it is contributing to, \textit{i.e.} more efficient for the Sun-like rotating SBR97n035 model.

\subsection{Force Balance}

\begin{figure*}
  \centering
  \includegraphics[width=0.49\linewidth]{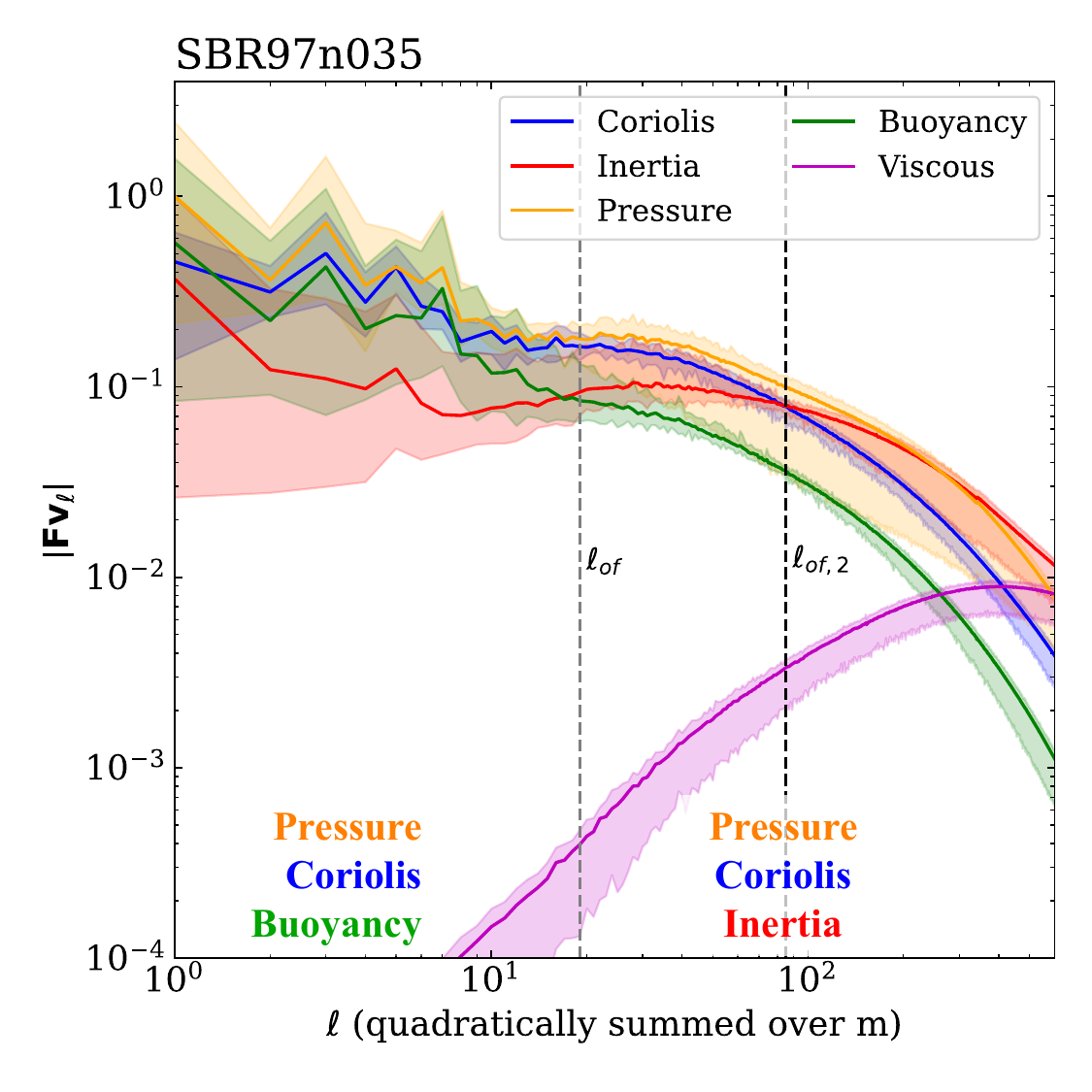}
  \includegraphics[width=0.49\linewidth]{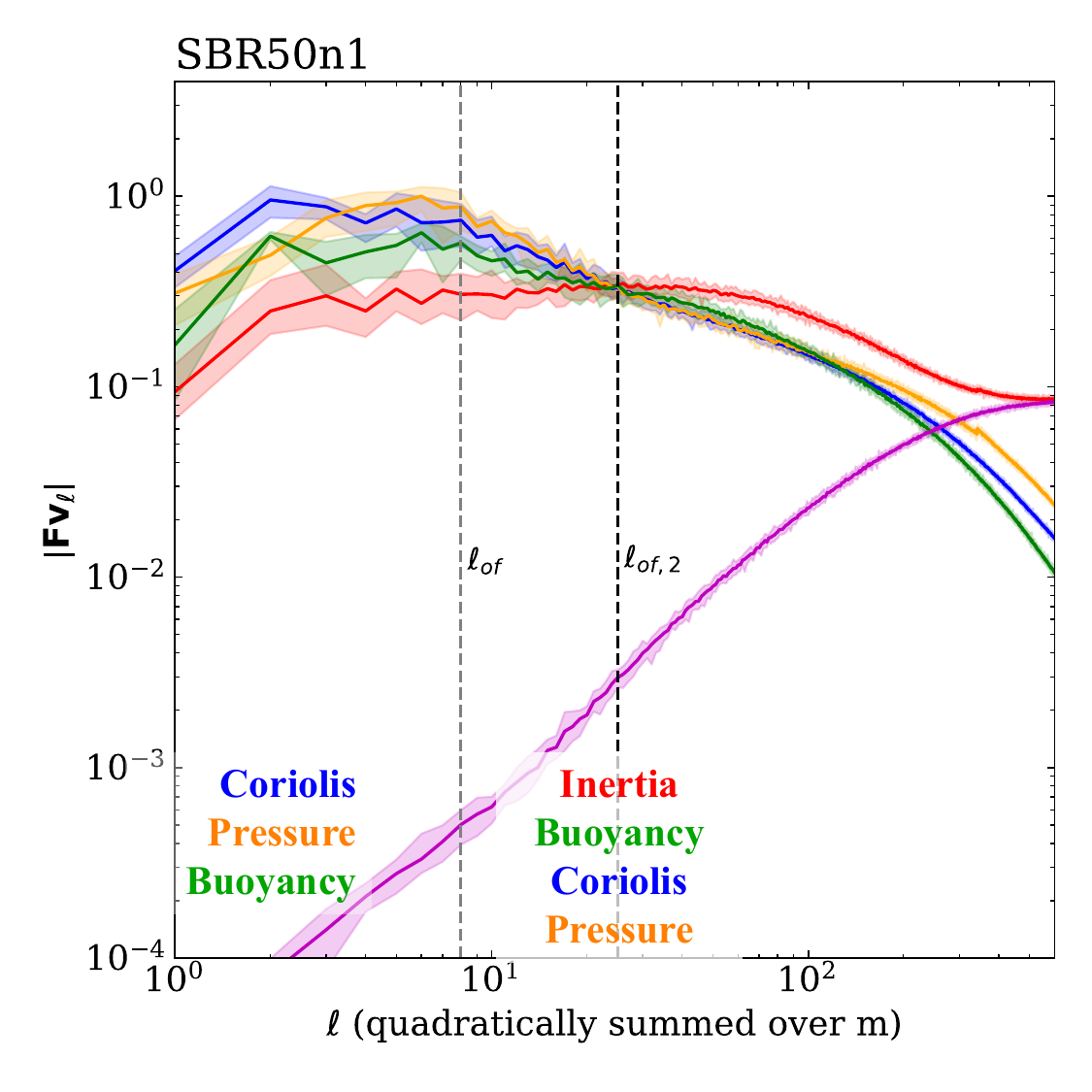}
  \caption{Non-axisymmetric spectrum of the force balance \ref{eq:ASHeq_v} in SBR97n035 (\textit{left}) and SBR50n1 (\textit{right}) models, represented as a function of the spherical harmonic degree $\ell$. Similarly to Figure~\ref{fig:RmsSPdiagKE}, this has been quadratically summed over $m>0$, and each spectrum is normalized to the peak amplitude of the pressure contribution (yellow). Each one is computed as a root-mean-square (rms), such that $|\textbf{F}_\ell|=\sqrt{\int_t\int_r \textbf{f}^2_\ell r^2 \text{d}t\text{d}r/\int_t\int_r r^2\text{d}t\text{d}r}$. We also consider radius from $0.75$ to $0.9$~R$_\odot$, and solid lines represent the spatio-temporal rms while corresponding shaded regions represent the min/max deviation in time from it. The dashed gray vertical line represents $\ell_{of}$, taken from Figure~\ref{fig:RmsSPdiagKE}, and the dashed black vertical line ($\ell_{of,2}$) targets the lowest scale where the Coriolis (blue) contribution is higher than the inertia (red) one.}
  \label{fig:RmsFBdiag}
\end{figure*}

After the characterization of the temporal evolution of energies occurring in our setup, we now consider the force balance, which gives information on which process actually creates the movement. Previously, we noted that understanding the role of the buoyancy force is key in order to interpret physically the differences between the solar-like rotating model SBR97n035 and the anti-solar one SBR50n1. Since buoyancy acts only in the vertical force balance, focusing on the vertical component allows isolating further its corresponding dynamics.

This can be computed by projecting the radial part of the momentum Equation~\ref{eq:ASHeq_v} on the spherical harmonics (see \textit{e.g.} \citealt{aubertSphericalConvectiveDynamos2017}) such that
\begin{eqnarray}
  \label{eq:FBanaSH}
  \overbrace{\left(\frac{\partial \mathbf{v}}{\partial t}\Big|_r + (\mathbf{v} \cdot \nabla)\mathbf{v}\Big|_r\right)^m_\ell}^{\footnotesize \mbox{\textcolor{red}{Inertia}}} 
  &=& \underbrace{- \left[\nabla_r\left(\frac{P_{\rm tot}}{\rho_{\rm tot}}\right)\right]^m_\ell}_{\footnotesize \mbox{\textcolor{orange}{Pressure}}} \overbrace{+ \left(\frac{s_{\rm tot}}{c_P} g\right)^m_\ell}^{\footnotesize \mbox{\textcolor{Green}{Buoyancy}}} \nonumber\\ &\,&\underbrace{- \left(2\mathbf{\Omega}_* \times \mathbf{v}\Big|_r\right)^m_\ell}_{\footnotesize \mbox{\textcolor{blue}{Coriolis}}} \overbrace{- \left(\frac{1}{\bar{\rho}} \nabla \cdot \mbox{\boldmath $\cal D$}\Big|_r\right)^m_\ell}^{\footnotesize \mbox{\textcolor{magenta}{Viscous}}}
\end{eqnarray}
where $c_P$, $g$ and \mbox{\boldmath $\cal D$} are respectively the heat capacity at constant pressure, the star's gravity field and the viscous stress tensor defined in Equation~\ref{eq:tenseurD}. Similarly to Section~\ref{sec:PowerBal}, we compute the \textit{rms} value in time and radii of the non-axisymmetric part ($m>0$) of the different terms in Equation~\ref{eq:FBanaSH}. We include here the local temporal derivative to the inertia contribution in order to assess a clear balance interpretation. Indeed, even when assuming a quasi-stationary state where this local derivative averages to zero over time, its \textit{rms} value does not vanish as the local derivative is never precisely zero at a given time in this diagnostics. We focus again our attention on the bulk of the CZ, and illustrate the overall force balance in Figure~\ref{fig:RmsFBdiag}.

First, we note a three-term balance occurring at the largest scales between the Coriolis, pressure and the buoyancy force in both models for $\ell<\ell_{of}$. At intermediate scales, we note that the buoyancy (green) contributions drops and becomes significantly smaller in SBR97n035, especially for $\ell\geq\ell_{of}=19$. We then note a three-term balance Pressure-Coriolis-Inertia in the range $50\leq\ell\leq 200$ for this model, by contrast to a four-term one Inertia-Buoyancy-Coriolis-Pressure in the range $10\leq\ell\leq 100$ of the anti-solar rotating model. Following \cite{teedSolenoidalForceBalances2023}, we drop pressure in our notation and conclude that the main force balance in the anti-solar model is a Coriolis-Inertia-Buoyancy/Archimedean (CIA), whereas a Coriolis-Inertia (CI) balance is achieved in the solar model.

Similarly to the energetic diagnostic in Section~\ref{sec:EkDiag}, the Coriolis spectra (blue) are rather similar between both models, however the shape of the inertia spectrum has changed in this force diagnostics, being significantly weaker in the balance at small scales for SBR97n035. To characterize it, we can define $\ell_{of,2}$ such that $Ro_{FB}(\ell_{of,2})=1$, where $Ro_{FB}(\ell)=\left(\frac{{\rm d} \mathbf{v}}{{\rm d} t} |_r\right)^{m>0}_\ell/\left(2\mathbf{\Omega}_* \times \mathbf{v} |_r\right)^{m>0}_\ell$. Similarly to $\ell_{of}$, we note $\ell_{of,2}$ is located at a smaller scale (higher $\ell$) in the SBR97n035 case (left panel, $\ell=85$ in comparison to 25 in SBR50n1).

It is interesting to note that the buoyancy term is mainly equal or smaller than the Coriolis force at all scales in these non-axisymmetric balance. This is consistent with Figure~\ref{fig:Rossby_comp}, where we saw that the convective Rossby number $Ro_{\rm c}$ is lower than 1 between 0.75 and 0.9~R$_\odot$, meaning that the Coriolis force dominates over buoyancy in both the axisymmetric and non-axisymmetric balances. The only exception to this statement happens for $\ell_{of,2}=25\leq\ell\leq 100$ in the anti-solar rotating case (right), where the buoyancy (green) exceeds Coriolis and becomes the second term of the balance after the inertia.\\

In summary, when comparing the different diagnostics of Section \ref{sec:SpectralAna}, we see that the control of Nu has allowed to control the buoyant driving of the convective scales, thereby extending the range of large scales influenced by the Coriolis force in case SBR97n035. The difference between diagnostics however lies on the scale range that is impacted: in the solar rotating case, the buoyancy injects/drives mostly at large scales (see Figure \ref{fig:RmsFBdiag}) and then nonlinear effects transfer energy along the turbulent cascade towards intermediate scales to establish a different balance there. In comparison, in the anti-solar case SBR50n1, buoyancy driving plays a dominant role in the energy/force balance for the majority of the considered scales. Replicating these analyses at various depths confirm that these trends are robust throughout the whole CZ, and that the range of rotationally-constrained scales increases with depth (\textit{i.e} $\ell_{of}$ and $\ell_{of,2}$ increase, as expected from Figure \ref{fig:Rossby_comp}).

\section{Discussions}\label{sec:discuss}

Global models presented here do not resolve the convection zone all the way up to the solar surface, but are getting within 1\%. Due to the steep decrease of density scale height in the last few percents of the solar radius, this would require numerical resources out of the scope of this study, and a fully compressible code such as \textit{Dyablo} \citep{delormeNOVELNUMERICALMETHODS2022}. We thus remind here that present models may not precisely depict the very near-surface dynamics and driving of entropy rain, which subsequent impact on deep convection is still subject to debate \citep{cossetteSUPERGRANULATIONLARGESTBUOYANTLY2016,hottaWeakInfluenceNearsurface2019,kapylaEffectsRotationSurface2023,hottaDynamicsLargeScaleSolar2023}. It is likely that even smaller scales will have to be resolved in order to decipher the surface dynamics \citep{kupkaModellingStellarConvection2017} and their mutual influence with deep convection (see \citealt{delormeNOVELNUMERICALMETHODS2022} and \citealt{popovasGlobalMHDSimulations2022}). As a compromise between including the most scales and limiting numerical resources, the top of SBR97n035 and SBR50n1 models presented here has been set to $0.9914$ R$_\odot$.

The \textit{Convective Conundrum} points out the limits of MLT \citep{1958ZA.....46..108B} to fully account for the complex 3D nonlinear interactions occurring in a rotating stellar shell. Various improvements to MLT are being explored to incorporate effects such as the Deardorff term \citep{brandenburgSTELLARMIXINGLENGTH2016} or rotational constraints \citep{vasilRotationSuppressesGiantscale}. In our study, we have modified the convective flux by reducing the Nusselt number ($Nu$) from its MLT prediction. This was achieved by increasing the radiative flux $F_{\rm rad}$ via adjustment of the radiative diffusion coefficient $\kappa_{\rm rad}$. It is crucial to clarify that altering $\kappa_{\rm rad}$ does not imply a change in the physical opacity $\kappa_{\rm op}$ of the medium, but merely serves here as a numerical recipe to govern the convective flux. This way, we can control the experiment without directly affecting the entropy transport in the non-axisymmetric balance, thus ensuring the consistency of the convective dynamics across the scales resolved in our simulations (see magnitudes of the radiative term in Figure~\ref{fig:RmsSPdiagS2}).

As mentioned in Section~\ref{sec:intro}, a transition towards an anti-solar DR profile is expected once a given Rossby number value is reached \citep{gastineSolarlikeAntisolarDifferential2014}. Parametric studies using the same ASH code already reported this transition value being within the interval $0.7\leq Ro_{\rm f}\leq 1.2$ \citep{mattConvectionDifferentialRotation2011,brunDifferentialRotationOvershooting2017}. In the present paper, we reproduce the ordering solar-type/anti-solar as a function of the Rossby number, however we have seen that this fluid Rossby number transition value is here in the interval $1.5\leq Ro_{\rm f}\leq 2.5$ for this set of models. Indeed, since vorticity peaks at small scales \citep{mieschStructureEvolutionGiant2008,candelaresiKineticHelicityNeeded2013}, the Rossby number value characterizing a given regime is likely to be Reynolds number dependent, and so will be the transition value towards the anti-solar differential rotation regime. Other parameters may also influence it, such as the density contrast \citep{kapylaEffectsStratificationSpherical2011}, the Prandtl number \citep{kapylaTransitionAntisolarSolarlike2023}, or the presence of a magnetic field \citep{brunPoweringStellarMagnetism2022,hottaGenerationSolarlikeDifferential2022}. Specific parametric studies are therefore required to characterize these dependencies. However, this is not a major issue in the present paper, where we have kept these parameters constant between SBR97n035 and SBR50n1 (see Tables \ref{tab:AB2vsAS1vsSunBusyAdim} and \ref{tab:AB2vsAS1vsSunBusy} for details).

The resulting dynamics of SBR97n035 turn out to be in qualitative agreement with the solar DR profile. Nonetheless, we recognize here that the models yields a more cylindrical rotation profile than the one inferred through helioseismology (see Figure~\ref{fig:Fig1}). Reasons for the break of Taylor-Proudman constraint in the Sun, \textit{i.e.} $\partial_z\langle v_\phi\rangle\neq 0$, is an active area of research. While certain studies suggest potential parametric adjustments to capture this complexity in the meantime (see \citealt{hottaDynamicsLargeScaleSolar2023} for a detailed bibliography), our research did not focus on meticulously calibrating the model to replicate this specific helioseismic feature. Our primary focus here is to gain understanding of the overlying trends regarding rotational constraints on the global convective dynamics and the resulting self-consistent rotation profile. Of course, a fine-tuning of $Nu$ at a given $Re$ would be possible to get even better agreement.

The apparition of a Deardorff zone (DZ) results from the deposition of low-entropy material at the bottom of the CZ. Different studies have emphasized the dependence of its extent to the Prandtl number (see \textit{e.g.} \citealt{1977GApFD...8...93G,bekkiConvectiveVelocitySuppression2017}). Indeed, characteristics of the deposition depend on the ratio between the characteristic travel time of a convective eddy $\tau_c=L/v$ and the characteristic timescale over which it diffuses/deposits its entropy signature $\tau_\kappa=L^2/\kappa$ (\textit{i.e.} the Peclet number $Pe=vL/\kappa=\tau_\kappa/\tau_c=Pr\times Re$). Further note that $Pe$ plays a key role in distinguishing overshooting (small $Pe$) from penetrative convection (large $Pe$) as explained in \cite{1991A&A...252..179Z}. However, it is interesting to note in Figure~\ref{fig:deltas} that SBR50n1 exhibits a wider DZ than SBR97n035, despite similar Prandlt $Pr=1/4$, and Peclet numbers $Pe\simeq 200$ for both models. Let's further remind here that $\kappa_{\rm rad}$ plays a negligible role in the non-axisymmetric diffusion for both models in comparison to $\kappa$ (see in Figure~\ref{fig:RmsSPdiagS2}). This means that the enhancement of the low-entropy deposition at the bottom of the CZ must be enhanced in SBR50n1 via another dynamical reason. Stronger correlations between the radial velocity $v_r$ and temperature perturbation $T$ may be a possibility for this DZ enlargement, and would be coherent with the larger enthalpy flux $F_{\rm en}$ of SBR50n1 in comparison to SBR97n035 (see Figure~\ref{fig:SBR97n035vsSBR50n1}). Knowing the exact scaling between the spatial extent of the DZ and $F_{\rm en}$ will however require a larger parametric study, with more models spanning the Nusselt $Nu$ parameter space, which goes beyond the scope of this study. Another possible contribution to explain such difference as been reported by \cite{kapylaConvectiveScaleSubadiabatic2024}, who observes that the DZ extend decreases as the rotational constraint is stronger. The comparison between SBR97n035 and SBR50n1 is then coherent with such a result, as SBR50n1 exhibits a larger Rossby number, hence a weaker rotational constraint.

The set of models we presented in this study does not consider magnetism yet, as we aim at investigating here some aspects of the current \textit{Convective Conundrum} from the viewpoint of hydrodynamic global 3D simulations. The exact impact of the magnetism on the angular momentum redistribution and the convective dynamics is currently subject to debate \citep{brunInteractionDifferentialRotation2004,kapylaConfirmationBistableStellar2014,karakMagneticallyControlledStellar2015,guerreroROLETACHOCLINESSOLAR2016,brunPoweringStellarMagnetism2022,hottaGenerationSolarlikeDifferential2022,warneckeSmallscaleLargescaleDynamos2024} and will require a dedicated numerical setup to be investigated.\\

\section{Conclusion}\label{sec:ccl}

During the last 20 years, a key goal of global modelling of convection in rotating spheres has been to approach more and more the stellar turbulent regime \citep{mieschStructureEvolutionGiant2008}. In particular, global models of solar-type turbulent convection have been developed with some success, in order to explain the possible origin of the observed solar differential rotation (DR) and surface magnetic flux, at the origin of the 11 years activity cycle, along with coherent scenarii of longer time-scale magneto-rotational evolution (see \textit{e.g.} \citealt{kapylaReynoldsStressHeat2011,brunMagnetismDynamoAction2017,strugarekReconcilingSolarStellar2017,brunPoweringStellarMagnetism2022,hottaDynamicsLargeScaleSolar2023,norazMagnetochronologySolartypeStar2024}). However, it appeared that the amplitude of giant convection cells in these global solar-type models are usually stronger than what is inferred from helioseismology \citep{hanasogeAnomalouslyWeakSolar2012,greerHELIOSEISMICIMAGINGFAST2015}. Subsequently, this leads to an over-estimation of the effective Rossby number $Ro$ of global models, and can lead to the transition towards an anti-solar rotating regime (retrograde equator) when increasing the turbulence degree \citep{gastineSolarlikeAntisolarDifferential2014,brunDifferentialRotationOvershooting2017,hindmanMorphologicalClassificationConvective2020,hottaGenerationSolarlikeDifferential2022}. This is known as the \textit{Convective Conundrum} \citep{omaraVelocityAmplitudesGlobal2016}, and acknowledges the current need for better understanding of the solar convective dynamics. In that sense, the differential rotation trends found as a function of $Ro$ appear robust \citep{gastineSolarlikeAntisolarDifferential2014,norazMagnetochronologySolartypeStar2024}, but the precise location of a given star, \textit{e.g.} the Sun, in the Rossby parameter space has to be currently considered with care.

In this context, we performed a numerical study aiming at understanding key force and energy balances in the solar convective envelope. To do so, we have constructed a theoretical path in parameter space and proposed a fluid mechanics experiment where we control $Ro$, while increasing the Reynolds number $Re$, hence the turbulence degree, and keeping solar parameters ($L_\odot, \Omega_\odot$). This is made possible by controlling the Nusselt number $Nu$, which quantifies the amount of energy transported by convection \citep{kapylaExtendedSubadiabaticLayer2017,kapylaEffectsSubadiabaticLayer2019}. In other words, we can then limit velocity amplitudes ($\sim$ constant $Ro$) while decreasing viscous dissipation (increasing $Re$), by controlling the amount of energy transported by convection vs diffusion processes (decreasing $Nu$). We were then able to construct a relatively highly-turbulent global Sun-like model ($Re\sim 800$) while ensuring a solar-like DR regime (prograde equator, case SBR97n035) and a solar rotation rate $\Omega_\odot$. We also constructed a control run, exhibiting a similar $Re$ but a larger $Nu$. This control run presents an anti-solar retrograde equator. We have compared the dynamics of the two models in Section~\ref{sec:Overview}, compared their dynamics to existing solar observation constraints in Section~\ref{sec:sol_comp}, and characterized them using spectral analyses in Section~\ref{sec:SpectralAna}.

We report that the morphology of the convective dynamics has significantly changed, even if both models share a similar turbulence degree. By limiting the amount of energy the convection has to transport (through the decrease of $Nu$), convective velocities have been globally decreased over the whole convective spectrum, and even more so at large scales. In the Sun-like rotating case SBR97n035, such a global translation has changed the force balance over the turbulence cascade, by expanding the range of scales significantly impacted by the Coriolis force. As a direct consequence, velocity amplitudes of SBR97n035 are even more damped at large scales, and lie now in the observational range inferred by helioseismology (between revisited \citetalias{hanasogeAnomalouslyWeakSolar2012} and \citetalias{greerHELIOSEISMICIMAGINGFAST2015}, see Figure~\ref{fig:SpComp}). This enhanced decrease of amplitude at large scales then results in a shift of the spectrum peak towards smaller-scale.

Our spectral analysis has revealed that both models can be distinguished by the nature of energy transfers across three distinct regimes: the large-scale \textit{quasi-geostrophic} (QG) balance, the \textit{buoyant-inertial} interplay at intermediate scales, and the dominance of \textit{dissipation} at small scales. For the SBR97n035 model, which exhibits Sun-like rotation characteristics, the QG equilibrium extends down to smaller scales compared to the anti-solar model. A closer examination of the kinetic energy transfer identifies significant alterations in the contributions of buoyancy and inertial effects (via Reynolds stress) to the spectrum. The model with a controlled Nusselt number ($Nu$) shows a marked suppression of buoyancy-driven energy at larger scales, thereby attenuating the subsequent transfer of kinetic energy through inertia. Detailed views into this advective processes allow understanding how differential rotation is energetically sustained. For SBR97n035, there is a noticeable shift in energy injections from non-axisymmetric convective scales toward larger scales, emphasizing the effects of stronger rotational constraints on the model’s dynamics. Analysis of the force balance further corroborates these findings, particularly accentuating the reduced influence of buoyancy forces. Hence, the maintenance of the prograde equatorial rotation in our model are predominantly supported by a delicate Coriolis-Inertia (CI) balance, as opposed to the typically considered Coriolis-Inertia-Archimedes (CIA) framework.

In the context of the convective conundrum, limiting $Ro\sim v/2\Omega R_*$ is usually done by either increasing the rotation rate $\Omega_*$ or decreasing the luminosity $L_*\sim v^3$. On one hand, altering $\Omega_*$ does affect the buoyancy, but not similarly than reducing the Nusselt number, as only the Coriolis force is changed. On the other hand, we expect a similar impact between reducing the Nusselt number and reducing the luminosity, as modifying the radiative diffusivity affects the axisymmetric flux balance but not directly the non-axisymmetric dynamics (\textit{cf.} Figure~\ref{fig:RmsSPdiagS2}). Therefore, reducing $Nu$ allows decreasing $Ro$ while maintaining observational constraints ($L_\odot, \Omega_\odot$) and preserving what we consider to be the key dynamical regimes of the system.

Using $\tau-\ell$ analysis, inspired from the geophysics community \citep{natafTurbulenceCore2015,natafDynamicRegimesPlanetary2023}. We have further analyzed properties of turbulent dynamics in the spectral space. A similar Kolmogorov scale has been found for both models, and differences in the Rhines scale confirms that the rotational constraint is stronger on SBR97n035 dynamics and shapes the convection into smaller convective columns aligned with the rotation axis in the bulk of the CZ. We also note an overall increase of buoyancy characteristic time-scales for SBR97n035, implying a morphology change of the inertia spectrum. An interesting point we report here is an inversion of transport timescales ordering between both models. Indeed, the characteristic timescale of entropy advection becomes smaller than the one of momentum for SBR97n035.

To synthesize the main messages when comparing the energetic Equation~\ref{fig:RmsSPdiagKE} and dynamic Equation~\ref{fig:RmsFBdiag} diagnostics, the difference between both lies on the scale range that is impacted. Indeed, the most significant drop-off of spectra in SBR97n035 occurs at large scales due to the rotational constraint in the energetics analysis. Conversely, the force diagnostics point to a drop-off happening at small scales due to the global dampening of buoyancy, and a subsequent decrease in the inertia power. This distinction underscores a key concept: examining energy transfers quantifies how the system changes across different scales, while force balance analysis underlines the actual mechanisms driving this change. Conceptually, the transition from CIA to CI raises then questions about the source of motions, typically attributed to Archimedean buoyancy. Specifically, there is large-scale injection or forcing (as seen in the force diagnostics, Figure \ref{fig:RmsFBdiag}) and a subsequent transfer along the cascade toward intermediate scales (as shown in the energy diagnostics, Fig. \ref{fig:RmsSPdiagKE}). This suggests that the buoyancy force has a lesser effect in case SBR97n035, especially at intermediate and smaller scales, while turbulent transfers dominate the energetics balance at these scales. In case SBR50n1, buoyancy however still dominates the dynamics locally over much of the scales considered.\\

When confronted with recent observations, we find that our Sun-like rotating model is compatible with helioseismology in terms of large-scale convective amplitudes (see Figure~\ref{fig:SpComp}), surface Rossby mode $m=1$, and inferred superadiabaticity profiles (see Section~\ref{sec:sol_comp}). This is encouraging as we now have a relatively turbulent model (SBR97n035, $Re \sim 800$) that exhibits solar-type differential rotation with a prograde equator, that agrees qualitatively with the constraints, while being also consistent with the Sun's rotation rate and luminosity. We do not claim here that $\ell=0$ diffusion transport in the convection zone reflects a physical truth about the Sun's energy transfer, but rather that the theoretical path we propose in the ($Ro$,$Re$,$Nu$) space is an opportunity to investigate and conserve what we believe to be the main force balances possibly operating in the Sun, even if current model are still far from the real solar turbulent regime.

As mentioned above, the Sun-like rotating model SBR97n035 exhibits columnar convective patterns in the bulk of the CZ (see Figure~\ref{fig:SBR97n035vsSBR50n1pv}), resulting from the stronger Taylor-Proudman constraint in this model. However, it is key to note that such patterns do not imprint the dynamics of the model near the surface (see Figure~\ref{fig:SpComp}). It is important to remind here that the detection of such thermal Rossby modes is still pending on the Sun. Therefore, the SBR97n035 model emphasizes here the possibility that these modes actually exist in the deep convection zone with limited amplitudes, but that their signal is hidden and concealed by near-surface small-scale convective granulation (see \textit{e.g.} \citealt{bessolazHUNTINGGIANTCELLS2011,guerreroDIFFERENTIALROTATIONSOLARLIKE2013,featherstoneEMERGENCESOLARSUPERGRANULATION2016}). 

Additionally, our models reveal the presence of interesting \textit{polar vorticity rings}, which are manifested as transient cyclonic features surrounding up-flows (\textit{cf.} Figure~\ref{fig:PolarView}). This phenomenon is distinct from previously observed polar cells or plumes \citep{hindmanMorphologicalClassificationConvective2020}. The transient nature of these vorticity rings, coupled with the small-scale down-flows that constitute them and their location in polar regions, may make their detection challenging. However, the potential existence of such vorticity rings in polar areas highlights the necessity of expanded observational coverage of the Sun, to probe the full spectrum of convective dynamics.

Because solar photospheric observations remain mainly in the ecliptic plane (modulo the $\beta_0$ angle of the Sun with respect to the ecliptic plane), the precise observations we can make of the Sun's surface are concentrated in regions away from the poles. Acquisition time for such observations is also limited by the solar rotation, and not having a full coverage of the solar surface limits techniques for inverting internal structure. It would then be interesting to observe the Sun's polar regions and see if a signal from internal convective columns or surface vorticity rings can be reported, something that has not been done yet. Unveiling the dynamics of these features may provide insights into the complex interplay between rotation, convection, and magnetism in the Sun. Such investigation could start with observations of the Solar Orbiter mission during its \textit{out-of-the-ecliptic} phase, and potentially followed by other missions (see \textit{e.g.} \textit{Solaris} \citealt{hasslerSolarisSolarPolar2020}, \textit{$4\pi$-HeliOS} \citealt{2022cosp...44.1530R}).

Finally, most numerical MHD models of stellar and planetary dynamo have reported the importance of the buoyancy contribution, with the so-called quasi-geostrophic Magneto-Archimedean-Coriolis (QG-MAC) balance \citep{davidsonScalingLawsPlanetary2013,aubertSphericalConvectiveDynamos2017,gastineStableStratificationPromotes2021,zaireTransitionMultipolarDipolar2022,natafDynamicRegimesPlanetary2023}. However, our current setup and recent studies \citep{hottaGenerationSolarlikeDifferential2022} suggest the need for a reduced buoyancy in order to yields a prograde equator at unprecedentedly high turbulent regimes. This questions the presence of such a balance in the solar interior. Instead, a QG-MIC could be a good candidate, with Inertia/advection replacing Archimede/buoyancy. This is a concept that we aim to explore in future work.\\

\textit{Q.N., A.S.B. and A.S. are thankful to H.C. Nataf, Y. Bekki, L. Gizon, A. Birch, A. Fournier, C. Blume and B. Hindman for useful discussions, as well as J. Schou for providing rotation profile data, and the anonymous referee for his/her careful and constructive review.  Q.N. thanks H. Hotta for providing HKS22 spectral data, and A.S. thanks J. Aubert for providing benchmark data to validate the implementation of the force balance analysis using outputs from the ASH code. All authors acknowledge financial support by ERC Whole Sun Synergy grant \#810218, INSU/PNST, CNES and Solar Orbiter funds. We thank GENCI via project 1623 for having provided part of the massive computing resources needed to perform this extensive study. Q.N. acknowledges funding from the Research Council of Norway through its Centres of Excellence scheme, project number 262622. A.S. acknowledges funding from the DIM Origines Ile-de-France program DynamEarths. This work benefited from discussions within the NORDITA program “Stellar Convection: Modelling, Theory and Observations ”.}

%

\vspace{5mm}
\facilities{TGCC (Très grand centre de calcul du CEA) - CINES (Centre Informatique National de l'Enseignement Supérieur) - IDRIS (Institut du développement et des ressources en informatique scientifique)}


\software{ASH \citep{cluneComputationalAspectsCode1999,brunGlobalScaleTurbulent2004}, Python \citep{VanRossumPython}, numpy \citep{harrisArrayProgrammingNumPy2020}, matplotlib \citep{HunterMatplotlib}, shtns \citep{SchaefferShtns}, pyshtools \citep{WieczorekPyshtools}, pyvista \citep{sullivan2019pyvista}
          }



\appendix

\section{Numerical method}
\label{sec:num_method}
In order to solve the equations governing the physics of stars, we use the ASH code (\textit{Anelastic Spherical Harmonic}, \citealt{cluneComputationalAspectsCode1999, mieschThreeDimensionalSpherical2000, brunGlobalScaleTurbulent2004}). The time resolution is provided by a semi-implicit Crank-Nicholson method for the linear terms, and by an explicit Adams-Bashford method of order 2 for the non-linear terms and the Coriolis term. The spatial resolution is provided by a pseudo-spectral method in spherical coordinates: equations are solved by projection onto the spherical harmonics along the $\theta$ and $\phi$ directions, and with a finite difference scheme in the radial direction. We choose a finite-difference radial resolution method for all the models discussed, to ensure the numerical stability of this study, while significantly increasing the radial resolution at strategic locations and keeping a limited number of points elsewhere. The strategic locations are the stable/convective zone transition, where changes in the entropy gradient and convective overshoot occur, as well as the near-surface region, where strong flux changes are expected as a result of our modifications on $\kappa_{\rm rad}$. We report the specificities of each grid in Table~\ref{tab:AB2vsAS1vsSunBusy}.

\subsection{Equations}
This resolution is done entirely in the anelastic approximation (\citealt{1962JAtS...19..173O}, \citealt{1969JAtS...26..448G}, see also the Appendix from \citealt{derosaDynamicsUpperSolar2001} for a detailed derivation). It allows lifting the incompressibility assumption proposed by methods like Boussinesq, without having to consider the sound waves ($\partial \rho/ \partial t = 0$). This allows to keep the important effects caused by the density stratification in the stellar interiors, while maintaining an acceptable CFL criterion. It is then based on convective motions rather than on the speed of sound, allowing much larger time steps when the Mach number is subsonic, and thus a reduced computation time for solar convective envelope modeling. 

We further choose here to use the LBR formulation (\textit{Lantz-Braginsky-Roberts}, \citealt{1992PhDT........78L,braginskyEquationsGoverningConvection1995}), which has the ability to effectively conserve energy in both unstable convective regions and stable radiative interiors \citep{brownENERGYCONSERVATIONGRAVITY2012,vasilENERGYCONSERVATIONGRAVITY2013}. More especially, this formulation neglects interactions between fluctuating pressure and stratification, by introducing a reduced pressure such as

\begin{equation}
  \boldsymbol{\nabla} \cdot (\bar{\rho}\vv) = 0,
  \label{eq:ASHeq_rho}
\end{equation}
\begin{equation}
    \frac{\partial \mathbf{v}}{\partial t}+(\mathbf{v} \cdot \nabla)\mathbf{v} = - [\nabla \omega + \nabla \bar{\omega} + \bar{\omega}\nabla\rm{ln}\bar{\rho} - \mathbf{g}] - \frac{s}{c_P} \mathbf{g} - 2\mathbf{\Omega}_* \times \mathbf{v} - \frac{1}{\bar{\rho}} \nabla \cdot \mbox{\boldmath $\cal D$}, 
    \label{eq:ASHeq_v}
\end{equation}
\begin{equation}
  \bar{\rho} \bar{T} \frac{\partial s}{\partial t} = -\bar{\rho} \bar{T} \vv \cdot {\boldsymbol{\nabla}}(\bar{s} +s) - \boldsymbol{\nabla} \cdot \mathbf{q} + \Phi_d,
  \label{eq:ASHeq_S}
\end{equation}

We note the velocity field $\mathbf{v} = (v_r,v_\theta,v_\phi)$, the reduced pressure $\omega=P/\bar{\rho}$ and the energy flux $\mathbf{q}$ defined as Equation~\ref{eq:EnergFlux}. $c_P$ and $\mathbf{g}$ are respectively the heat capacity at constant pressure and the star's gravity field, followed by ${\bf \Omega}_*=\Omega_*\hat{\bf e}_z$ the angular velocity in the rotating frame, with $\hat{\bf e}_z$ as the unit vector oriented along the axis of rotation. \mbox{\boldmath $\cal D$} is the viscous stress tensor defined in Equation~\ref{eq:tenseurD}, and $\Phi_d$ the dissipation term expressed as

\begin{equation}
  \Phi_d = {\cal D}_{ij}\frac{\partial v_{ij}}{\partial x_j} = 2\bar{\rho} \nu \left[ e_{ij} e_{ij} - \frac{1}{3} (\boldsymbol{\nabla} \cdot \mathbf{v})^{2} \right],
\label{eq:dissipationNu}
\end{equation}

where $e_{ij}=1/2(\partial_jv_i+\partial_iv_j)$ is the stress tensor, with $\delta_{i,j}$ the Kronecker symbol. To close the system of equations, we consider a perfect gas equation of state, \textit{i.e.} $\bar{P}=R\bar{\rho}\bar{T}$ where $R=c_P(\gamma-1)/\gamma$ and $\gamma$ is the adiabatic exponent. The linearization of thermodynamics variable under the anelastic assumption then gives
\begin{equation}
  \frac{\rho}{\bar{\rho}}=\frac{P}{\bar{P}}-\frac{T}{\bar{T}}=\frac{P}{\gamma\bar{P}}-\frac{s}{c_P}.
  \label{eq:GPlin}
\end{equation}

\subsection{Numerical setup}\label{sec:app_num_set}

\begin{figure*}
  \centering
  \includegraphics[width=\linewidth]{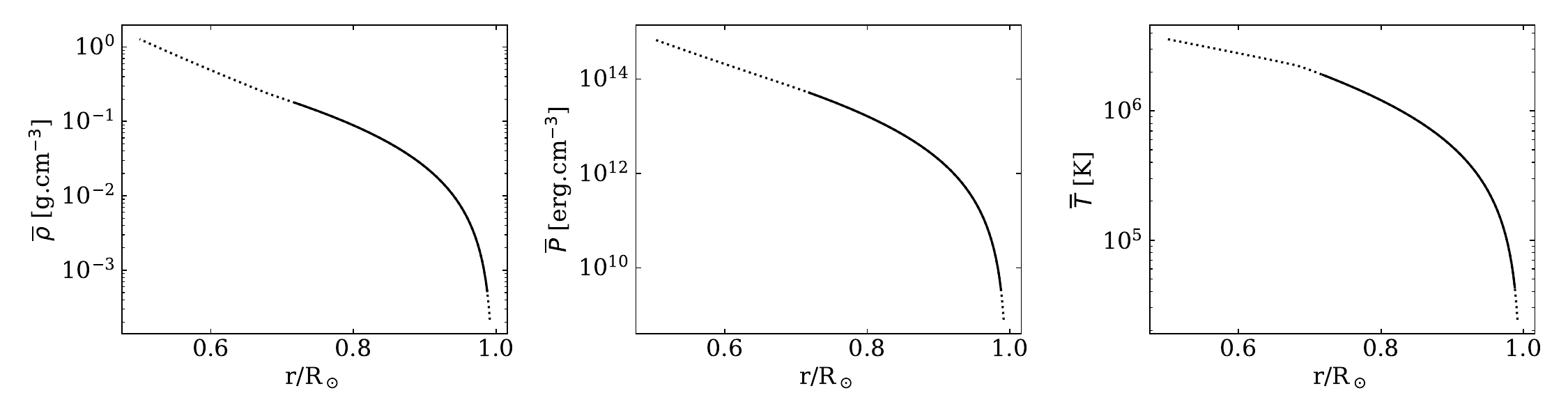}
  \caption{Radial profile of the mean density, pressure and temperature used as initial background state in the 3D models of the present paper. Solid lines represent the 1D structure used in AS1 and dotted lines the structure for SBR97n035 and SBR50n1.}\label{fig:RefStateApp}
\end{figure*}

To determine the one-dimensional mean state in model initialization, we first extract the entropy gradient $d\bar{s}/dr$ and gravity $g$ profiles from a solar reference model with the CESAM code \citep{morelCESAMCodeStellar1997,brunSeismicTestsSolar2002}. The reference density profile $\Bar{\rho}$ satisfying hydrostatic equilibrium as a solution is then obtained by using a Newton-Raphson method on the following equation,
\begin{equation}
  \frac{d\bar{\rho}}{dr} + \frac{g}{\gamma} \bar{\rho}^{2-\gamma}e^{-\gamma \frac{\bar{s}}{c_P}}+\frac{\bar{\rho}}{c_P}\frac{d\bar{s}}{dr} = 0.
  \label{eq:NRaph}
\end{equation}
We can then deduce the pressure profile $\Bar{P}$ and temperature profile $\Bar{T}$ via the equation of state \ref{eq:GPlin}, which gives us here profiles presented in Figure~\ref{fig:RefStateApp}.

In order to fully define the set of anelastic equations presented, we choose boundary conditions conserving angular momentum. To this end, we choose them to be impenetrable and free of friction/torque at the top and bottom, such that
\begin{equation}
    \label{eq:CL_imp}
    v_r=\frac{\partial}{\partial r}\left( \frac{v_{\theta}}{r} \right) = \frac{\partial}{\partial r}\left( \frac{v_{\phi}}{r} \right)= 0|_{r=r_{\rm top},r_{\rm bot}}.
\end{equation}
The energy flow is imposed via the entropy gradient, such that
\begin{equation}
    \label{eq:CL_dSdr}
    \frac{\partial \bar{s}}{\partial r}=1.34\times 10^{-2}{|_{r=r_{\rm bot}}}=1.5\times 10^{-7}{|_{r=r_{\rm top}}}\;\text{cm s$^{-2}$K$^{-1}$},
\end{equation}
which further implies that the fluctuating entropy gradient $\partial \langle s\rangle /\partial r$ is conserved at zero at the base and surface. As no radiative zone is included in AS1 model, the bottom gradient entropy value is set to $\frac{\partial \bar{s}}{\partial r}{|_{r=r_{\rm bot}}}=-7.5\times 10^{-8}$~cm~s$^{-2}$~K$^{-1}$. The number of radial grid points per density scale height is of the order of 100 close to the surface in both SBR97n035 and SBR50n1 models.

All diffusive coefficients are assumed to be stationary, spherically symmetrical and possess a radial dependence. The viscosity profile $\nu$ is prescribed such as
\begin{equation}
  \label{eq:profilDiff}
  \nu(r) = \nu_{\rm bot} + \nu_{\rm top} f_{step}(r),
\end{equation}
\begin{equation}
  \text{with}\;\;f_{step}(r)=(\bar{\rho}/\bar{\rho}_{top})^{\alpha}[1 - \beta]f(r), \nonumber
\end{equation}
\begin{equation}
  \text{where}\;\;f(r)=0.5(\tanh((r-r_t)/\sigma_t)+1)\;\;\text{and}\;\;\beta = \nu_{\rm bot}/\nu_{\rm top}. \nonumber
\end{equation}
We impose a viscosity contrast $\beta=2.5\times 10^{-4}$ between the top of the convective zone and the bottom of the radiative zone we simulate. This transition takes into account the additional effective diffusive transport brought by the mixing of convective motions, and parameters are $r_t=0.68$~R$_\odot$ and $\sigma_t=1.43\times 10^{-2}$~R$_\odot$. The decay coefficient in the convective zone is set at $\alpha=-1/3$ and values of $\nu_{\rm top}$ are listed in Table~\ref{tab:AB2vsAS1vsSunBusy}. The $\kappa$ profile can then be deduced via the Prandtl number $Pr=\nu/\kappa$, which is kept here constant at 1/4 within a simulation and for the different models. The $\kappa_0$ diffusive profile is localized near the surface, and adjusted to transport the energy flux out of the domain. It decreases with distance from the surface, as enthalpy is transported by larger-scale convective cells. For both SBR97n035 and SBR50n1, it is prescribed such as
\begin{equation}
  \label{eq:profilDiff0}
  \kappa_0(r) = \kappa_{\rm 0, bot} + \kappa_{\rm 0, top} f(r)/(\bar{\rho}\bar{T}\frac{d\bar{s}}{dr}),
\end{equation}
where $r_t=0.9957$~R$_\odot$ and $\sigma_t=2.87\times 10^{-3}$~R$_\odot$ for the function $f(r)$, $\kappa_{\rm 0, bot}=1$~cm$^2$s$^{-1}$ and $\kappa_{\rm 0, top}=1.34\times 10^{12}$~cm$^2$s$^{-1}$.\\
For AS1 model, $\kappa_0$ is prescribed such as 
\begin{equation}
  \label{eq:profilDiff0AS1}
  \kappa_0(r) = [\kappa_{\rm 0, bot} + \alpha * \tanh((r - r_{\rm top})/\sigma_t)] / (1+\exp((r_b-r)/\sigma_b)),
\end{equation}
\begin{equation}
  \text{with}\;\; \alpha = (2 * \kappa_{\rm 0, bot} - \kappa_{\rm 0, top}) / \tanh((r_b - r_{\rm top})/\sigma_t), \nonumber
\end{equation}
where $r_b=0.7184$~R$_\odot$, $\sigma_b=2.87\times 10^{-2}$~R$_\odot$, $\sigma_t=6\times 10^{-4}$~R$_\odot$, $\kappa_{\rm 0, bot}=2\times 10^{5}$~cm$^2$s$^{-1}$ and $\kappa_{\rm 0, top}=5\times 10^{15}$~cm$^2$s$^{-1}$.

\begin{table}
  \begin{center}
    \begin{tabular}{lccccccc}
      \hline
      \hline
      Name & N$_r$ & N$_\theta$ & $r_{\rm bot}$ & $r_{\rm top}$ & $\nu_{\rm top}$ & $\kappa_{\rm top}$ & $N_\rho$\\
      & & & (${\rm R_\odot}$) & (${\rm R_\odot}$) &  ($10^{12}$ cm$^2$s$^{-1}$) & ($10^{12}$ cm$^2$s$^{-1}$) & \\[0.5ex]
      \hline
      \hline
      AS1      & 673  & 1024 & 0.72 & 0.9876 & $4.0$ & $16$ & 5.9\\[0.5ex]
      SBR97n035 & 2000 & 1536 & 0.5  & 0.9914 & $0.35$ & $1.4$ & 6.7\\[0.5ex]
      SBR50n1  & 2000 & 1024 & 0.5  & 0.9914 & $1.0$ & $4.0$ & 6.7\\[0.5ex]
      \hline
      \hline
    \end{tabular}
    \caption{Control parameters of the 3 solar models discussed. We fix their luminosity, and rotation rate in the solar regime, maintaining a Prandlt number $Pr=\nu/\kappa=1/4$ in the whole domain. From the left, we list the name of the model, the radial and latitudinal dimension of the simulated grid (for reminder $N_\phi=2 N_\theta$), the lower (bot) and upper (top) radius of the simulated domain, the viscosity and thermal conductivity prescribed at the top, as well as the density contrast $N_\rho=ln(\bar{\rho}(r_{\rm BCZ})/\bar{\rho}(r_{\rm top}))$ considered. Considering a given structure, here the solar one, $N_\rho$ will directly be impacted by the radii where we decided to crop the simulated domain. As a reminder, $N_{\rho,\odot}=ln(\bar{\rho}(r_{\rm BCZ})/\bar{\rho}(R_\odot))=13.5$ in the Sun's convection zone, because the density profile is decreasing steeply in the last Mm before reaching the photosphere. For the first model considered in this table, the base of the convective zone (BCZ) is at $r_{\rm BCZ}=r_{\rm bot}=0.72$~R$_\odot$. For cases where $r_{\rm bot}=0.5$~R$_\odot$, a radiative zone is coupled and $r_{\rm BCZ}=0.715$~R$_\odot$.}
  \end{center}
  \label{tab:AB2vsAS1vsSunBusy}
\end{table}

\section{Spectral decomposition of transfers equations}
\label{app:spec_decomp}
In order to decompose the different transfer equations on every spatial scales of the system, we project the evolution equations of the kinetic energy and the square of the entropy on the RST vector spherical harmonics basis, defined from \cite{Rieutord:1987go} and \cite{Mathis:2005kz}. This basis is indeed very practical for decomposing spectrally the various operators in the problem.

\subsection{Definitions}

The RST basis is defined such that
\begin{equation}
  \label{eq:RST}
  \left\{
  \begin{array}{lcl}
    \Rlm{} &=& \Ylm{} \er \\
    \Slm{} &=& \gradperp \Ylm{} = \dth\Ylm{}\ethe + \frac{1}{\sin{\theta}}\dphi\Ylm{}\ephi\\
    \Tlm{} &=& \gradperp\times\Rlm{} = \frac{1}{\sin{\theta}}\dphi\Ylm{}\ethe  -\dth\Ylm{}\ephi,
  \end{array}
  \right.
\end{equation}
which have the following properties :
\begin{eqnarray}
  \label{eq:R.R}
  \int_S \Rlm{1} \cdot \left( \Rlm{2} \right)^* \dint{\Omega}{} &=&
  \delta_{l_1,l_2}\delta_{m_1,m_2} \\
  \label{eq:S.S_T.T}
  \int_S \Slm{1} \cdot \left( \Slm{2} \right)^* \dint{\Omega}{} &=&
  \int_S \Tlm{1} \cdot \left( \Tlm{2} \right)^* \dint{\Omega}{} =
  l_1(l_1+1)\delta_{l_1,l_2}\delta_{m_1,m_2} 
\end{eqnarray}
with $S$ the spherical shell surface and $\dint{\Omega}{}$ the solid angle. We also note $(\Slm{})^* = (-1)^m\mathbf{S}^{m}_l$ as these relations are easily directly derived from Laplace spherical
harmonics definition:
\begin{equation}
  \label{eq:norm}
  Y_l^m = \sqrt{\frac{(2l+1)}{4\pi}\frac{(l-m)!}{(l+m)!}}P^m_l(\cos \theta)e^{im\varphi}.
\end{equation}
Such a basis is then well suited for spectral decomposition of a vector field $\mathbf{X}(\theta,\phi)$ on the surface S of a sphere, which can be expressed as
\begin{equation}
  \mathbf{X} = \sumlm{} \mathcal{A}^\ell_m\Rlm{} + \mathcal{B}^\ell_m\Slm{}
  + \mathcal{C}^\ell_m\Tlm{},
  \label{eq:xSH}
\end{equation}
where ${A}^\ell_m$, ${B}^\ell_m$ and ${C}^\ell_m$ are the decomposition coefficient of $X$ on vector of the RST basis.

\subsection{Operators}
The RST basis can then be used to write the various mathematical operators on a scalar field\\ $\psi=\sumlm{}\psi_m^lY_l^m$, such as
\begin{eqnarray}
  \label{eq:gradient}
  \grad\psi&=& \sumlm{} \dr\psi_m^l\Rlm{} + \frac{\psi_m^l}{r}\Slm{}\\
  \label{eq:lapla}
  \Div\grad\psi &=& \sumlm{} \Delta_l\psi_m^lY_l^m
\end{eqnarray}
where $\Delta_l=\drr + \frac{2}{r}\dr-\frac{l(l+1)}{r^2}$. Now for a vector $\mathbf{X}=\sumlm{}\mathcal{A}^l_m\Rlm{} + \mathcal{B}^l_m\Slm{} + \mathcal{C}^l_m\Tlm{}$, we obtain
\begin{eqnarray}
    \label{eq:divergence}
    \Div\mathbf{X} &=&
    \sumlm{}\left[\frac{1}{r^2}\dr(r^2\MAlm{})-l(l+1)\frac{\MBlm{}}{r}\right]Y_l^m \\
    \label{eq:curl}
    \rota\mathbf{X} &=&
    \sumlm{}\left[l(l+1)\frac{\MClm{}}{r}\right]\Rlm{} +
    \left[\frac{1}{r}\dr(r\MClm{})\right]\Slm{}  +
    \left[\frac{\MAlm{}}{r}-\frac{1}{r}\dr(r\MBlm{})\right]\Tlm{} \\
    \label{eq:lapla_vect}
    \nabla^2\mathbf{X} &=&
    \sumlm{}\left[\Delta_l\MAlm{}-\frac{2}{r^2}(\MAlm{}-l(l+1)\MBlm{})\right]\Rlm{} +
    \left[\Delta_l\MBlm{}+2\frac{\MAlm{}}{r^2}\right]\Slm{}  + \left[\Delta_l\MClm{}\right]\Tlm{}.
\end{eqnarray}
We can explain here the general procedure for calculating an integral over these quantities, by illustrating the example of computing the surface integral $\int_{\partial V} \nab\cdot\mathbf{X}\dint{\Omega}{}$, such that
\begin{eqnarray}
  \int_{\partial V} \nab \cdot \mathbf{X}\dint{\Omega}{} &=& \sumlm{}
  \int_{\partial V} \left[\frac{1}{r^2}\dr(r^2\MAlm{})-\ell(\ell+1)\frac{\MBlm{}}{r}\right]Y_\ell^m \dint{\Omega}{}\\
  &=& \sumlm{} \int_{\partial V} \frac{1}{r^2}\dr(r^2\MAlm{}) Y_\ell^m
  \underbrace{\sqrt{4\pi}Y_0^0}_{=1} \dint{\Omega}{} \\
  &=& \sqrt{4\pi} \sumlm{} \frac{1}{r^2}\dr(r^2\MAlm{}) \int_{\partial
    V} Y_\ell^m Y_0^0\dint{\Omega}{} \\
  &=& \sqrt{4\pi}\frac{1}{r^2}\dr(r^2\mathcal{A}^0_0).
\end{eqnarray}
Using these tools, it is then possible to fully develop the spectral decomposition of the equations~\ref{eq:KEanaSH} and \ref{eq:S2anaSH} as follows.

\subsection{Kinetic energy evolution}
It is first possible to decompose spatially the evolution of the kinetic energy shown in Equation~\ref{eq:KEanaSH} by considering the spectrum of the velocity vector field $\vv_\ell$ over the RST basis. The equation for the evolution of radial kinetic energy density $E_\ell^K$ is then obtained by projecting the Navier-Stokes equation~\ref{eq:ASHeq_v} on the RST basis, multiplying it by $\vv_\ell$, and then integrating over spherical surface $S$ at a given radius $r$, such as
\begin{equation}
  \p_t E_\ell^K = \sum\limits_{\substack{\ell_1,\ell_2\\|\ell_1-\ell_2|\leq \ell \leq \ell_1+\ell_2}} \big[{\cal R}_\ell(\ell_1,\ell_2)\big] + {\cal C}_\ell(\ell-1,\ell+1) + {\cal H}_\ell + {\cal B}_\ell  + {\cal V}_\ell,
\end{equation}
where the various terms on the right-hand side are respectively the contribution of non-linear advection via the Reynolds tensor, the impact of the Coriolis force, the pressure gradient, the buoyancy force, and the viscous contribution, defined as follows
\begin{equation}
  \label{eq:advecSH}
  {\cal R}_\ell(r,\ell_1,\ell_2)=\bar{\rho}\int_S[  \vv_{\ell_1}\cdot \nab (\vv_{\ell_2})]_\ell\cdot\vv_\ell{\rm d}\Omega,
\end{equation}
\begin{equation}
  \label{eq:corioSH}
  {\cal C}_\ell(r,\ell-1,\ell+1)=-2\bar{\rho}\int_S(\oom_0\times\vv_{\ell-1}+\oom_0\times\vv_{\ell+1)})\cdot\vv_\ell{\rm d}\Omega,
\end{equation}
\begin{equation}
  \label{eq:gradpSH}
  {\cal H}_\ell(r)=-(1-\delta_{\ell,0})\int_S\nab P_\ell\cdot\vv_\ell{\rm d}\Omega,
\end{equation}
\begin{equation}
  \label{eq:buoyaSH}
  {\cal B}_\ell(r)=(1-\delta_{\ell,0})\int_S\rho {\bf g}_\ell\cdot\vv_\ell{\rm d}\Omega,
\end{equation}
\begin{equation}
  \label{eq:viscSH}
  {\cal V}_\ell(r)=-\int_S(\nab\cdot{\cal D})_\ell\cdot\vv_\ell{\rm d}\Omega.
\end{equation}
We refer the reader to Appendixes of \cite{strugarekMAGNETICENERGYCASCADE2013} and \cite{strugarekModelingTurbulentStellar2016} for the detailed development of each term. In Section~\ref{sec:DRscales}, we consider a specific part of the Reynolds tensor \ref{eq:advecSH}, defined such as

\begin{equation}
  \label{eq:R30}
  \overline{ {\cal R}_{\ell=3}^{m=0}}(\ell_{1,2}) = \frac{1}{(2\ell +1){\cal R}_\ell^{m>0}}\sum\limits_{\substack{\ell_{2,1}\\|\ell_1-\ell_2|\leq \ell \leq \ell_1+\ell_2\\m_1+m_2=0 ; m_1>0, m_2>0}} |{\cal R}_{\ell=3}(\ell_1,\ell_2)|,
\end{equation}

\begin{equation}
  \label{eq:R302}
  \text{where}\;{\cal R}_\ell^{m>0} = \sum\limits_{\substack{\ell_1,\ell_2\\|\ell_1-\ell_2|\leq \ell \leq \ell_1+\ell_2\\m_1+m_2=0 ; m_1>0, m_2>0}} {\cal R}_{\ell=3}(\ell_1,\ell_2).
\end{equation}
This term quantifies mean energy transfers between non-axisymmetric component of convective motions correlations and the $\ell=3$ axisymmetric component of large-scale flows (here the differential rotation.)

\subsection{Squared entropy evolution}
A similar development can be done for the radial density of quadratic entropy ${\cal E}_\ell^S$ evolution shown in Equation~\ref{eq:S2anaSH}, by considering this time the spectral decomposition of the internal energy equation, multiplied by the spectrum of the entropy scalar field $S_\ell$ and integrated over the spherical surface $S$. This then yields to the following budget
\begin{equation}
  \p_t {\cal E}_\ell^S = \sum\limits_{\substack{\ell_1,\ell_2\\|\ell_1-\ell_2|\leq \ell \leq \ell_1+\ell_2}} \big[{\cal A}_\ell(\ell_1,\ell_2)\big] + {\cal S}_\ell^b + {\cal T}_\ell + {\cal K}_\ell + {\cal V}^S_\ell  + {\cal Q}_0\delta_{\ell,0} ,
\end{equation}
where the various terms on the right-hand side are respectively the contribution of entropy advection, the advective impact of the reference state $\sba$, radiative transfer, thermal dissipation, viscous dissipation, and a spherically symmetric contribution. These terms are defined such as
\begin{equation}
  \label{eq:SadvecSH}
  {\cal A}_\ell(r,\ell_1,\ell_2)=\int_S -  \left[ \vv_{\ell_1}\cdot \nab (s_{\ell_2}) \right]_\ell \cdot s_\ell{\rm d}\Omega ,
\end{equation}
\begin{equation}
  \label{eq:SadvecBSH}
  {\cal S}_\ell^b = \int_S - \vv_\ell \cdot \nab (\bar{s}) \cdot s_\ell{\rm d}\Omega ,
\end{equation}
\begin{equation}
  \label{eq:SradSH}
  {\cal T}_\ell(r)=\frac{1}{\rb\bar{T}}\int_S\nab\cdot \left[ \kappa_{\rm rad}\bar{\rho}c_P\nab(T_\ell) \right] \cdot s_\ell{\rm d}\Omega ,
\end{equation}
\begin{equation}
  \label{eq:SdissSH}
  {\cal K}_\ell(r)=\frac{1}{\rb\bar{T}}\int_S\nab\cdot \left[\kappa\bar{\rho}\bar{T}\nab (s_\ell) \right] \cdot s_\ell {\rm d}\Omega,
\end{equation}
\begin{equation}
  \label{eq:SviscSH}
  {\cal V}^S_\ell(r)=\frac{2\nu}{\bar{T}}\int_S \left[ e_{ij}e_{ij}-\frac{1}{3}(\nab\cdot\vv)^2 \right]_\ell \cdot s_\ell {\rm d}\Omega,
\end{equation}
\begin{equation}
  \label{eq:S0SH}
  {\cal Q}_0(r) = \frac{\sqrt{4\pi}}{\rb\bar{T}} \left(\nab\cdot \left[\kappa_{\rm rad}\bar{\rho}c_P\nab(\bar{T}) + \kappa_{0}\bar{\rho}\bar{T}\nab (\bar{s}+\langle s\rangle) \right] \cdot s_0 + \bar{\rho}\epsilon \cdot s_0\right),
\end{equation}
where $\langle s\rangle=s_{0,0}$ is the spherical average of the specific entropy perturbation ($\ell=0$, $m=0$).

\section{Computing the critical Rayleigh number}\label{sec:app_critic}

Numerical experiments have been able to confirm key results from analytical works about rotating convection \citep{1961hhs..book.....C} along the last decades (starting with \citealt{robertsThermalInstabilityRotatingfluid1968}, \citealt{busseThermalInstabilitiesRapidly1970}, \citealt{1975JAtS...32.1331G,1977GApFD...8...93G}). In particular, it has been shown that characteristics of the convective instability, such as the temporal and spatial frequency of the most unstable mode, as long as critical states, depend on the Taylor number $Ta$, quantifying the amplitude of the rotational constraint by the Coriolis force over the one of the viscous dissipation (see \citealt{jonesLinearTheoryCompressible2009} and reference therein). In particular, \cite{takehiroAssessmentCriticalConvection2020} recently confirmed that the critical Rayleigh number scales such that $Ra_{\rm c}\propto Ta^{2/3}$ for different anelastic numerical experiments \citep{jonesLinearTheoryCompressible2009,brunDifferentialRotationOvershooting2017}. This critical $Ra_c$ value characterizes the critical state of the convective instability, which we aim at quantifying for our models presented here.

Taking results from \cite{takehiroAssessmentCriticalConvection2020}, we compute the following linear regression $Ra_{\rm c} = Ra_{{\rm c,B17},\odot}  (\Omega_*/\Omega_\odot)^{1.74 \pm 0.19} (M_*/M_\odot)^{-4.39 \pm 0.73}$, which yields $Ra_{{\rm c},\odot} = (2.8 \pm 0.9)\times 10^4$ for a solar rotation rate and mass in \cite{brunDifferentialRotationOvershooting2017} (B17) experiment. Similarly, we find a linear regression for the Taylor number $Ta = Ta_{{\rm B17},\odot} (\Omega_*/\Omega_\odot)^{2.94 \pm 0.07} (M_*/M_\odot)^{-7.54 \pm 0.26}$, giving $Ta_{{\rm B17},\odot} = (1.7 \pm 0.2 )\times 10^5$. Now using the scaling $Ra_{\rm c}\propto Ta^{2/3}$ to decipher $Ra_{{\rm c},\odot}$ values of our models, in which we decreased viscosity to increase the turbulence degrees and hence increased Taylor numbers, we finally find $Ra_{{\rm c}}=2.5\times 10^5$, $9.9\times 10^6$ and $2.5\times 10^6$ for AS1, SBR97n035 and SBR50n1 respectively (see Table~\ref{tab:AB2vsAS1vsSunBusyAdim}).

\section{Angular momentum evolution and transport}\label{sec:AmomBal}

\begin{figure*}
  \centering
  \includegraphics[width=\linewidth]{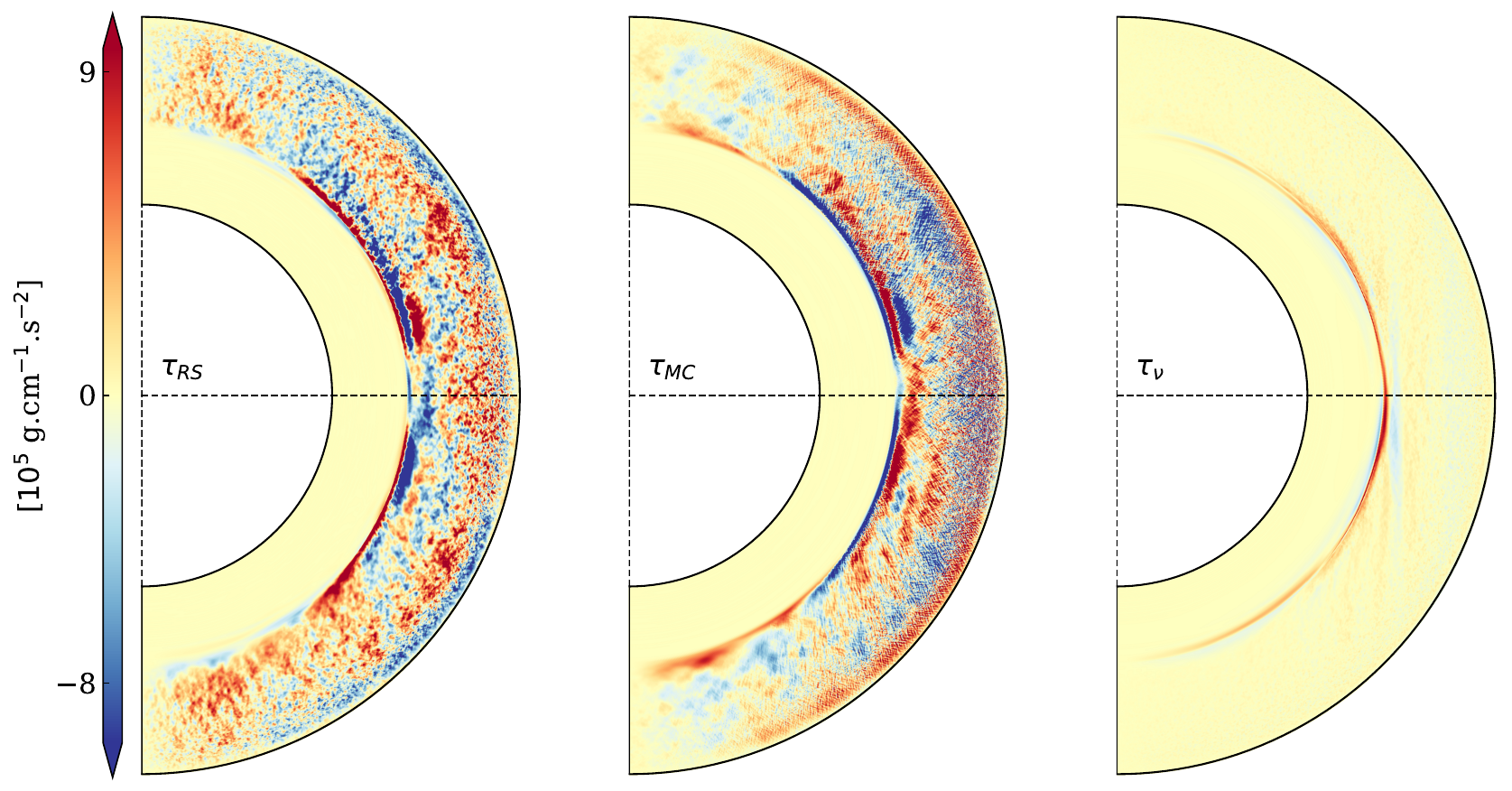}
  \includegraphics[width=\linewidth]{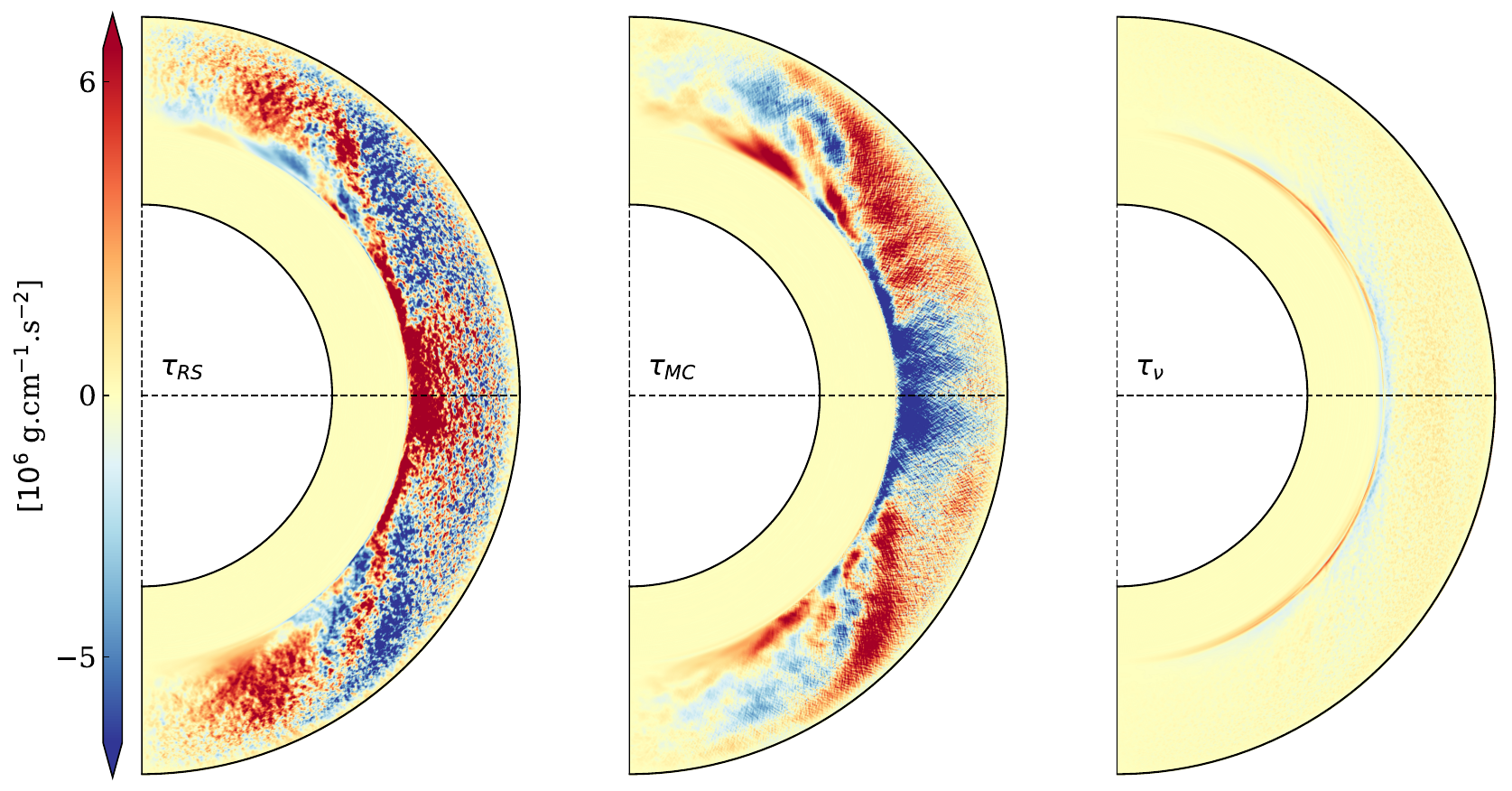}
  \caption{ Angular momentum evolution in the solar-rotating case (SBR97n035, top) and the anti-solar case (SBR50n1, bottom). The red (respect. blue) color means that angular momentum is deposited/increased (respect. extracted/decreased) locally. The left, middle and right column show the contribution from the Reynolds stress, meridional flow and viscosity, respectively.}\label{fig:AmomBal}
\end{figure*}

In order to understand processes that redistribute angular momentum in the simulation, and more especially in the bulk of the CZ, we can compute the evolution of the angular momentum. Following previous studies (see \textit{e.g.} \citealt{Rüdiger+1989,elliottTurbulentSolarConvection2000,brunGlobalScaleTurbulent2004,kapylaReynoldsStressHeat2011,hottaGenerationSolarlikeDifferential2022}), we can express the different contributions by averaging the longitudinal component of the momentum Equation \ref{eq:ASHeq_v} in time and longitude ($\langle\rangle$)
\begin{equation}
  \label{eq:AmomBal}
  \bar{\rho}\frac{\partial {\cal L}}{\partial t} = \tau_{\rm RS} + \tau_{\rm MC} + \tau_\nu,
\end{equation}
with
\begin{equation}
  \label{eq:TRS}
  \tau_{\rm RS} = -\nab\cdot (\bar{\rho} \lambda\langle \mathbf{v}_mv_\phi\rangle), \nonumber
\end{equation}
\begin{equation}
  \label{eq:TMC}
  \tau_{\rm MC} = -\nab\cdot (\bar{\rho} \langle\mathbf{v}_m\rangle {\cal L}), \nonumber
\end{equation}
\begin{equation}
  \label{eq:Tnu}
  \tau_\nu = -\nab\cdot (-\bar{\rho}\nu \lambda^2\nab\Omega) , \nonumber
\end{equation}
where ${\cal L}=\lambda (\langle v_\phi\rangle + \lambda\Omega_\odot)$ is the specific angular momentum, $\Omega=\Omega_\odot+\lambda^{-1}\langle v_\phi\rangle$ is the total angular velocity and $\lambda=r\sin\theta$. Then $\tau_{\rm RS}$, $\tau_{\rm RS}$ and $\tau_{\nu}$ are the Reynolds stress, meridional flow and viscous contribution to angular momentum transport respectively. We illustrate them in Figure \ref{fig:AmomBal} for both SBR97n035 (top row) and SBR50n1 (bottom row).

In both models, we can see fine structures in spatial fluctuations along with a negligible contribution of the viscosity ($\tau_\nu$, right column) due to the highly turbulent nature of the flow. The mirror trend between the Reynolds stress from the turbulence ($\tau_{\rm RS}$, left column) and the component from the meridional flow ($\tau_{\rm MC}$, middle column) then indicate that a stationary state has been reached ($\bar{\rho}\frac{\partial {\cal L}}{\partial t}=\sum\tau\simeq 0$). Our choice of anelastic approximation further imply that we experience a low Mach number regime. This in turn ensures that ${\bf \nabla.(\bar{\rho}.\mathbf{v}_m)\sim 0}$ is approximately satisfied, which finally leads to the following equation, known as gyroscopic pumping \citep{mieschGYROSCOPICPUMPINGSOLAR2011,featherstoneMeridionalCirculationSolar2015}
\begin{equation}
  \label{eq:GyroPump}
  \bar{\rho}\langle \mathbf{v}_m\rangle\cdot\nab {\cal L} \sim \tau_{\rm RS} + \tau_\nu,
\end{equation}
Gyroscopic pumping is a way of understanding which mechanism is sustaining the meridional flow. Hence, this indicates that the Reynolds stress sustains the meridional circulation, as all other mechanisms (here the viscous one) are negligible.

As we choose stress-free and torque free boundary conditions, no external torque is applied to the system and the angular momentum is globally conserved. Looking at the expression of the different torques $\tau$ from Equation \ref{eq:AmomBal}, we see that they can be directly interpreted as the local deposition and extraction of angular momentum from a corresponding flux. In Figure \ref{fig:AmomBal}, the flux therefore extracts it from blue regions and deposits it in red regions. In the anti-solar case (bottom row), the Reynolds stress from turbulence (left column) transfers the angular momentum from regions close to the surface, as well as mid-latitude CZ-bulk, towards the polar regions and the BCZ at the equator, leading then to the anti-solar profile. In the solar-rotating case, it extracts angular momentum from regions close to the surface and deposits it in the bulk of the CZ and the equatorial part, helping to sustain the solar-like prograde equator of SBR97n035. Interestingly, we see a noticeable deposition of angular momentum (red) from $\tau_{\rm RS}$ in polar regions, balanced subsequently by the meridional flow extracting it (blue in $\tau_{\rm MC}$). This means that it's the meridional circulation that helps to maintain the retrograde rotation of SBR97n035 poles.

\section{Polar dynamics and middle of the convection zone}

\begin{figure*}
  \centering
  \includegraphics[width=\linewidth]{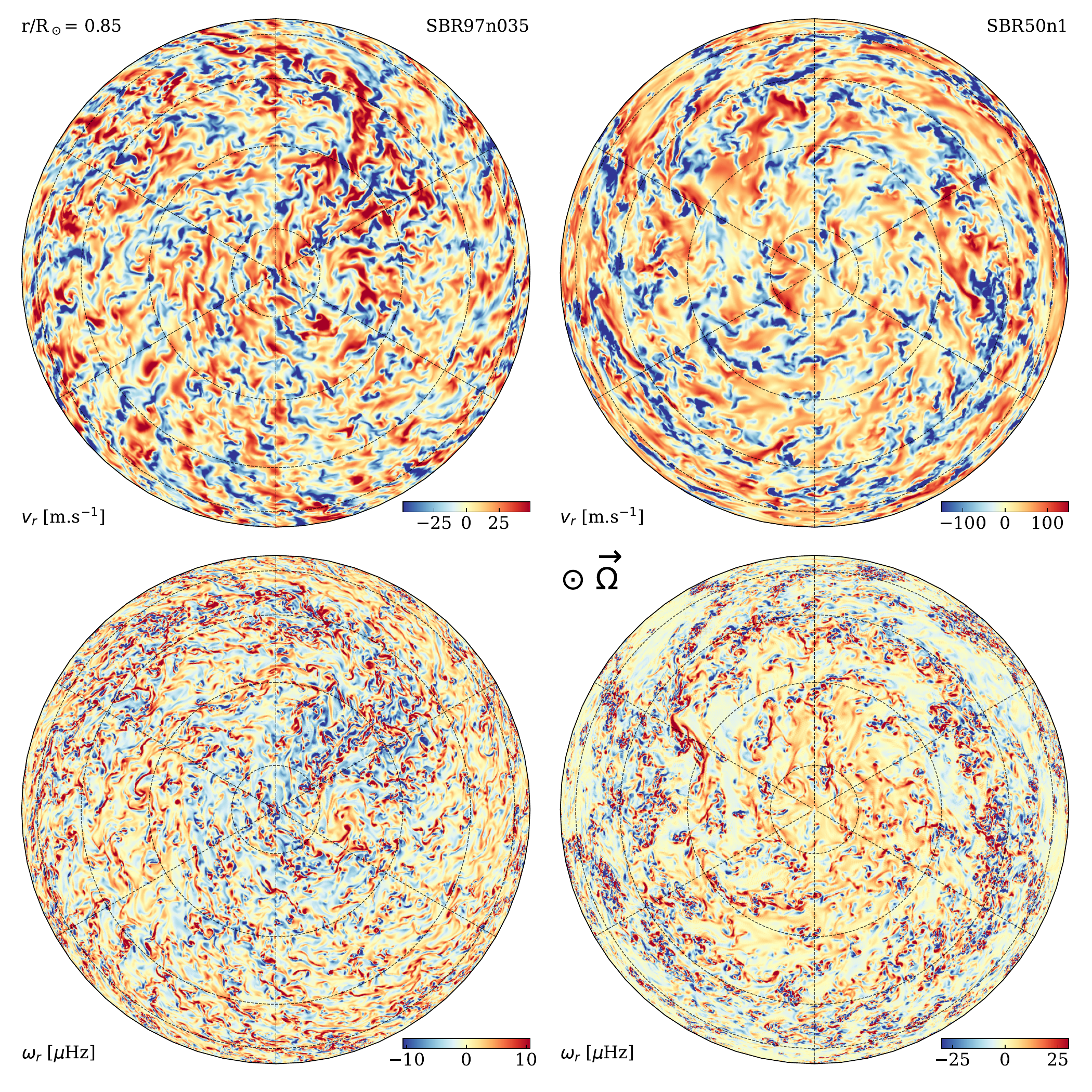}
  \caption{Similar to Figures~\ref{fig:PolarView} and \ref{fig:PolarView3}, North-Pole view, looking now at the middle of the convection zone $r=0.85$ R$_\odot$.
   Animations are available at\dataset[doi:10.5281/zenodo.14650437]{https://doi.org/10.5281/zenodo.14650437} \citep{norazCcZenodo25}.}\label{fig:PolarView2}
\end{figure*}

\begin{figure*}
  \centering
  \includegraphics[width=\linewidth]{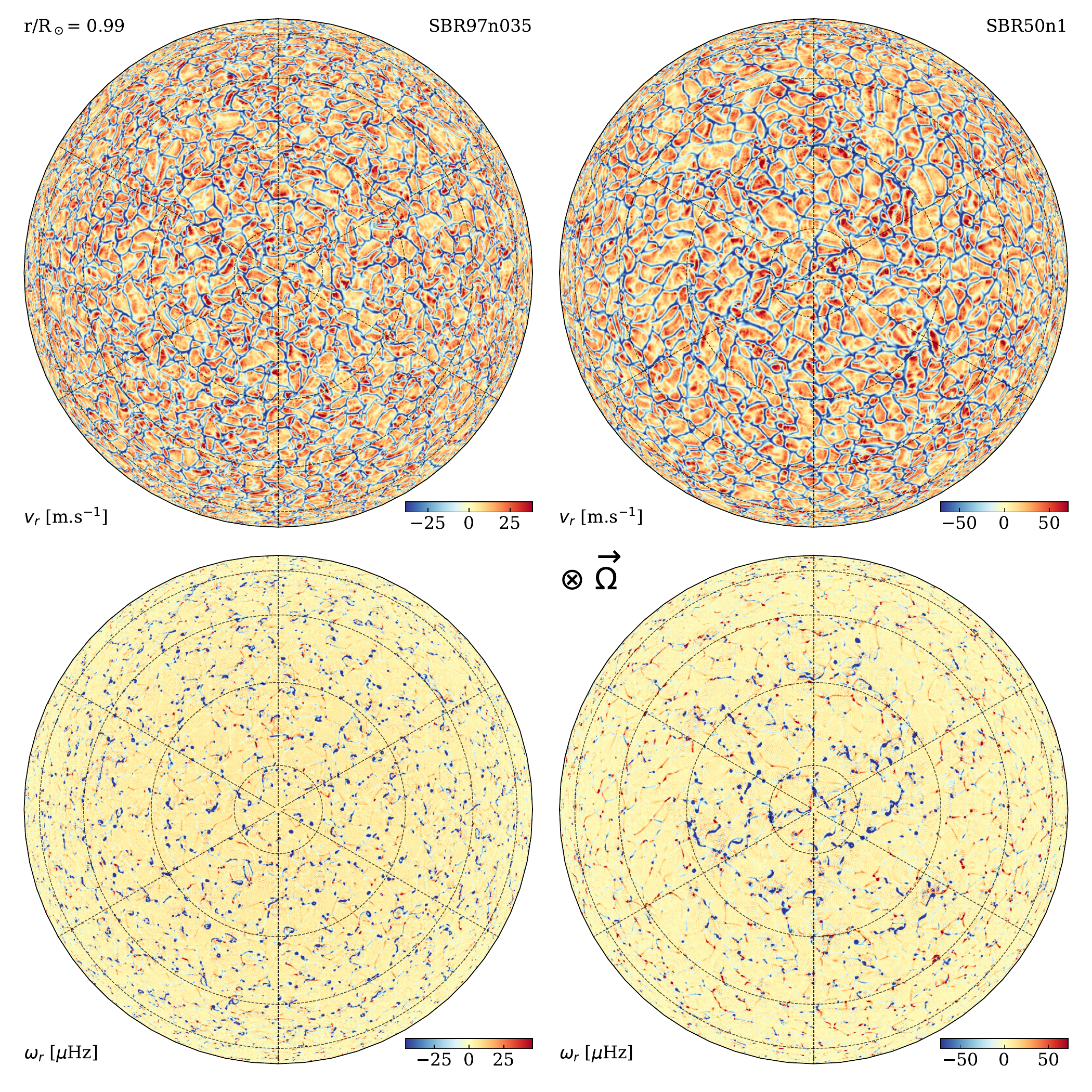}
  \caption{South-Pole view of the near-surface dynamics ($0.99$ R$_\odot$, 10 grid-points below the top) in SBR97n035 (left panels) and SBR50n1 (right panels) models. Similarly to Figure~\ref{fig:PolarView}, we show maps of the radial velocity $v_r$ and radial vorticity $\omega_r$ on top and bottom panels, respectively. The maximum value of a color bar corresponds to twice the standard deviation of the map it corresponds to. Please note that all observations made in Section~\ref{sec:PolDyn} also apply to this Figure, with the understanding that "anti-clockwise" and "red" mentions should be reversed to "clockwise" and "blue," respectively, due to the sign inversion of the Coriolis force when transitioning from one hemisphere to the other. Animations are available at\dataset[doi:10.5281/zenodo.14650437]{https://doi.org/10.5281/zenodo.14650437} \citep{norazCcZenodo25}.}\label{fig:PolarView3}
\end{figure*}

\begin{figure*}
  \centering
  \includegraphics[width=\linewidth]{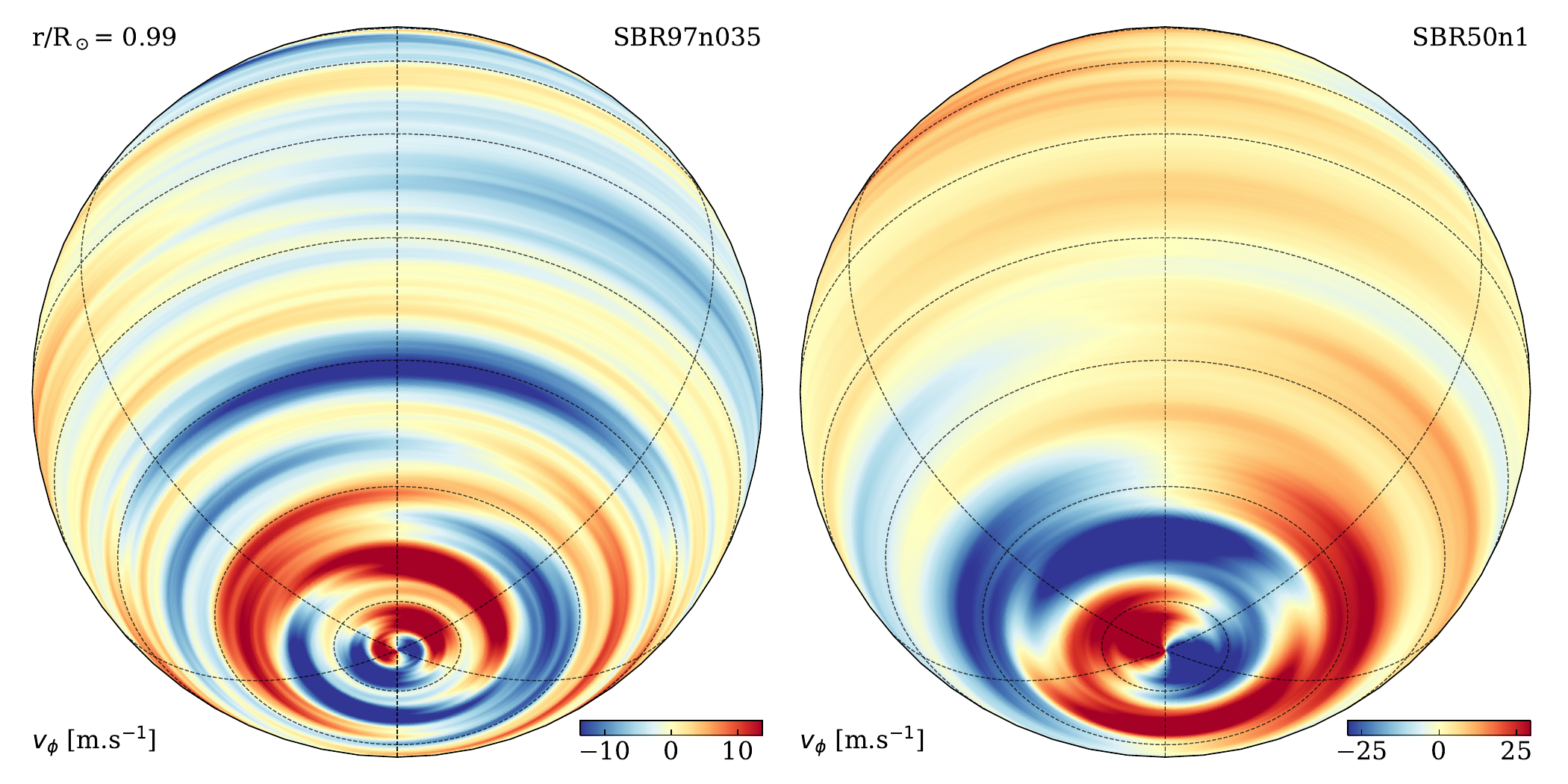}
  \caption{Similar to Figure~\ref{fig:Vphim1}, from the South-Pole point of view. The color bar maximum corresponds to two standard deviations of the given map. Although there is not a clear global symmetry with respect to the Northern Hemisphere, we see here that at least the direction of the spiral starting from the South-pole is symmetric in comparison to the North-pole one in both models (see Figure~\ref{fig:Vphim1}). 
  }
  \label{fig:Vphim1South}
\end{figure*}

As a complement of the analysis in Sections~\ref{sec:PolDyn} and \ref{sec:sol_comp}, we illustrate in Figure~\ref{fig:PolarView2} the polar dynamics from the North-Pole perspective and at the middle of the CZ ($0.85$~R$_\odot$), and show in Figures~\ref{fig:PolarView3} and \ref{fig:Vphim1South} the polar dynamics at the near-surface ($0.99$~R$_\odot$) from the South-Pole perspective. All Figures illustrate the Sun-like rotating model SBR97n035 in left panels, and the anti-solar rotating one SBR50n1 in right panels. Figures~\ref{fig:PolarView3} and \ref{fig:PolarView2} show radial velocity and radial vorticity maps in top and bottom panels, respectively. Figure~\ref{fig:Vphim1South} illustrates the $m=1$ mode extracted from $v_\phi$ velocity fields. 

\bibliography{CCpap}
\bibliographystyle{aasjournal}



\end{document}